\documentclass[british,final,format=manuscript,screen=true,natbib=true,authorversion=true,nonacm=true]{acmart}
\pdfoutput=1
\usepackage{xparse}
\usepackage{etoolbox}
\usepackage{environ}
\usepackage{graphicx}

\makeatletter
\newcommand{\IfLabelExistsTF}[3]{\@ifundefined{r@#1}{#3}{#2}}
\newcommand{\IfLabelExistsT}[2]{\IfLabelExistsTF{#1}{#2}{}}
\newcommand{\IfLabelExistsF}[2]{\IfLabelExistsTF{#1}{}{#2}}
\newcommand{\IfDefTF}[3]{\@ifundefined{#1}{#3}{#2}}
\newcommand{\IfDefT}[2]{\IfDefTF{#1}{#2}{}}
\newcommand{\IfDefF}[2]{\IfDefTF{#1}{}{#2}}
\makeatother

\def\proofltsskipamount{1.7\smallskipamount}
\def\termltsskipamount{1\smallskipamount}
\providecommand\rulescalebox[2][]{#2} 

\usepackage{subfiles}
\usepackage{xr}
\usepackage{currfile}

\ifdefined\docroot\else
	\let\docroot\currfilepath
\fi

\newcommand\IfDocRootTF[2]{%
	\ifcurrfilepath{\docroot}{#1}{#2}%
}
\newcommand\IfDocRootT[1]{\IfDocRootTF{#1}{}}
\newcommand\IfDocRootF[1]{\IfDocRootTF{}{#1}}

\usepackage[utf8]{inputenc}
\usepackage{microtype}

\usepackage{xcolor}

\colorlet{type}{blue!60!black}
\colorlet{process}{red!60!black}

\colorlet{textcolor}{.}
\newcommand\usecolor[2]{\begingroup%
	\colorlet{uc@saved}{.}
	\color{#1}#2\color{uc@saved}
\endgroup}
\newcommand\cboxed[2]{\begingroup%
  \colorlet{cb@saved}{.}%
  \color{#1}%
  \boxed{\color{cb@saved}#2}%
\endgroup}

\usepackage[british]{babel}
\usepackage[all]{foreign}
\defnotforeign[aka]{{\xperiodafter{a.k.a}}}
\defnotforeign[wrt]{{\xperiodafter{w.r.t}}}
\defnotforeign[wl]{{\xperiodafter{w.l.o.g}}}
\defasforeign[mutatismutandis]{mutatis mutandis}
\defasforeign[Mutatismutandis]{Mutatis mutandis}
\defasforeign[criteria]{criteria}
\defasforeign[criterium]{criterium}
\defasforeign[etsimilia]{et similia}
\defasforeign[perse]{per se}
\defasforeign[perdefinitionem]{per definitionem}
\defasforeign[perdef]{\xperiodafter{per def}}

\usepackage{amsmath}
\usepackage{amsthm}
\usepackage{amsfonts}
\usepackage{amssymb}
\usepackage{centernot}
\usepackage{scalerel}
\usepackage{stmaryrd}
\SetSymbolFont{stmry}{bold}{U}{stmry}{m}{n}
\usepackage{mathtools}
\usepackage{esvect}

\allowdisplaybreaks

\def\shuffle{\mathrel{{\sqcup\mspace{-3mu}\sqcup}}}

\makeatletter
\@ifundefined{minaw@}{\newdimen\minaw@}{}
\def\rightarrowtrianglefill@#1{\m@th\setboxz@h{$#1\relbar$}\ht\z@\z@
  $#1\copy\z@\mkern-6mu\cleaders
  \hbox{$#1\mkern-2mu\box\z@\mkern-2mu$}\hfill
  \mkern-6mu\mathord{\smash\rightarrowtriangle\vphantom{\rightarrow}}$}
\newcommand{\xrightarrowtriangle}[2][]{%
  \mathrel{\mathop{%
    \setbox\z@\vbox{\m@th
      \hbox{$\scriptstyle\;{#1}\;\;\,$}%
      \hbox{$\m@th\scriptstyle\;{#2}\;\;\,$}%
    }%
    \hbox to\ifdim\wd\z@>\minaw@\wd\z@\else\minaw@\fi{%
      \rightarrowtrianglefill@\displaystyle}}%
  \limits^{#2}\@ifnotempty{#1}{_{#1}}}%
}
\renewcommand{\xRightarrow}[2][]{\ext@arrow 0359\Rightarrowfill@{#1}{#2}}
\makeatother

\DeclareFontFamily{U}{matha}{\hyphenchar\font45}
\DeclareFontShape{U}{matha}{m}{n}{
     <5> <6> <7> <8> <9> <10> gen * matha
     <10.95> matha10 <12> <14.4> <17.28> <20.74> <24.88> matha12
     }{}
\DeclareSymbolFont{matha}{U}{matha}{m}{n}
\DeclareMathSymbol{\varleftarrow}{3}{matha}{"D0}
\DeclareMathSymbol{\varrightarrow}{3}{matha}{"D1}
\DeclareMathSymbol{\varleftrightarrow}{3}{matha}{"D8}

\def\lclens{{[\llap(}}
\def\rclens{{\rlap)]}}

\DeclareFontEncoding{LS2}{}{\noaccents@}
\DeclareFontSubstitution{LS2}{stix}{m}{n}
\DeclareSymbolFont{stix@largesymbols}{LS2}{stixex}{m}{n}
\SetSymbolFont{stix@largesymbols}{bold}{LS2}{stixex}{b}{n}
\DeclareMathDelimiter{\lBrace}{\mathopen} {stix@largesymbols}{"E8}%
                                          {stix@largesymbols}{"0E}
\DeclareMathDelimiter{\rBrace}{\mathclose}{stix@largesymbols}{"E9}%
                                          {stix@largesymbols}{"0F}

\makeatletter
\newcommand{\raisemath}[1]{\mathpalette{\raisem@th{#1}}}
\newcommand{\raisem@th}[3]{\raisebox{#1}{$#2#3$}}
\makeatother

\usepackage{array}
\usepackage{xtab}
\usepackage{booktabs}
\usepackage{multicol}
\usepackage{dashrule}

\makeatletter
\def\crulefill#1{\leavevmode\leaders\hrule\@height#1 height 0.7ex depth \dimexpr0.4pt-0.7ex\hfill\kern0pt}
\makeatother

\def\headertext#1{
	\vspace{2pt}
	\crulefill{\lightrulewidth}\ \textsf{\small \emph{#1}}\ \crulefill{\lightrulewidth}
  \\[-2.7ex]%
}

\newcommand*{\footer}[1][1.6\belowdisplayskip]{%
	\rule[#1]{\textwidth}{\lightrulewidth}%
	\vspace{-1.6\smallskipamount}%
}

\usepackage{placeins}

\usepackage[inline]{enumitem}

\usepackage[capitalise,nameinlink,noabbrev]{cleveref}
\makeatletter
\newcommand{\customlabel}[4][0]{%
	\protected@write\@auxout{}{\string\newlabel{#3}{{#4}{\thepage}{#4}{#3}{}}}%
	\protected@write\@auxout{}{\string\newlabel{#3@cref}{{[#2][#1][#1]#4}{\thepage}}}%
}
\newcommand{\phantomlabel}[4][0]{%
	\customlabel[#1]{#2}{#3}{#4}%
	\hypertarget{#3}{}%
}
\newcommand{\crefv}[1]{%
	\begingroup\@cref@compressfalse\@cref@sortfalse\cref{#1}\endgroup%
}
\newcommand{\Crefv}[1]{%
	\begingroup\@cref@compressfalse\@cref@sortfalse\Cref{#1}\endgroup%
}
\newcommand{\crefabbrev}[1]{%
	\begingroup\@cref@abbrevtrue\cref{#1}\endgroup%
}
\makeatother

\AtEndPreamble{\hypersetup{
	final,
}}

\usepackage{proof-dashed}

\newcommand{\rname}[1]{\ensuremath{\textsc{\MakeLowercase{#1}}}}
\newcommand{\rlabel}[2]{{%
	\customlabel{infrule}{#2}{#1}%
	\hypertarget{#2}{#1}}}
\crefname{infrule}{rule}{rules}
\Crefname{infrule}{Rule}{Rules}

\newcommand*\delayedrule[2][]{\rname{${:}\ref*{#2}#1$}}
\newcommand*\selfsyncrule[2][]{\rname{${|}\ref*{#2}#1$}}

\AtEndPreamble{
	\theoremstyle{acmplain}
  
	\newtheorem{fact}[theorem]{Fact}
	\theoremstyle{acmdefinition}
	\newtheorem{remark}[theorem]{Remark}
}

\crefname{proof}{proof of}{proofs of}
\Crefname{proof}{Proof of}{Proofs of}

\makeatletter
\newcounter{proofatend@statemet}

\patchcmd{\@addmarginpar}{\ifodd\c@page}{\ifodd\c@page\@tempcnta\m@ne}{}{}
\reversemarginpar

\newcommand*\fixstatement[2][Proof of]{%
	\expandafter\preto\csname end#2\endcsname{%
		\global\def\proofatend@proofof{#1}%
		\stepcounter{proofatend@statemet}%
		\label{proofatend:\theproofatend@statemet}%
		\if\proofatend@mode\proofatend@postpone%
		\IfLabelExistsTF{proofatend:proof-\theproofatend@statemet}{%
			\marginpar{\hyperref%
				[proofatend:proof-\theproofatend@statemet]%
				{\rotatebox[origin=c]{-90}{$\triangle$}}
			}%
		}{}%
		\fi%
}}

\def\proofatend@toks{1000}
\newcounter{proofatend@stored}
\newcounter{proofatend@printed}

\let\proofatend@omit0
\let\proofatend@inplace1
\let\proofatend@postpone2

\newcommand{\omitproofs}{\global\let\proofatend@mode\proofatend@omit}
\newcommand{\postponeproofs}{\global\let\proofatend@mode\proofatend@postpone}
\newcommand{\keepproofs}{\global\let\proofatend@mode\proofatend@inplace}

\postponeproofs

\NewEnviron{proofatend}[1][Proof]{\ignorespaces%
	\if\proofatend@mode\proofatend@inplace%
		\begin{proof}[#1]\BODY\end{proof}%
	\else\if\proofatend@mode\proofatend@postpone%
		\edef\next{%
			\noexpand\begin{proof}[{\proofatend@proofof~\noexpand\cref{proofatend:\theproofatend@statemet}}]%
			\noexpand\phantomsection%
			\noexpand\label{proofatend:proof-\theproofatend@statemet}%
			\unexpanded\expandafter{\BODY}}%
		\global\toks\numexpr\proofatend@toks+\value{proofatend@stored}\relax=\expandafter{\next\end{proof}}%
		\stepcounter{proofatend@stored}%
	\fi\fi\ignorespacesafterend}

\NewDocumentCommand{\omittedproofs}{o}{%
	\ifnum\value{proofatend@printed}<\value{proofatend@stored}%
		\IfNoValueF{#1}{#1}%
		\count@=\value{proofatend@printed}%
		\loop%
			\the\toks\numexpr\proofatend@toks+\count@\relax%
			\stepcounter{proofatend@printed}%
			\ifnum\count@<\value{proofatend@stored}%
				\advance\count@\@ne%
		\repeat%
	\fi%
}
\makeatother

\AtEndPreamble{
	\fixstatement{theorem}
	\fixstatement{proposition}
	\fixstatement{corollary}
	\fixstatement{lemma}
	\fixstatement{fact}
  \fixstatement[Solution to]{exercise}
}

\newcommand*{\CP}{{\upshape CP}\xspace}
\newcommand*{\HCP}{{\upshape HCP}\xspace}

\newcommand*{\CLL}{{\upshape CLL}\xspace}
\newcommand*{\lts}{lts\xspace}

\DeclareMathOperator{\cn}{cn} 
\DeclareMathOperator{\fn}{fn} 
\DeclareMathOperator{\bn}{bn} 
\DeclareMathOperator{\np}{np} 

\makeatletter
\DeclareMathOperator{\bigtensor}{{\textstyle\bigotimes}}
\DeclareMathOperator{\bigparr}{\scalerel*{\parr}{\bigtensor}}
\makeatother

\def\crel#1{\mathrel{\mathcal{#1}}}

\newcommand*\separate[3]{#2 \ast_{#1} #3}
\newcommand*\together[3]{#2 \circledast_{#1} #3}

\let\defeq\triangleq

\edef\aleq{\mathbin{=_{\alpha}}}

\let\bis\approx
\let\sbis\sim
\let\bcong\approxeq

\let\deneq\bumpeq

\let\reducesto\rightarrow

\makeatletter
\def\lto#1{\xrightarrow{\lbox{#1}}} 
\def\slto#1{\xRightarrow{\lbox{#1}}} 
\def\notlto#1{\centernot{\xrightarrow{\lbox{#1}}}}
\def\notslto#1{\centernot{\xRightarrow{\lbox{#1}}}}
\newcommand\lbox[2][1em]{
	\sbox0{$\scriptstyle #2$}
	\ifdim \wd0 < #1%
		\mathmakebox[#1]{\usebox0}%
	\else%
		\mathmakebox[\wd0]{\usebox0}%
	\fi%
}

\newcommand{\plto}{%
	\@ifstar{\plto@v\downarrow}{\plto@h\lto}%
}
\newcommand{\pslto}{%
	\@ifstar{\plto@v\Downarrow}{\plto@h\slto}%
}
\def\plto@b#1{\cboxed{black!20}{\mspace{-10mu}\array{c}#1\endarray\mspace{-10mu}}}
\def\plto@h#1#2#3#4{%
	\plto@b{#2}%
	#1{#3}%
	\plto@b{#4}%
}
\def\plto@v#1#2#3#4{%
	\array[b]{l}%
		\plto@b{#2}%
		\\%
		\quad%
		\cboxed{white}{\scriptstyle\left#1\cboxed{white}{\scriptstyle#3}\right.}\\%
		\plto@b{#4}%
	\endarray%
}
\newcommand{\notplto}[2]{%
	\plto@b{#1} \notlto{#2}%
}
\makeatother

\newcommand\defbarb[3][0pt]{%
	\expandafter\newcommand\csname#2\endcsname[2][]{{#3^{##1}_{\raisemath{#1}{##2}}}}
}
\defbarb{barb}{{\Downarrow}}
\defbarb{notbarb}{{\centernot\Downarrow}} 
\defbarb{sbarb}{{\downarrow}}
\defbarb{notsbarb}{{\centernot\downarrow}} 
\def\expbarb#1{\barb[#1]}

\let\useleaf\diamond
\let\displeaf\dagger
\def\tnode#1{\langle#1\rangle}
\newcommand*\usenode[1][\useleaf]{\tnode{#1}}
\def\dispnode{\tnode{\displeaf}}
\def\dupnode#1#2{\tnode{#1,#2}}

\newcommand{\dendom}[1]{\lBrace#1\rBrace}
\newcommand{\densem}[1]{\left\llbracket{#1}\right\rrbracket}
\let\denout\triangleleft
\let\denin\triangleright
\newcommand{\antider}[3][]{{{\textstyle\int^{#1}_{#2}}{#3}}}
\newcommand{\der}[3][]{{\textstyle{\mathrm{d}}^{#1}_{#2}{#3}}}

\def\tdot{\mathbin{\mspace{-2mu}{\cdot}\mspace{-2mu}}}

\def\iddom{\mathbb{N}}
\def\wpf{\wp_f}
\def\ids{\mathop{\mathrm{ids}}}
\def\id{\mathop{\mathrm{id}}}
\def\dendual{\mathrel{\triangleleft\mspace{-1mu}\triangleright}}
\def\deldep#1#2{#1{\setminus}{#2}}

\let\phl\relax 
\let\thl\relax 
\let\nohl\relax 
\newcommand{\hl}[1]{\highlight#1\endhighlight}
\newenvironment{highlight}
	{\ignorespaces%
	 \def\phl##1{\usecolor{process}{##1}}%
	 \def\thl##1{\usecolor{type}{##1}}%
	 \def\nohl##1{\usecolor{textcolor}{##1}}}%
	{\let\phl\relax%
	 \let\thl\relax%
	 \let\nohl\relax%
	 \ignorespacesafterend}

\newcommand*{\bang}{\mathord{!}}
\newcommand*{\query}{\mathord{?}}
\newcommand*{\tensor}{\otimes}
\newcommand*{\parr}{\mathbin{\bindnasrepma}}
\newcommand*{\with}{\mathbin{\binampersand}}
\newcommand*{\one}{1}

\newcommand*{\dual}[1]{#1^{\bot}}

\newcommand*{\cht}[2]{\phl{#1}\colon\mspace{-2mu}\thl{#2}} 
\newcommand*{\emptyseq}{\bullet}
\newcommand*{\emptyhyp}{\varnothing}
\newcommand*{\hyp}[1]{\mathcal{#1}}

\let\seq\vdash
\newcommand{\judge}[2]{\phl{#1} \vdash \thl{#2}}

\newcommand{\pjudge}[2]{\phl{#1} \Vdash \thl{#2}}

\newcommand*{\strip}[1]{\left\lfloor{#1}\right\rfloor}

\newcommand*{\minl}{\mathsf{i\mspace{-.7mu}n\mspace{-.7mu}l}}
\newcommand*{\minr}{\mathsf{i\mspace{-.7mu}n\mspace{-.7mu}r}}
\newcommand*{\sends}[2]{{#1[#2]}}
\newcommand*{\recvs}[2]{{#1(#2)}}
\newcommand*{\closes}[1]{\sends{#1}{}}
\newcommand*{\waits}[1]{\recvs{#1}{}}
\newcommand*\fixspace[2][2mu]{\mspace{#1}{#2}\mspace{#1}}
\newcommand*{\inls}[1]{{#1}\fixspace{\triangleleft}{\minl}}
\newcommand*{\inrs}[1]{{#1}\fixspace{\triangleleft}{\minr}}
\newcommand*{\coinls}[1]{{#1}\fixspace{\triangleright}{\minl}}
\newcommand*{\coinrs}[1]{{#1}\fixspace{\triangleright}{\minr}}
\newcommand*{\clientuses}[2] {\query\sends{#1}{#2}}
\newcommand*{\serveruses}[2] {\bang \recvs{#1}{#2}}
\newcommand*{\clientdisps}[1]{\query\sends{#1}{}}
\newcommand*{\serverdisps}[1]{\bang \recvs{#1}{}}
\newcommand*{\clientdups}[3] {\query\sends{#1}{#2,#3}}
\newcommand*{\serverdups}[3] {\bang \recvs{#1}{#2,#3}}
\newcommand*{\forwardop}{{\varleftrightarrow}}
\newcommand*{\forwards}[3][]{{#2 \forwardop^{#1} #3}}
\newcommand{\nil}{\boldsymbol{0}}
\let\forward\forwards
\newcommand*{\pp}{\mathbin{\boldsymbol{|}}}

\newcommand*{\res}[2]{{(\boldsymbol{\nu}{#1})\mspace{2mu}}{#2}}
\newcommand*{\prefixed}[2]{{#1.#2}}
\newcommand*{\send}[3]{\prefixed{\sends{#1}{#2}}{#3}}
\newcommand*{\recv}[3]{\prefixed{\recvs{#1}{#2}}{#3}}
\newcommand*{\close}[2]{\prefixed{\closes{#1}}{#2}}
\newcommand*{\wait}[2]{\prefixed{\waits{#1}}{#2}}
\newcommand*{\inl}[2]{\prefixed{\inls{#1}}{#2}}
\newcommand*{\inr}[2]{\prefixed{\inrs{#1}}{#2}}
\NewDocumentCommand{\choice}{s m m m}{
	\IfBooleanTF{#1}
	{{#2}\fixspace{\triangleright}\left\{\mspace{-10mu}\array{rl}%
		\minl\colon&\mspace{-15mu}#3\\%
		\minr\colon&\mspace{-15mu}#4%
	\endarray\mspace{-8mu}\right\}}%
	{{#2}\fixspace{\triangleright}\left\{%
		\minl\colon\mspace{-2mu}#3;%
		\minr\colon\mspace{-2mu}#4\right\}}}
\NewDocumentCommand{\choicebit}{s m m m}{
	\IfBooleanTF{#1}
	{{#2}\fixspace{\triangleright}\left\{\mspace{-10mu}\array{rl}%
		0\mapsto&\mspace{-15mu}#3\\%
		1\mapsto&\mspace{-15mu}#4%
	\endarray\mspace{-8mu}\right\}}%
	{{#2}\fixspace{\triangleright}\left\{%
		0\mapsto\mspace{-2mu}#3;%
1\mapsto\mspace{-2mu}#4\right\}}}

\newcommand*{\clientdisp}[2]{\prefixed{\clientdisps{#1}}{#2}}
\newcommand*{\clientuse}[3]{\prefixed{\clientuses{#1}{#2}}{#3}}
\newcommand*{\clientdup}[4]{\prefixed{\clientdups{#1}{#2}{#3}}{#4}}
\newcommand*{\server}[3]{\prefixed{\bang{#1(#2)}}{#3}}

\NewDocumentCommand{\hole}{o}{\IfNoValueTF{#1}{\Box}{\boxed{#1}}}

\newcommand*{\lsync}[2]{(#1 \parallel #2)}

\newcommand*{\lsend}[2]{{\sends{#1}{#2}}}
\newcommand*{\ltensor}[2]{{#1 \tensor #2}}
\newcommand*{\lrecv}[2]{{\recvs{#1}{#2}}}
\newcommand*{\lparr}[2]{{#1 \parr #2}}
\newcommand*{\lclose}[1]{{\closes{#1}}}
\newcommand*{\lone}{{\one}}
\newcommand*{\lwait}[1]{{\waits{#1}}}
\newcommand*{\lbot}{{\bot}}

\newcommand*{\linl}[1]{{\inls{#1}}}
\newcommand*{\loplusl}[2]{{#1 \oplus_1 #2}}
\newcommand*{\lcoinl}[1]{{\coinls{#1}}}
\newcommand*{\lwithl}[2]{{#1 \with_1 #2}}
\newcommand*{\linr}[1]{{\inrs{#1}}}
\newcommand*{\loplusr}[2]{{#1 \oplus_2 #2}}
\newcommand*{\lcoinr}[1]{{\coinrs{#1}}}
\newcommand*{\lwithr}[2]{{#1 \with_2 #2}}
\newcommand*{\lforward}[2]{{\forwards{#1}{#2}}}
\newcommand*{\laxiom}[1]{{\ref*{rule:axiom}\,{#1}}}
\newcommand*{\luse}[2]{{\clientuses{#1}{#2}}}
\newcommand*{\lquery}[1]{{\query #1}}
\newcommand*{\lcouse}[2]{{\serveruses{#1}{#2}}}
\newcommand*{\lbang}[1]{{\bang #1}}
\newcommand*{\ldup}[3]{{\clientdups{#1}{#2}{#3}}}
\newcommand*{\lcontract}[1]{{\query_{\ref*{rule:contract}} #1}}
\newcommand*{\lcodup}[4][]{{\serverdups{#2}{#3}{#4}_{#1}}}
\newcommand*{\lcocontract}[1]{{\bang_{\ref*{rule:contract}} #1}}
\newcommand*{\ldisp}[1]{{\clientdisps{#1}}}
\newcommand*{\lweaken}[1]{{\query_{\ref*{rule:contract}} #1}}
\newcommand*{\lcodisp}[2][]{{\serverdisps{#2}_{#1}}}
\newcommand*{\lcoweaken}[1]{{\bang_{\ref*{rule:contract}} #1}}

\newcommand{\plbl}[2]{\phl{#1}\colon\mspace{-2mu}\thl{#2}}

\title[A Fully-Abstract Semantics for Classical Processes]{Better Late Than Never}
\subtitle{A Fully-Abstract Semantics for Classical Processes}

\def\affIMADA{
  \department{Department of Mathematics and Computer Science}
  \institution{University of Southern Denmark}
  \streetaddress{Campusvej 55}
  \city{Odense}
  \postcode{5230}
  \country{Denmark}
}

\author[W.~Kokke]{Wen Kokke}
\orcid{0000-0002-1662-0381}
\affiliation{
  \department{Laboratory for Foundations of Computer Science}
  \institution{University of Edinburgh}
  \streetaddress{10 Crichton Street}
  \city{Edinburgh}
  \state{Scotland}
  \postcode{EH8 9AB}
  \country{United Kingdom}
}
\email{wen.kokke@ed.ac.uk}

\author[F.~Montesi]{Fabrizio Montesi}
\orcid{0000-0003-4666-901X}
\affiliation{
  \affIMADA
}
\email{fmontesi@imada.sdu.dk}

\author[M.Peressotti]{Marco Peressotti}
\orcid{0000-0002-0243-0480}
\affiliation{
	\affIMADA
}
\email{peressotti@imada.sdu.dk}

\begin{abstract}
We present Hypersequent Classical Processes (\HCP), a revised interpretation of the ``Proofs as 
Processes'' correspondence between linear logic and the $\pi$-calculus initially proposed by 
\citet{A94}, and later developed by \citet{BS94}, \citet{CP10}, and \citet{W14}, among others.
\HCP mends the discrepancies between linear logic and the syntax and observable semantics of 
parallel composition in the $\pi$-calculus, by conservatively extending linear logic to 
hyperenvironments (collections of environments, inspired by the hypersequents by \citet{A91}).
Separation of environments in hyperenvironments is internalised by $\tensor$ and corresponds to 
parallel process behaviour. Thanks to this property, for the first time we are able to extract a 
labelled transition system (\lts) semantics from proof rewritings.
Leveraging the information on parallelism at the level of types, we obtain a logical 
reconstruction of the delayed actions that \citet{MS04} formulated to model non-blocking I/O in the 
$\pi$-calculus. We define a denotational semantics for processes based on Brzozowski derivatives, 
and uncover that non-interference in \HCP corresponds to Fubini's theorem of double antiderivation.
Having an \lts allows us to validate \HCP using the standard toolbox of behavioural theory. We 
instantiate bisimilarity and barbed congruence for \HCP, and obtain a full abstraction result: 
bisimilarity, denotational equivalence, and barbed congruence coincide.
\end{abstract}



\begin{document}

\maketitle

\section{Introduction}
\label{sec:introduction}

\paragraph{Background}
Since its introduction by \citet{G87}, linear logic has been tremendously influential in the study of concurrency.
\citet{A94}, and later \citet{BS94}, kickstarted the search for a direct correspondence between proofs in linear logic and processes in (a fragment of) the $\pi$-calculus. This direction is appealing because it carries the hope of providing \emph{canonical} foundations for concurrency, ideally as firm as those provided by the Curry-Howard correspondence between natural deduction and the simply-typed $\lambda$-calculus for functional programming.
These initial efforts inspired seminal typing disciplines for the $\pi$-calculus, \eg, session 
types by \citet{HVK98} and linear types by \citet{KPT99}.

\citet{CP10} recently revitalised this research line, by developing a correspondence between a 
variant of the session-typed $\pi$-calculus and intuitionistic linear logic: processes correspond to 
proofs, session types (communication protocols) to propositions, and communication to cut 
elimination.
\citet{W14} revisited the correspondence for Classical Linear Logic (\CLL) and developed the 
calculus of Classical Processes (\CP).

\paragraph{The problem}
Despite these recent successes, it is still unclear how we can obtain a unified foundation for 
concurrency based on linear logic and the $\pi$-calculus.
This is due to a series of discrepancies between the two theories, both 
on the levels of syntax and semantics---ultimately, we will see that bridging these gaps leads to a 
reconciliation of ``Proofs as Processes'' with the behavioural theory of the $\pi$-calculus, in 
terms of a full abstraction result.
As base for our investigation, we use Wadler's calculus \CP.
\CP is convenient to study the discrepancies of interest, because its design is ``guided'' by 
linear logic: the syntax of processes in \CP corresponds to the structure of the rules of linear 
logic, and the semantics of these processes is extracted from the traditional steps of cut 
elimination.

Some discrepancies are syntactic.
Parallel composition $P\pp Q$ is a central construct in most process calculi, but only appears combined with output and restriction in \CP.
The term for output of a linear name in \CP is $\send{x}{y}{(P\pp Q)}$, read ``send $y$ over $x$ and
proceed as $P$ in parallel to $Q$''. Notice that the term constructor for output here actually
takes $x$, $y$, $P$ and $Q$ as parameters at the same time---whereas in process calculi, only one 
continuation is typically necessary. This discrepancy is caused by the structure of rule $\tensor$ 
in \CLL, which \CP uses to type 
output: the typing rule checks that the process
using $y$ ($P$) and the one using $x$ ($Q$) share no resources, by
taking two premises ($P$ and $Q$).
In general, there is no independent parallel term $P\pp Q$ in the grammar of \CP; \citet{W14} hints 
at the possibility of typing $P\pp Q$ using rule $\rname{Mix}$ by \citet{G87}, but this rule
does not allow $P$ and $Q$ to synchronise as in the $\pi$-calculus. Synchronisation in \CP is governed instead by the restriction operator $\res{xy}{(P\pp Q)}$, which connects the names $x$ at $P$ and $y$ at $Q$ to enable communication (this restriction term, where $x$ and $y$ represent the two endpoints of a bidirectional channel, was adopted in the latest presentation of \CP \citep{CLMSW16} and was originally introduced by \citet{V12} for the session-typed $\pi$-calculus).
Again, parallel is mixed with another operator (restriction), but now it means that $P$ and $Q$ will communicate.

The discrepancies carry over from syntax (and typing) to semantics.
Consider the rule for reducing an output with a compatible input in \CP, below.
\[
\rulescalebox{
\res {xy}{\left(\send {x}{x'}{(P\pp Q)} \pp \recv{y}{y'}{R}\right)}
\reducesto
\res{x'y'}{
	\left(P  \pp
	\res{xy}{(Q\pp R)}
	\right)
}}
\]
Notice how the rule needs to inspect the structure of the continuation of the output term ($P \pp
Q$) to produce a typable structure for the resulting network, by nesting restrictions appropriately.

An important consequence of these discrepancies is that \CP is still missing a labelled transition system (\lts) semantics.
Keeping with our example, it is difficult to define a transition axiom for output, as in $x[y].(P \pp Q) \lto{x[y]} P\pp Q$, because it is not possible to type $P \pp Q$. Even if it were,
we hit another problem when attempting to recreate the reduction above using transitions. Ideally,
we would define a rule that does not inspect the structure of processes, but only their
observables:
\[\rulescalebox{
\infer{
	\res{xy}{(P \pp Q)}
	\lto{\tau}
	\res{xy}
	{
		\res {x'y'}{(P' \pp Q')}
	}
}{
	P \lto{\sends{x}{x'}} P'
	&
	Q \lto{\recvs{y'}{y'}} Q'
}
}
\qquad \text{.}\]
However, this is not possible because the resulting restriction term is not typable in (nor is in the syntax of) \CP. This problem was already noticed by \citet{CP10}, whose correspondence between $\tau$-transitions and proof normalisation relies on intermediate rewritings that are allowed in the $\pi$-calculus, but are not supported by the logic.

Having an \lts for \CP would be desirable, because it would allow us to study its behavioural theory using the solid toolbox of process calculi based on observable transitions---\eg, bisimilarity (and variations thereof).
Also, there is reason to believe that such a study would be interesting.
\citet{A17} informally argued that bisimilarity would be incomplete for \CP: for example, \CP has 
no (well-typed) context that can distinguish the processes $\recv{x}{x'}{\recv{y}{y'}{P}}$ and $\recv{y}{y'}{\recv{x}{x'}{P}}$, 
since typing would force $x$ and $y$ to be connected to different parallel processes. However, 
bisimilarity would distinguish these two processes. Motivated by this informal argument, Atkey 
developed a denotational semantics for processes in \CP derived from the relational semantics of 
linear logic \citep{B91}. There are still no indications of how this line of work can be reconciled 
with the standard observational equivalences of process calculi.

Therefore, while the foundations of \CP are certainly validated from the side of logic, we are still far from validating them from the side of process calculi.

\paragraph{This paper}
We present Hypersequent Classical Processes (\HCP), a calculus that mends the discrepancies we discussed between linear logic and the $\pi$-calculus.
The key twist from linear logic to \HCP is to generalise classical linear logic from sequents with 
single environments to sequents with collections of environments, called hyperenvironments. 
Hyperenvironments essentially represent independent sequents, inspired by the theory of 
hypersequents by \citep{A91}, thus the name of \HCP.
The idea is that whenever two sequents $\seq \Gamma$ and $\seq \Delta$ can be proven separately ($\Gamma$ and $\Delta$ are typing environments), then they can be composed as in $\seq \Gamma \pp \Delta$, where $\Gamma \pp \Delta$ is a hyperenvironment. Intuitively, each environment in a hyperenvironment can be proven independently---in parallel, if you like.
From a logical perspective, the operator ``$\pp$'' for composing hyperenvironments is internalised 
by the $\tensor$ connective of linear logic (if $\seq \Gamma,A \pp \Delta,B$, then $\seq 
\Gamma,\Delta, A \tensor B$), just like ``,'' for composing environments is internalised by the 
$\parr$ connective (if $\seq \Gamma,A,B$, then $\seq \Gamma, A \parr B$).
This new symmetric treatment of $\tensor$ and $\parr$ is the foundation of all our contributions, which we believe represent a concrete step forward in Abramsky's original programme of ``Proofs as Processes''.
Our first contributions deal with the design of \HCP, whereas the others with its validation. Best 
comes last: our entire development is validated by the titular result of this paper, a full 
abstraction result that ties together bisimilarity, denotational equivalence, and contextual 
equivalence for \HCP.

\begin{enumerate}[noitemsep]
\item
\HCP reconciles the structure of proofs with the syntax of processes. On the process calculus 
side, term constructors have the expected modularity of process algebras, \eg, parallel composition 
and restriction are respectively the usual abelian monoid $P\pp Q$ and the term $\res {xy} P$ of the 
session-typed $\pi$-calculus \citep{V12}.
We formalise that \HCP is grounded in classical linear logic (\CLL) by proving that the two 
systems are equally powerful: we can internalise the new ingredient of environment composition 
using the connective $\tensor$.

\item \HCP supports sound proof rewritings that correspond to transition rules for processes, 
which we use to extract an \lts.
Our \lts mends the discrepancy we discussed about semantics, and extends the Curry-Howard 
correspondence of ``Proofs as Processes'' to the SOS style by \citet{P04}, by viewing proofs as 
states and our new proof rewritings as transitions.
Well-typed processes enjoy progress in our \lts.

\item Thanks to the fact that hyperenvironments allow us to see independence at the level of types 
(the ``$\pp$'' operator for composing environments), \HCP supports new proof rewritings that yield 
a logical reconstruction of the \lts originally 
studied by \citet{MS04} for the $\pi$-calculus with non-blocking I/O (delayed actions).

\item Our \lts bridges the gap between the research lines of ``Proofs as Processes'' and of 
behavioural theory for process calculi. As the first step on this bridge, we instantiate standard 
bisimilarity for \HCP. Bisimilarity gives us two immediate confirmations that our \lts is sound:
well-typed processes that are bisimilar are also type equivalent;
and bisimilarity is a congruence.

Courtesy of delayed actions, bisimilarity relates Atkey's problematic processes $\recv{x}{x'}{\recv{y}{y'}{P}}$ and 
$\recv{y}{y'}{\recv{x}{x'}{P}}$.
Even further, bisimilarity characterises (coincides with) contextual equivalence (for \HCP, this is 
typed barbed congruence).
While the completeness of bisimilarity is not a requirement, it is certainly desirable---and somewhat expected, for a first-order process calculus \citep{SW01}.

\item We define a denotational semantics for \HCP, by reformulating the one for \CP by \citet{A17}.
Atkey's denotations are inspired by the relational semantics of \CLL by \citet{B91}.
We rediscover (a refinement of) these denotations from a different angle, by defining Brzozowski derivatives \citep{B64} \wrt the observable actions in our \lts.
This has three benefits.
First, it gives a formal and direct connection between the operational and denotational semantics of \HCP.
Second, it shows that the denotational semantics of \HCP agrees with a standard notion of observability. Third, it reveals that non-interference, usually a topic of operational semantics, can be stated for 
\HCP in denotational terms: Fubini's theorem of double antiderivation holds in our setting 
\citep{F07}, formalising the intuition that the order of independent actions is not discriminated.
In a sense, Fubini's theorem for \HCP explains from a denotational perspective why 
delayed actions are operationally sound.

\item
As we anticipated, \HCP enjoys full abstraction, in the sense that all three semantic equivalences we present coincide:
bisimilarity $=$ denotational $=$ contextual equivalence.
\end{enumerate}

\citet{W14} ended his presentation of Classical Processes by stating:
\begin{quote}
``As $\lambda$-calculus provided foundations for functional programming in the last century, may we hope for this emerging calculus to provide foundations for concurrent programming in the coming century?''
\end{quote}
From the riverbank of behavioural theory for process calculi, delaying the execution of actions seems to be an important aspect for this agenda. Better late than never.

\section{Hypersequent Classical Processes}
\label{sec:hcp-language}

We start our formal development by presenting the process syntax and proof theory of Hypersequent Classical Processes (\HCP).

\subsection{Processes}
In \HCP, programs are processes (\hl{$\phl{P}$,$\phl{Q}$,$\phl{R}$,\dots}) that communicate using names (\hl{$\phl{x}$,$\phl{y}$,$\phl{z}$,\dots}). A name represents one of the two endpoints of a bidirectional channel. This style was introduced to the session-typed $\pi$-calculus by \citet{V12}, and later adopted in the latest presentation of Classical Processes by \citet{CLMSW16}.
Process terms are given by the following grammar.
\\[\abovedisplayskip]%
\begin{highlight}%
  \sbox0{$\phl P, \phl Q \Coloneqq {} $}%
  \sbox1{$\choice{x}{P}{Q}$}%
  \begin{xtabular}{%
      >{\raggedleft$}m{\wd0}<{$}%
      >{\raggedright$\phl\begingroup}m{\wd1}<{\endgroup$}%
      >{\quad\it}l<{}}
    \phl P, \phl Q \Coloneqq {}
            & \send{x}{y}{P}       & output $y$ on $x$ and continue as $P$\\
    \mid {} & \recv{x}{y}{P}       & input $y$ on $x$ and continue as $P$ \\
    \mid {} & \close{x}{P}         & output (empty message) on $x$ and continue as $P$\\
    \mid {} & \wait{x}{P}          & input (empty message) on $x$ and continue as $P$\\
    \mid {} & \inl{x}{P}           & select left on $x$ and continue as $P$ \\
    \mid {} & \inr{x}{P}           & select right on $x$ and continue as $P$ \\
    \mid {} & \choice{x}{P}{Q}     & offer a binary choice between $P$ (left) or Q (right) on $x$\\
    \mid {} & \server{x}{y}{P}     & offer a service\\
    \mid {} & \clientuse{x}{y}{P}  & consume a service\\
    \mid {} & \clientdup{x}{x_1}{x_2}{P} & duplicate a service\\
    \mid {} & \clientdisp{x}{P}    & dispose of a service\\
    \mid {} & \res{xy}{P}           & name restriction, ``cut''\\
    \mid {} & P \pp Q              & parallel composition of processes $P$ and $Q$\\
    \mid {} & \nil                 & terminated process \\
    \mid {} & \forward{x}{y}       & link $x$ and $y$
  \end{xtabular}%
\end{highlight}%
\\[\belowdisplayskip]%
Term $\send{x}{y}{P}$ allocates a fresh name $y$, outputs $y$ over $x$ and then proceeds as $P$. Dually, term $\recv{x}{y}{P}$ inputs a name $y$ over $x$ and then proceeds as $P$.
Both output and input terms bind the transmitted name ($y$) to the respective continuation $P$.
Terms $\close x{P}$ and $\wait{x}{P}$ respectively model output and input with no content.
Terms $\inl{x}{P}$ and $\inr{x}{P}$ respectively send on $x$ the selection of the left or right branch of a (binary) offer available on the other end of the channel before proceeding as $P$. Dually, term $\choice{x}{P}{Q}$ offers on $x$ a choice between proceeding as $P$ (left branch) or $Q$ (right branch).
Term $\server{x}{y}{P}$ is a server that offers on $x$ a service implemented by the replicable process $P$, where $y$ is bound in $P$.
A server term can be used by clients any number of times. Accordingly, we have three client terms to interact with a server.
The client term $\clientuse{x}{y}{P}$ requests exactly one copy of the service provided by the server on $x$, and then proceeds by communicating with the service on channel $y$.
The client term $\clientdisp{x}{P}$ disposes the server on $x$---the service is used zero times.
The client term $\clientdup{x}{x_1}{x_2}{P}$ requests that the server on $x$ is duplicated in two new instances, respectively available on the new channels $x_1$ and $x_2$.
A restriction term $\res{xy}{P}$ forms a channel by connecting and binding the two endpoints $x$ and $y$ in $P$, enabling communications from $x$ to $y$ and \viceversa. Restriction hides the endpoints $x$ and $y$ from the context.
Terms $P \pp Q$ and $\nil$ are the standard terms for the parallel composition of two processes and the terminated process.
Term $\forward{x}{y}$ is a forwarding proxy: inputs on $x$ are forwarded as outputs on $y$ and \viceversa.

In the remainder, we use $\pi$ to range over term prefixes: $\sends xy$, $\recvs xy$, $\closes x$, $\waits x$, $\inls{x}$, $\inrs{x}$, $\serveruses xy$, $\clientuses xy$, $\clientdups{x}{x_1}{x_2}$, and $\clientdisps x$.
Free and bound names of processes and prefixes are defined as expected, as well as $\alpha$-conversion. We write $\fn(P)$, $\bn(P)$, $\cn(P)$, for the set of free, bound, and all channel names in $P$, respectively, and likewise for prefixes. We write $P \aleq Q$ if $P$ and $Q$ are $\alpha$-equivalent.

\begin{example}
  \label{ex:and-server}
We write a server that computes the logical AND of two bits, adapting an example by \citet{ALM16} to \HCP.
We use selections to model sending bits.
Since \HCP is pretty low-level as a programming language, we use the following syntactic sugar.
\begin{equation*}
\send x0P \defeq \send{x}{x'}{\inl{x'}{\close {x'}P}}
\qquad
\send x1P \defeq \send{x}{x'}{\inr{x'}{\close {x'}P}}
\qquad
\choicebit xPQ \defeq \choice x{\wait xP}{\wait xQ}
\end{equation*}
With these abbreviations, we can write a server that offers a service for computing logical AND.
\begin{equation*}
Server_y \defeq 
	\server{y}{y'}
	{\recv{y'}{p}
		{\recv{y'}{q}
			{\choicebit*{p}
				{\choicebit{q}
					{\send{y'}{0}{\close{y'}{\nil}}}
					{\send{y'}{0}{\close{y'}{\nil}}}}
				{\choicebit{q}
					{\send{y'}{0}{\close{y'}{\nil}}}
					{\send{y'}{1}{\close{y'}{\nil}}}}}}}
\end{equation*}

We now define a compatible client, $Client^{b_1b_2}_{xz}$, which sends bits $b_1$ and $b_2$ (0 or 1) to a server that accepts two bits on $x$ (the client abstracts from the concrete operation that the server computes).
The client uses the result to decide whether to select left or right on another channel $z$.
\begin{equation*}
Client^{b_1b_2}_{xz} =
\clientuse x{x'}{
\send{x'}{b_1}{
\send{x'}{b_2}{\choicebit{x'}{
\wait {x'}{\inl z{\close z\nil}}
}{
\wait {x'}{\inr z{\close z\nil}}
}
}}}
\end{equation*}
\end{example}

\paragraph{Relation to other calculi}
The main difference between the syntax of \HCP and its predecessors in the research line
of ``Proofs as Processes'' is that parallel composition $P\pp Q$ is a term in its own right instead of being an inseparable subcomponent of other terms, as we discussed in the Introduction.
Our restriction and output terms have the familiar arities of the $\pi$-calculus: output $\send xyP$ has a single continuation, and likewise restriction $\res {xy}P$ binds $xy$ to a single process (instead of two).
Of course, designing an ``expected'' syntax for a session-typed process calculus is not hard---otherwise, it would not be expected! The real challenge is designing a proof theory based on linear logic where the structures of proofs match this syntax precisely, as we will do in \cref{sec:typing}.

Our client terms for explicit server management are inspired by \citet{W14}, who presented them as an alternative notation for Classical Processes (\CP). In \CP, server duplication and disposal do not have terms: these actions are handled by the semantics of \CP by looking at the typing proofs of processes. We chose the explicit terms for \HCP because, as we will see, server duplication and disposal are \emph{observable} actions. Thus, to define an \lts in the usual SOS style, it is desirable that these observables arise from corresponding syntactic terms.

From the perspective of $\pi$-calculus, \HCP is essentially a fragment of the internal $\pi$-calculus by \citet{S96}, with two differences. First, the explicit management of servers (our client terms, which we will see correspond to the rules for the exponential connective ``?'').
Second, the fact that channels are formed explicitly by the restriction term as proposed later by \citet{V12}, rather than implicitly by using the same name in different processes.
The hallmark of the internal $\pi$-calculus is that output always sends a fresh name, as in \HCP. This makes the theory of the calculus more convenient (output and input are symmetrical). The usual $\pi$-calculus term for outputting a free name can be recovered as syntactic sugar by using links \citep{ALM16}.
\begin{displaymath}
\begin{array}{rcl}
x\langle y \rangle.P & \defeq & \send xz{(\forward yz \pp P)}
\end{array}
\end{displaymath}
Similar considerations apply to polyadic communications \cite{SW01}.

\subsection{Typing}
\label{sec:types}
\label{sec:typing}

\paragraph{Types}
\HCP uses propositions from Classical Linear Logic (\CLL) as types for (endpoint) names.
Types (\hl{$\thl{A}$,$\thl{B}$,$\thl{C}$,\dots}) are defined by the following grammar.
\\[\abovedisplayskip]
\begin{highlight}
  \xentrystretch{-0.3}%
  \sbox0{$\thl{A},\thl{B}  \Coloneqq {} $}
  \sbox1{$A \tensor B$}
  \begin{xtabular}{%
      >{\raggedleft$}m{\wd0}<{$}%
      >{\raggedright$\thl\begingroup}m{\wd1}<{\endgroup$}%
      >{\hspace{.9ex}\it}l<{}%
      >{\hspace{1.2ex}$}r<{$}%
      >{\raggedright$\thl\begingroup}m{\wd1}<{\endgroup$}%
      >{\hspace{.9ex}\it}l<{}}
    \thl{A},\thl{B} \Coloneqq {}
            & \thl{A \tensor B} & send $A$, proceed as $B$&
    \mid {} & {A \parr B} & receive $A$, proceed as $B$\\
    \mid {} & {A \oplus B} & select $A$ or $B$&
    \mid {} & {A \with B} & offer $A$ or $B$\\
    \mid {} & {\one} & unit for $\tensor$&
    \mid {} & {\bot} & unit for $\parr$\\
    \mid {} & {\query A} & client request&
    \mid {} & {\bang A} & server accept
  \end{xtabular}
\end{highlight}
\\[\belowdisplayskip]
Types on the left-hand column are for outputs and types in the right-hand column for inputs.
Connectives on the same row are respective duals, \eg, $\tensor$ and $\parr$ are dual of each other.
We assume the standard notion of duality of \CLL, writing $\dual A$ for the dual of $A$. Duality proceeds homomorphically and replaces connectives with their duals, for example $\dual{(A \tensor B)} = \dual A \parr \dual B$.

\paragraph{Environments and Hyperenvironments}
Let $\Gamma$, $\Delta$, $\Theta$ range over unordered environments, which associate names to types.
\begin{highlight}
\[
\thl{\Gamma, \Delta, \Theta} ::= \thl{\cht{x_1}{A_1},\ldots,\cht{x_n}{A_n}} \quad \emph{environment}
\]
\end{highlight}
We write $\emptyseq$ for the empty environment. Given an environment $\Gamma = x_1:A_1,\ldots,x_n:A_n$, we write $\cn(\Gamma)$ for the set $\{x_1,\ldots,x_n\}$ of names in $\Gamma$. Names in the same environment must be distinct. Two environments can be composed only if they do not share names: whenever we write $\Gamma,\Delta$, this implies $\cn(\Gamma)\cap\cn(\Delta) = \emptyset$.

Environments are collected in unordered hyperenvironments, ranged over by $\hyp G$, $\hyp H$.
\begin{highlight}
\[
\thl{\hyp G}, \thl{\hyp H} ::= \thl{\Gamma_1 \pp} \cdots \thl{\pp \Gamma_n} \quad \emph{hyperenvironment}
\]
\end{highlight}
The idea is that all environments in a hyperenvironment can be proven independently.
We write $\emptyhyp$ for the empty hyperenvironment and $\cn(\hyp H)$ for the set of names appearing in (all the environments in) $\hyp H$.
As for environments, we require all names in hyperenvironments to be distinct: $\hyp G \pp \hyp H$ implies $\cn(\hyp G) \cap \cn(\hyp H) = \emptyset$.
Environments and hyperenvironments are equated up to exchange: $\Gamma,\Delta = \Delta,\Gamma$ and 
$\hyp G \pp \hyp H = \hyp H \pp \hyp G$.

\paragraph{Judgements and Typing}
Typing judgements assign processes to hyperenvironments and have the form:
\begin{highlight}
$
\judge{P}{\thl{\hyp G}}
$
\end{highlight}.
The rules for deriving judgements are displayed in \cref{fig:hcp-typing}.
We say that a process $P$ is \emph{well-typed} whenever there exists some $\hyp G$ such that $\judge{P}{\hyp{G}}$.

\begin{remark}[Alternative notation]
An alternative notation for our judgements could be $P ::\ \seq \Gamma_1 \pp \cdots \pp \seq \Gamma_n$ because, as we will show later, each sequent $\seq \Gamma_i$ is always guaranteed to be independently provable in classical linear logic.
Thus, our judgements can be seen as collections of sequents, recalling the hypersequents by \citet{A91}. This is the reason behind the name of \HCP.
We chose our notation to reduce eyestrain.
\end{remark}

\begin{figure}[t]
  \begin{highlight} 
    \headertext{Structural rules}%
		\[
      \infer[\rlabel{\rname{Ax}}{rule:axiom}]
      {\judge{\forward{x}{y}}{\cht{x}{A^\bot}, \cht{y}{A}}}
      {}
      \qquad
      \infer[\rlabel{\rname{H-Cut}}{rule:cut}]
      {\judge{\res{xy}{P}}{\hyp{G} \pp \Gamma, \Delta}}
      {\judge{P}{\hyp{G}\pp\Gamma, \cht{x}{A} \pp \Delta, \cht{y}{\dual{A}}}}
      \qquad
      \infer[\rlabel{\rname{H-Mix}}{rule:mix}]
      {\judge{P \pp Q}{\hyp{G} \pp \hyp{H}}}
      {\judge{P}{\hyp{G}}&\judge{Q}{\hyp{H}}}
      \qquad
      \infer[\rlabel{\rname{H-Mix$_0$}}{rule:mix_0}]
      {\judge{\nil}{\emptyhyp}}
      {}
    \]
 		\footer
     
    \headertext{Logical rules}%
		\begin{spreadlines}{.5\bigskipamount}%
		\begin{gather*}
      \infer[\rlabel{\rname{$\tensor$}}{rule:tensor}]
      {\judge{\send{x}{y}{P}}{\hyp{G} \pp \Gamma,\Delta,\cht{x}{A \tensor B}}}
      {\judge{P}{\hyp{G} \pp \Gamma,\cht{y}{A} \pp \Delta,\cht{x}{B}}}
      \qquad
      \infer[\rlabel{\rname{$\one$}}{rule:one}]
      {\judge{\close{x}{P}}{\hyp G \pp \cht{x}{\one}}}
      {\judge{P}{\hyp G}}
      \qquad
      \infer[\rlabel{\rname{$\parr$}}{rule:parr}]
      {\judge{\recv{x}{y}{P}}{\hyp{G} \pp \Gamma, \cht{x}{A \parr B}}}
      {\judge{P}{\hyp{G} \pp \Gamma, \cht{y}{A}, \cht{x}{B}}}
      \qquad
      \infer[\rlabel{\rname{$\bot$}}{rule:bot}]
      {\judge{\wait{x}{P}}{\hyp{G} \pp \Gamma, \cht{x}{\bot}}}
      {\judge{P}{\hyp{G} \pp \Gamma}}
      \\
      \infer[\rlabel{\rname{$\oplus_1$}}{rule:oplus_1}]
      {\judge{\inl{x}{P}}{\hyp{G} \pp \Gamma, \cht{x}{A \oplus B}}}
      {\judge{P}{\hyp{G} \pp \Gamma, \cht{x}{A}}}
      \qquad
      \infer[\rlabel{\rname{$\oplus_2$}}{rule:oplus_2}]
      {\judge{\inr{x}{P}}{\hyp{G} \pp \Gamma, \cht{x}{A \oplus B}}}
      {\judge{P}{\hyp{G} \pp \Gamma, \cht{x}{B}}}
      \qquad
      \infer[\rlabel{\rname{$\with$}}{rule:with}]
      {\judge{\choice{x}{P}{Q}}{\Gamma, \cht{x}{A \with B}}}
      {\judge{P}{\Gamma, \cht{x}{A}} &
        \judge{Q}{\Gamma, \cht{x}{B}}}
      \\
      \infer[\rlabel{\rname{$\bang$}}{rule:!}]
      {\judge{\server{x}{y}{P}}{\query\Gamma, \cht{x}{\bang A}}}
      {\judge{P}{\query\Gamma, \cht{y}{A}}}
      \qquad
      \infer[\rlabel{\rname{$\query$}}{rule:?}]
      {\judge{\clientuse{x}{y}{P}}{\hyp{G} \pp \Gamma, \cht{x}{\query A}}}
      {\judge{P}{\hyp{G} \pp \Gamma, \cht{y}{A}}}
	    \qquad
      \infer[\rlabel{\rname{W}}{rule:weaken}]
      {\judge{\clientdisp{x}{P}}{\hyp{G} \pp \Gamma, \cht{x}{\query A}}}
      {\judge{P}{\hyp{G} \pp \Gamma}}
      \qquad
      \infer[\rlabel{\rname{C}}{rule:contract}]
      {\judge{\clientdup{x}{x'}{x''}{P}}{\hyp{G} \pp \Gamma, \cht{x}{\query A}}}
      {\judge{P}{\hyp{G} \pp \Gamma, \cht{x'}{\query A}, \cht{x''}{\query A}}}
	  \end{gather*}%
	  \end{spreadlines}%
	  \footer%
  \end{highlight}%
  \caption{\HCP, typing rules.}
  \label{fig:hcp-typing}
\end{figure}

Typing rules associate types to names by looking at how endpoints are used in process terms. Rule selection is structural on the syntax of processes, in the sense that it depends only on the outermost constructor of a process term.
In \cref{rule:!}, we write $\query\Gamma$ for an environment of the form $\query A_1, \dots, \query A_n$ (possibly empty).

Most of our rules---with the exception of \ref{rule:cut} (restriction), \ref{rule:tensor} (output), \ref{rule:mix} (parallel composition), and \ref{rule:mix_0} (terminated process)---are exactly those presented for Classical Linear Logic (\CLL) by \citet{G87}, but extended to hyperenvironments. Dual terms are typed with dual types.

The most important new rules are the structural \cref{rule:mix,rule:cut}.
\Cref{rule:mix} types the parallel composition of two processes, by combining their 
hyperenvironments.
Previous work proposed a different rule for mixing environments, 
given below \citep{G87,W14}.
\begin{highlight}
\[\rulescalebox{
\infer[\rname{Mix}]
{\judge{P \pp Q}{\Gamma,\Delta}}
{\judge{P}{\Gamma}&\judge{Q}{\Delta}}
}\]
\end{highlight}
Notice the key difference: our rule keeps the information that the resources in the two premises come from independent proofs.
This information allows us to reformulate cut as \cref{rule:cut}, which uses a single premise.
\Cref{rule:cut} types the a restriction $\res {xy}{P}$ by checking that the channel is used by parallel components (separate environments) in $P$ in a dual way (as usual in \CLL).
In general, the key novelty of \HCP is that parallelism is guaranteed by separation of hyperenvironments.
By contrast, the standard cut rule of linear logic requires two separate \emph{proofs} as premises, yielding the restriction term constructor $\res{x}{(P \pp Q)}$ that we discussed in the \nameref{sec:introduction}.
\Cref{rule:mix,rule:cut} and hyperenvironments form thus the key to the desired decoupling of restriction and the parallel operator.
\Cref{rule:mix_0} types $\nil$, the unit of parallel composition for processes, as $\emptyhyp$, the unit of composition for hyperenvironments.

Our \cref{rule:tensor} is reformulated from \CLL using the same intuition for \cref{rule:cut}. The original rule requires two separate proofs for $A$ and $B$ respectively, whereas ours has a single premise requiring that $A$ and $B$ are in separate environments. In other words, $\tensor$ internalises $\pp$ in propositions, which yields a logical reconstruction of the output term from the internal $\pi$-calculus \cite{S96}.

The other rules are straightforward adaptations to hyperenvironments of the rules in 
\citep{W14} for Classical Processes.
\Cref{rule:axiom} types a link (forwarder), checking that the connected endpoints have dual types. 
This ensures that any message on $x$ can be safely forwarded to $y$, and \viceversa.
All logical rules enforce linear usage, except for client requests (typed with the exponential 
connective $\query$), for which contraction and weakening are allowed.
Contraction (\cref{rule:contract}) allows for multiple client requests for the same server, and 
weakening (\cref{rule:weaken}) for not using a server.

Types are preserved under $\alpha$-conversion, in the sense that whenever $P \aleq Q$, $\judge{P}{\hyp{G}}$ iff $\judge{Q}{\hyp{G}}$.

\begin{example}
\label{ex:and-typing}
Define the types for sending and receiving a bit, respectively.
\[
\begin{array}{lcl}
Bit = \one \oplus \one \quad \emph{send a bit} & \qquad\qquad & \dual{Bit} = \bot \with \bot \quad \emph{receive a bit}
\end{array}
\]
Then, we can type the server and client terms from \cref{ex:and-server} with dual types, as follows.
\[
\judge{Server_y}{\cht{y}{\bang(\dual{Bit} \parr \dual{Bit} \parr Bit \tensor \one)}} \qquad
\judge{Client^{b_1b_2}_{xz}}{
\cht{x}{\query ( Bit \tensor Bit \tensor \dual{Bit} \parr \bot), \cht{z}{Bit}}
}
\]
Thus, by \cref{rule:cut,rule:mix} we can type
their composition for all distinct names $x$, $y$ and 
$z$, and any bits $b_1$ and $b_2$, \eg, to compute the logical AND of 0 and 1:
$
\judge{\res {xy}{\left( Client^{01}_{xz} \pp Server_y\right)}}{
\cht{z}{Bit}
}
$.
\end{example}

For some processes, there are different acceptable ways of distributing free names in hyperenvironments.
For example, the process $\wait{x}{\close{y}{\close{z}{\nil}}}$ is typed by 
both ${\cht{x}{\bot},\cht{y}{\one} 
\pp \cht{z}{\one}}$ and ${\cht{y}{\one} \pp \cht{x}{\bot},\cht{z}{\one}}$, the only difference 
being that the name $x$ appears in a different component (environment).
In general, given $\judge P{\hyp G}$, for any $P$ and $\hyp G$, if we erase types from $\hyp G$ then we obtain a partition of the free names of $P$. We write $\strip{\hyp{G}}$ for the name partition obtained by removing types from $\hyp G$. For example, 
$\strip{\cht{x}{\bot},\cht{y}{\one} \pp \cht{z}{\one}}$ is $x,y \pp z$ (corresponding 
to $\{\{x,y\},\{z\}\}$ in standard set notation).
Intuitively, name partitions describe which names are used by each parallel component of a process.
We write a judgement $\pjudge{P}{G}$ to say that a process $P$ supports the partition $G$ on the 
set of its free names. The rules for deriving partitioning judgements are obtained by erasing all 
types ($A$, $B$, and connectives) from the typing rules 
displayed in \cref{fig:hcp-typing} and replacing ``$\seq$'' with ``$\Vdash$'' (we omit these rules for 
conciseness).
Thus, name partitions are independent of typing.
Computing all the possible name partitions for a process is trivially decidable: the set of free names of a process is always finite, giving a bound on the number of possible partitions.
Any derivation for $\judge{P}{\hyp{G}}$ is also a derivation for $\pjudge{P}{\strip{\hyp{G}}}$ once 
we erase channel types but not \viceversa: just consider $P = \close{x}{\nil}$ and $\hyp{G} = 
\cht{x}{\one \oplus \one}$ (thus $\strip{\hyp G} = x$).
We write $\np(P)$ for the set $\{G \mid \pjudge{P}{G}\}$ of name partitions induced by $P$.

We say that two hyperenvironments $\hyp{G}$ and $\hyp{G}'$ are one the \emph{shuffling} of the 
other, written $\hyp{G} \shuffle \hyp{G'}$, whenever they count the same number of non-empty 
environments and $\cht{x}{A}$ is in $\hyp{G}$ iff $\cht{x}{A}$ is in $\hyp{G}'$.

\begin{theorem}
	\label{thm:shuffling}
	If $\judge{P}{\hyp{G}}$, $\pjudge{P}{\strip{\hyp{G}'}}$, and $\hyp{G} \shuffle \hyp{G'}$ then, 
$\judge{P}{\hyp{G}}'$.
\end{theorem}

\begin{proofatend}
	We observe that because rule selection in \HCP is structural on the syntax of processes, 
derivations for $\judge{P}{\hyp{G}}$ and $\pjudge{P}{\strip{\hyp{G}'}}$ have the same structure 
(tree obtained by reading off them only rule names).
  We proceed by induction on the structure of the derivations for $\judge{P}{\hyp{G}}$ and 
$\pjudge{P}{\strip{\hyp{G}'}}$.
  The base case corresponds to \crefv{rule:axiom,rule:mix_0}: in hyperenvironments in the conclusion 
of both rules have no names that can be shuffled.
  The inductive case comprises two subcases: one for \crefv{rule:bot,rule:weaken} and one for the 
remaining rules.
  The first subcase correspond to rules that add a new name, say $x$, to an existing environment and 
which environment is to be used is not determined by the structure of the process term: each rule is 
about only one channel name ($x$) and this appears only in the conclusion. Therefore, we can shuffle 
$x$ as needed. Remaining names can be shuffled as needed by hypothesis.
  For the second case observe that environment selection is determined by the structure of the 
process and by induction hypothesis names in the premises can be shuffled to match the two 
partitions.
\end{proofatend}

\subsection{Relation with Classical Linear Logic}

If we erase processes and names from our typing rules and judgements, we essentially get a linear proof theory and sequents based on hyperenvironments. We write $\seq \hyp G$ when working under this erasure, abusing notation ($\hyp G$ does not contain names in this case).

We root \HCP in \CLL by relating their proof theories.
We start from the easier direction: all proofs in \CLL can be encoded into proofs in \HCP. Intuitively, this is because all rules in \CLL but \rname{Cut} and $\tensor$ are present also in \HCP (taking $\hyp G$ as empty). It is straightforward to reconstruct the missing rules by combining \rname{H-Mix} with $\tensor$ and \rname{H-Cut}.

\begin{theorem}\label{thm:cp2hcp}
If $\judge{}{\Gamma}$ in \CLL then $\judge{}{\Gamma}$ in \HCP.
\end{theorem}

\begin{proofatend}
By induction on the derivation of $\judge{}{\thl{\Gamma}}$. All rules in \CLL but \rname{Cut} and $\tensor$ are present also in \HCP (taking $\hyp G$ as empty).
It is straightforward to derive these rules in \HCP.
For example, we rewrite \rname{Cut} as follows. $\tensor$ is handled similarly.
\begin{highlight}\[
  \array{c}
  \infer[\rname{cut}]
    {\judge{}{\Gamma,\Delta}}
    {\judge{}{\Gamma, A}
    &\judge{}{\Delta, \dual{A}}}
  \endarray
  \Longrightarrow
  \array{c}
  \infer[\ref{rule:cut}]
    {\judge{}{\Gamma,\Delta}}
    {\infer[\ref{rule:mix}]
      {\judge{}{\Gamma,A \pp \Delta,\dual{A}}}
      {\judge{}{\Gamma,A}
      &\judge{}{\Delta,\dual{A}}}}
  \endarray\vspace{-3ex}
\]\end{highlight}
\end{proofatend}

If we consider processes, from the proof of \cref{thm:cp2hcp} we extract the expected encoding from 
the latest version of Wadler's Classical Processes (\CP, which uses \CLL as typing discipline) by 
\citet{CLMSW16} to visually identical terms in \HCP, \eg
$
  \phl{\lclens\res{xy}{(P \pp Q)}\rclens} = \phl{\res{xy}{( \lclens P \rclens \pp \lclens Q \rclens)}}
  \text{.}
$
This means that all well-typed processes in \CP are well-typed also in \HCP.

The opposite direction, from \HCP to \CLL, is not as straightforward because \HCP supports proof structures that do not appear in \CLL.
From a process perspective, there are behaviours that cannot be translated directly from \HCP to \CP. For example, the process $\choice{x}{P \pp Q}{P' \pp Q'}$, where $x$ appears in $Q$ and $Q'$, is typable in \HCP but cannot be written/typed in \CP. The choice sent on ${x}$ will affect the choice between ${P}$ and ${P'}$, even though neither has access to ${x}$.

Instead, we will prove that \HCP supports the same propositions as \CLL. This is the same as saying that \HCP and \CP inhabit the same types, or that the associated logical systems derive the same theorems. We use a standard method for proving the soundness of hypersequent calculi: hyperenvironments are internalised as propositions in \CLL.

We observe that all proofs in \HCP can be ``disentangled'', by moving applications of \cref{rule:mix} deeper in the proof tree. We can use this property to rewrite any derivation to a form in which all mixes are either attached to their respective cuts or tensors, or at the top-level. These consecutive applications can be rewritten as rule applications of cut and $\tensor$ from \CLL.

\begin{lemma}[Disentanglement]\label{lem:hcp-disentangle}
  If there exists a derivation $\rho$ of $\judge{}{\Gamma_1 \pp \dots \pp \Gamma_n}$ in \HCP, then there exist derivations $\rho_1$, \dots, $\rho_n$ of $\judge{}{\Gamma_1}, \dots,\judge{}{\Gamma_n}$ in \CLL.
\end{lemma}

\begin{proofatend}
  We move all applications of \cref{rule:mix} downwards in the derivation. There are three cases:
  \begin{enumerate}
  \item
    if a mix gets stuck under a cut, it forms a \CLL cut;
  \item
    if a mix gets stuck under a \ref{rule:tensor}, it forms a \CLL $\tensor$;
  \item
    otherwise, it moves all the way to the top.
  \end{enumerate}
  Furthermore, all applications of \cref{rule:one} are followed by an application of \cref{rule:mix_0}, forming a \CLL $\one$.
  Once this rewriting completes, we have a series of derivations $\rho_1$,\dots, $\rho_n$, composed by applications of \cref{rule:mix}.
\end{proofatend}

We define an encoding of hyperenvironments in \HCP into propositions in \CLL.
\begin{highlight}
  \begin{align*}
    \thl{\bigparr(\emptyseq)}            = {} & \thl{\bot} &
    \thl{\bigtensor(\emptyhyp})          = {} & \thl{\one} &
    \thl{\bigparr(\Gamma,A)}             = {} & \thl{\bigparr(\Gamma) \parr A} &
    \thl{\bigtensor(\hyp{G} \pp \Gamma)} = {} & \thl{\bigtensor(\hyp{G})\tensor\bigtensor(\Gamma)}
  \end{align*}
\end{highlight}

\begin{lemma}\label{lem:cp-bigparr}
    If $\judge{}{\Gamma}$ in \HCP, then $\judge{}{\bigparr\Gamma}$ in \CLL.
\end{lemma}

\begin{proofatend}
  By repeated application of $(\parr)$.
\end{proofatend}

By \cref{lem:cp-bigparr} and repeated applications of $\tensor$ in \CLL, we obtain the following theorem.
\begin{theorem}
  \label{thm:hcp2cp-bigtens}
    If $\judge{}{\hyp{G}}$ in \HCP, then $\judge{}{\bigtensor\hyp{G}}$ in \CLL.
\end{theorem}

\begin{proofatend}
    By case analysis on the structure of the hyperenvironment $\thl{\hyp{G}}$. If $\thl{\hyp{G}} = \thl{\emptyhyp}$. If $\thl{\hyp{G}} = \thl{\Gamma_1 \pp \dots \pp \Gamma_n}$, we apply \cref{lem:hcp-disentangle} to obtain proofs of $\judge{}{\Gamma_1}$, \dots, $\judge{}{\Gamma_n}$ in \CLL, then we apply \cref{lem:cp-bigparr} to each of those proofs to obtain proofs of $\judge{}{\bigparr\Gamma_1}$, \dots, $\judge{}{\bigparr\Gamma_n}$, and join them using $\tensor$ to obtain a single proof of $\judge{}{\bigtensor\hyp{G}}$ in \CP.
\end{proofatend}

\section{Operational Semantics}
\label{sec:hcp-transition}

\begin{figure}[t]
\begin{highlight}\begingroup
		\def\rulescaling{1}\small
    \headertext{Action labels ($l$, $l'$, \ldots)}%
    \begin{spreadlines}{-.05ex}%
    \setlength{\abovedisplayskip}{0pt}%
    \setlength{\belowdisplayskip}{0pt}%
    \def\sep{\hspace{1ex}}
		\begin{align*}
		\phl{\lclose{x}} & \sep \text{close $x$} &
		\phl{\lwait{x}} & \sep \text{wait on $x$} &
		\phl{\lsend{x}{y}} & \sep \text{output $y$ on $x$} &
		\phl{\lrecv{x}{y}} & \sep \text{input $y$ on $x$} \\
		\phl{\linl{x}} & \sep \text{select left} &
		\phl{\lcoinl{x}} & \sep \text{offer left} &
		\phl{\linr{x}} & \sep \text{select right}  &
		\phl{\lcoinr{x}} & \sep \text{offer right} \\
		\phl{\luse{x}{y}} & \sep \text{request $y$ on $x$} &
		\phl{\ldisp{x}} & \sep \text{request dispose $x$} &
		\phl{\ldup{x}{x_1}{x_2}} & \sep \text{request duplicate $x$} & 
		\phl{\lforward{x}{y}} & \sep \text{forward} \\
		\phl{\lcouse{x}{y}} & \sep \text{accept $y$ on $x$} &
		\phl{\lcodisp{x}} & \sep \text{accept dispose $x$} &
		\phl{\lcodup{x}{x_1}{x_2}} & \sep \text{accept duplicate $x$} & &
		\end{align*}%
		\end{spreadlines}%

		\headertext{Actions}
		\begin{spreadlines}{\termltsskipamount}
		\setlength{\belowdisplayskip}{0pt}%
		\begin{gather*}
    \rulescalebox{
      \deduce
        {\phl{\forward xy} \lto{\phl{\lforward yx}} \phl{\nil}}
        {\phl{\forward xy} \lto{\phl{\lforward xy}} \phl{\nil}}
    }
    \quad
		\rulescalebox{
			\deduce
			{\phl{\choice xPQ} \lto{\phl{\lcoinr x}} \phl{Q}}
			{\phl{\choice xPQ} \lto{\phl{\lcoinl x}} \phl{P}}
		}
		\quad
		\rulescalebox{
			\infer[\rlabel{\rname{Disp}}{rule:p-disp}]
				{\phl{\server{x}{x'}P} \lto{\phl{\lcodisp{x}}} \phl{\clientdisp{z_1}{\cdots\clientdisp{z_n}{\close{x}{\nil}}}}}
				{\phl{\fn(P) = \{x', z_1, \ldots, z_n\}}}
		}\quad
    \rulescalebox{
 		\deduce{
 			\phl{\ldup{x}{x_1}{x_2}.P} \lto{\phl{\ldup{x}{x_1}{x_2}}} \phl{x_1(x_2).P}
 		}{
 			\phl{\query x[].P} \lto{\phl{\query x[]}} \phl{\wait xP}
 			}
 		}
    \\
    \rulescalebox{
    	\infer[\rlabel{\pi}{rule:p-pre}]
    		{\phl{\pi.P} \lto{\phl{\pi}} \phl{P}}
        {\phl{\pi \neq \query x[],\ldup{x}{x_1}{x_2}}}
 		}
    \quad
		\rulescalebox{
				\infer[\rlabel{\rname{Dup}}{rule:p-dup}]
					{\phl{\server{x}{x'}P} \lto{\phl{\lcodup{x}{x_1}{x_2}}}
					  \phl{\clientdup{z_1}{z_1\sigma_1}{z_1\sigma_2}{\dots\clientdup{z_n}{z_n\sigma_1}{z_n\sigma_2}{\send{x_1}{x_2}{(\server{x_1}{x'\sigma_1}{P_1} \pp \server{x_2}{x'\sigma_2}{P_2})}}}}}
					{\phl{P_1 = P\sigma_1}
					& \phl{P_2 = P\sigma_2}
					& \phl{\fn(P_1) \cap \fn(P_2) = \emptyset}
					& \phl{\fn(P) = \{x',z_1,\ldots,z_n\}}}
			}
		\end{gather*}%
		\end{spreadlines}%

		\headertext{Structural}
		\begin{spreadlines}{\termltsskipamount}
		\setlength{\belowdisplayskip}{0pt}%
		\begin{gather*}
		\rulescalebox{
    	\infer[\rlabel{\ref{rule:mix-l}}{rule:p-mix-l}]
					{\phl{P \pp Q} \lto{\phl{l}} \phl{P' \pp Q}}
					{\phl{P} \lto{\phl{l}} \phl{P'}
					&\phl{\bn(l) \cap \fn(Q) = \emptyset}}
			}\qquad\rulescalebox{
				\infer[\rlabel{\ref{rule:mix-r}}{rule:p-mix-r}]
					{\phl{P \pp Q} \lto{\phl{l}} \phl{P \pp Q'}}
					{\phl{Q} \lto{\phl{l}} \phl{Q'}
					&\phl{\bn(l) \cap \fn(P) = \emptyset}}
			}
			\qquad
			\rulescalebox{
				\infer[\rlabel{\ref{rule:cut-forward}}{rule:p-cut-forward}]
					{\phl{\res{xy}{P}} \lto{\phl\tau} \phl{P'\{x/z\}}}
					{\phl{P} \lto{\phl{\lforward{y}{z}}} \phl{P'}}
			}
			\\\rulescalebox{
        \infer[\rlabel{\ref{rule:mix-sync}}{rule:p-mix-sync}]
					{\phl{P \pp Q} \lto{\lsync{\phl{l}}{\phl{l'}}} \phl{P' \pp Q'}}
					{\phl{P} \lto{\phl{l}} \phl{P'}
					&\phl{Q} \lto{\phl{l'}} \phl{Q'}
					&\phl{\bn(l) \cap \bn(l') = \emptyset}
					}
			}\qquad\rulescalebox{
        \infer[\rlabel{\ref{rule:cut-res}}{rule:p-cut-res}]
					{\phl{\res{xy}{P}} \lto{\phl{l}} \phl{\res{xy}{P'}}}
					{\phl{P} \lto{\phl{l}} \phl{P'}
          &\phl{x,y \notin \cn(l)}
          & \phl{\separate{P'}{x}{y}}
          }
			}\qquad\rulescalebox{
				\infer[\rlabel{\ref{rule:alpha}}{rule:p-alpha}%
					\IfLabelExistsF{sec:omitted-rules}%
					{\phantomlabel{infrule}{rule:alpha}{\rname{$\aleq$}}}]
					{\phl{P} \lto{\phl{l}} \phl{R}}
					{\phl{P \aleq Q}
					&\phl{Q} \lto{\phl{l}} \phl{R}}
			}
		\end{gather*}%
		\end{spreadlines}%

		\headertext{Communications}
		\begin{spreadlines}{\termltsskipamount}
		\setlength{\belowdisplayskip}{0pt}%
		\begin{gather*}
      \rulescalebox{
				\infer[\rlabel{\ref{rule:cut-tensor}}{rule:p-cut-tensor}]
					{\phl{\res{xy}{P}} \lto{\phl\tau} \phl{\res{xy}{\res{x'y'}{P'}}}}
					{\phl{P} \lto{\lsync{\phl{\lsend{x}{x'}}}{\phl{\lrecv{y}{y'}}}} \phl{P'}}
      }\qquad
	\rulescalebox{
			\infer[\rlabel{\ref{rule:cut-oplus_1}}{rule:p-cut-oplus_1}]
				{\phl{\res{xy}{P}} \lto{\phl{\tau}} \phl{\res{xy}{P'}}}
				{\phl{P} \lto{\lsync{\phl{\linl{x}}}{\phl{\lcoinl{y}}}} \phl{P'}}
 			}\quad\rulescalebox{
 			\infer[\rlabel{\ref{rule:cut-oplus_2}}{rule:p-cut-oplus_2}]
 				{\phl{\res{xy}{P}} \lto{\phl\tau} \phl{\res{xy}{P'}}}
 				{\phl{P} \lto{\lsync{\phl{\linr{x}}}{\phl{\lcoinr{y}}}} \phl{P'}}
			}
	\\
	\rulescalebox{
        \infer[\rlabel{\ref{rule:cut-one}}{rule:p-cut-one}]
          {\phl{\res{xy}{P}} \lto{\phl\tau} \phl{P'}}
          {\phl{P} \lto{\lsync{\phl{\lclose{x}}}{\phl{\lwait{y}}}} \phl{P'}}
      }\qquad
	\rulescalebox{
				\infer[\rlabel{\ref{rule:cut-?}}{rule:p-cut-?}]
					{\phl{\res{xy}{P}} \lto{\phl\tau} \phl{\res{x'y'}{P'}}}
					{\phl{P} \lto{\lsync{\phl{\luse{x}{x'}}}{\phl{\lcouse{y}{y'}}}} \phl{P'}}
			}\qquad
	\rulescalebox{
				\infer[\rlabel{\ref{rule:cut-weaken}}{rule:p-cut-weaken}]
					{\phl{\res{xy}{P}} \lto{\phl{\phl{\tau}}}
					 \phl{\res{xy}{P'}}}
					{\phl{P} \lto{\lsync{\phl{\ldisp{x}}}{\phl{\lcodisp{y}}}} \phl{P'}}
			}\qquad\rulescalebox{
				\infer[\rlabel{\ref{rule:cut-contract}}{rule:p-cut-contract}]
					{\phl{\res{xy}{P}} \lto{\phl{\phl{\tau}}} 
\phl{\phl{\res{x_1y_1}{P'}}}}
					{\phl{P} 
\lto{\lsync{\phl{\ldup{x}{x_1}{x_2}}}{\phl{\lcodup{y}{y_1}{y_2}}}} \phl{P'}}
			}
    \end{gather*}
		\end{spreadlines}%

		\headertext{Delayed Actions and Self-synchronisations}
		\begin{spreadlines}{\termltsskipamount}
		\setlength{\belowdisplayskip}{0pt}%
		\begin{gather*}
			\rulescalebox{
				\infer[\rlabel{\rname{$:\!\pi_1$}}{rule:pre_1-delayed}]
				{\phl{\pi.P} \lto{\phl{l}} \phl{\pi.P'}}
				{\phl{\pi \in \{\lclose x, \linl{x}, \linr{x}, \luse{x}{y}\}}
				& \phl P \lto{\phl l} \phl {P'}
				& \phl{\cn(\pi)\cap\cn(l) = \emptyset}
				}
			}\quad \rulescalebox{
			\infer[\rlabel{\rname{$:\!\pi_2$}}{rule:pre_2-delayed}]
					{\phl{\pi.P} \lto{\phl{l}} \phl{\pi.P'}}
					{\phl{\pi \in \{ \lwait{x}, \query x[] \}} &
					\phl{P} \lto{\phl{l}} \phl{P'}
					&\phl{\fn(P') \neq \emptyset}}
			}\\
			\rulescalebox{
				\infer[\rlabel{\ref{rule:tensor-delayed}}{rule:p-tensor-delayed}]{
					\phl{\send x{x'}P}
					\lto{\phl l}
					\phl{\send x{x'}{P'}}
				}{
					\phl{P} \lto{\phl l} \phl{P'}
					&
					\phl{x,x' \not\in \cn(l)}
					&
					\phl{\separate{P'}{x}{x'}}
				}
			}
			\quad
			\rulescalebox{
				\infer[\rlabel{\rname{$:\!\pi_3$}}{rule:pre_3-delayed}]{
					\phl{\pi.P}
					\lto{\phl l}
					\phl{\pi.P'}
				}{
					\phl{\pi \in \{\lrecv{x_1}{x_2}, \ldup{x}{x_1}{x_2}\}}
					&
					\phl{P} \lto{\phl l} \phl{P'}
					&
					\phl{\cn(\pi) \cap \cn(l) = \emptyset}
					&
					\phl{\together{P'}{x_1}{x_2}}
				}
			}
			\\
			\rulescalebox{
				\infer[\rlabel{\rname{$|\pi$}}{rule:pre-sync}]
				{\phl{\pi.P} \lto{\phl{\lsync{\pi}{l}}} \phl{P'}}
				{\phl{\pi \neq \query x[],\ldup{x}{x_1}{x_2}}
				& \phl{\pi.P} \lto{\phl l} \phl{\pi.P'}
				& \phl{\separate{\pi.P}{\fn(\pi)}{\fn(l)}}
				}
			}
			\quad
			\rulescalebox{
				\infer[\rlabel{\ref{rule:weaken-sync}}{rule:p-weaken-sync}]{
					\phl{\query x[].P} \lto{\phl{\lsync{\query x[]}{l}}} \phl{x().P'}
				}{
				\phl{\query x[].P} \lto{\phl l} \phl{\query x[].P'}
				& \phl{\separate{\query x[].P}{x}{\fn(l)}}
				}
			}
			\\
			\rulescalebox{
				\infer[\rlabel{\ref{rule:contract-sync}}{rule:p-contract-sync}]{
					\phl{\ldup x{x_1}{x_2}.{P}} \lto{\phl{\lsync{\ldup{x}{x_1}{x_2}}{l}}} 
\phl{\recv{x_1}{x_2}{P'}}
				}{
					\phl{\ldup x{x_1}{x_2}.P} \lto{\phl l} \phl{\ldup x{x_1}{x_2}.P'}
					&
					\phl{\separate{\ldup x{x_1}{x_2}.{P}}{x}{\fn(l)}}
				}
			}
		\end{gather*}%
		\end{spreadlines}%
		\footer%
\end{highlight}\endgroup
\caption{Labelled transition system of \HCP processes.}
\label{fig:hcp-lts}
\end{figure}

\HCP supports new proof rewritings \wrt~\CLL, which correspond to transition rules for processes. 
We use this property to define a semantics for \HCP in terms of a labelled transition system 
(\lts). Our semantics follows Plotkin's SOS style \citep{P04}, by viewing:
\begin{itemize}[nosep,noitemsep]
	\item the inference rules of our type system as operations of a (sorted) signature;
	\item proofs as terms generated by this signature;
	\item (labelled) proof transformations as (labelled) transitions;
	\item and a specification of rules for deriving proof transformations as an SOS specification.
\end{itemize}
Then, a semantics for \HCP processes in terms of an SOS specification is obtained simply by reading off how the SOS specification of proof transformations manipulate the processes that they type.

To illustrate the intuition for transitions, consider the proof for a judgement $\judge{\wait{x}{P}}{\hyp{G} \pp \Gamma, \cht{x}{\bot}}$. By the correspondence between term constructors and typing rules, the proof has the following shape.
\begin{highlight}
\vspace{-1ex}\[\rulescalebox{
\infer[\ref{rule:bot}]
	    {\judge{\wait xP}{\hyp{G} \pp \Gamma, \cht{x}{\bot}}}
	    {
	    \stackrel{\vdots}{\judge{P}{\hyp{G} \pp \Gamma}}
	    }
}\]
\end{highlight}
We can view \cref{rule:bot} as the outermost operation used in the proof.
Then, the proof of $\judge{P}{\hyp{G} \pp \Gamma}$ is the only argument of the operation and $x$ a parameter (operations are on proofs). This corresponds to the term constructor $\wait{x}{(-)}$ in the syntax of \HCP processes---which in this case takes $P$ as argument, \ie, the term corresponding to the proof of the premise.
Thus, this operation is the proof equivalent of the term constructor $\wait{x}{(-)}$ in the syntax of \HCP processes, which denotes an observable action.
Term constructors like this, also called action prefixes, are typically assigned a transition rule in process calculi where the target (\aka~\emph{derivative}) is the operator argument and the label is the prefixed operation. This correspondence points at the transition rule below---for readability, we \cboxed{black!20}{\mbox{box}} proofs and omit proof trees above premises in the remainder.
\begin{highlight}\[\rulescalebox{
\plto
	{\infer[\ref{rule:bot}]
	    {\judge{\wait xP}{\hyp{G} \pp \Gamma, \cht{x}{\bot}}}
	    {\judge{P}{\hyp{G} \pp \Gamma}}}
	{\plbl{\lwait{x}}{\lbot}}
	{\judge{P}{\hyp{G} \pp \Gamma}}
}\]\end{highlight}%
The label identifies the prefix constructor (\ie, rule name and parameter) and its syntax is inspired by common syntax for labels of action prefixes in process calculi.
By reading proof terms (processes) off the rule above we obtain the axiom below for processes.
\begin{highlight}\[
	\phl{\wait{x}{P}} \lto{\phl{\lwait{x}}} \phl{P}
\]\end{highlight}%
This axiom defines the expected semantics of the constructor $\wait{x}{(-)}$, a promising sign!

Following this methodology for all of our typing rules, we obtain the \lts on \HCP processes given 
by the SOS specification in \cref{fig:hcp-lts}, where $l$ ranges over transition 
labels (we abuse notation and use $l$ for the red part of transition labels when referring to 
processes only, and both the red and blue parts when referring to proof transitions).
We describe each transition rule in the remainder of this section, by discussing the proof 
transformations that they originate from. In transitions, we identify $\alpha$-equivalent processes.

In the sequel we write $\separate{P}{x}{y}$ (resp.~$\together{P}{x}{y}$) whenever there is a partition $G \in \np(P)$ that separates (resp.~does not separate) $x$ and $y$.
We write $\separate{P}{x}{y_1,\dots,y_n}$ for $\bigwedge_{i=1}^{n}\separate{P}{x}{y_i}$.

\subsection{Multiplicatives and Mix}
We start by giving a semantics to the multiplicative fragment of \HCP, which suffices to show all the key ideas behind our \lts.

The multiplicative fragment of \HCP is formed by the \crefv{rule:tensor,rule:parr,rule:one,rule:bot}, together with the structural \crefv{rule:mix,rule:mix_0,rule:cut}.
Observe that rules from the first group have the ``action prefix'' form described above.

\paragraph{Actions}
The transition rules for multiplicative prefixes are those below, plus the rule for $\wait xP$ already given.
\begin{highlight}
\begin{spreadlines}{\proofltsskipamount}
\begin{gather*}\rulescalebox{
\plto
	{\infer[\ref{rule:tensor}]
    {\judge{\send x{x'} P}{\hyp{G} \pp \Gamma,\Delta,\cht{x}{A \tensor B}}}
    {\judge{P}{\hyp{G} \pp \Gamma,\cht{x'}{A} \pp \Delta,\cht{x}{B}}}}
	{\plbl{\lsend{x}{x'}}{\ltensor{A}{B}}}
	{\judge{P}{\hyp{G} \pp \Gamma,\cht{x'}{A} \pp \Delta,\cht{x}{B}}}
}\\\rulescalebox{
\plto
	{\infer[\ref{rule:parr}]
    {\judge{\recv x{x'} P}{\hyp{G} \pp \Gamma, \cht{x}{A \parr B}}}
    {\judge{P}{\hyp{G} \pp \Gamma, \cht{x'}{A}, \cht{x}{B}}}}
	{\plbl{\lrecv{x}{x'}}{\lparr{A}{B}}}
	{\judge{P}{\hyp{G} \pp \Gamma, \cht{x'}{A}, \cht{x}{B}}}
}\qquad\rulescalebox{
\plto
	{\infer[\ref{rule:one}]
		{\judge{\close{x}P}{\hyp{G}\pp\cht{x}{\one}}}
	  {{\judge{P}{\hyp{G}}}}}
	{\plbl{\lclose{x}}{\lone}}
	{\judge{P}{\hyp{G}}}
}\end{gather*}
\end{spreadlines}
\end{highlight}%

\paragraph{Structural rules}
There are three transition rules for \cref{rule:mix}: two for executions where only one component is transformed (\cref{rule:mix-l,rule:mix-r}) and one where both components are transformed synchronously (\cref{rule:mix-sync}). 
(We omit \cref{rule:mix-r}, which is symmetric to \cref{rule:mix-l}.)%
\IfLabelExistsF{sec:omitted-rules}{%
	\phantomlabel{infrule}{rule:mix-r}{\rname{Par$_2$}}%
}%
\begin{highlight}
\begin{spreadlines}{\proofltsskipamount}
\begin{gather*}\rulescalebox{
\def\regh{\vphantom{\hyp{G}'\pp}}
\infer[\rlabel{\rname{Par$_1$}}{rule:mix-l}]
	{\plto
	  {\infer[\ref{rule:mix}]
	  	{\judge{P \pp Q}{\hyp{G} \pp \hyp{H}}\regh}
		  {\judge{P}{\hyp{G}}&\judge{Q}{\hyp{H}}\regh}}
	  {l}
	  {\infer[\ref{rule:mix}]
	  	{\judge{P' \pp Q}{\hyp{G}' \pp \hyp{H}}\regh}
		  {\judge{P'}{\hyp{G}'}&\judge{Q}{\hyp{H}}\regh}}}
	{\plto{\judge{P}{\hyp{G}}\regh}{l}{\judge{P'}{\hyp{G}'}\regh}
  &\phl{\bn(l) \cap \fn(Q) = \emptyset}}
}\\
\rulescalebox{
\def\regh{\vphantom{\hyp{G}'\pp\hyp{H}'}}
\infer[\rlabel{\rname{Syn}}{rule:mix-sync}]
	{\plto
	  {\infer[\ref{rule:mix}]
	  	{\judge{P \pp Q}{\hyp{G} \pp \hyp{H}}\regh}
		  {\judge{P}{\hyp{G}}&\judge{Q}{\hyp{H}}\regh}}
	  {\lsync{l}{l'}}
	  {\infer[\ref{rule:mix}]
	  	{\judge{P' \pp Q'}{\hyp{G}' \pp \hyp{H}'}\regh}
		  {\judge{P'}{\hyp{G}'}
      &\judge{Q'}{\hyp{H}'}\regh}}}
	{\plto{\judge{P}{\hyp{G}}\regh}{l}{\judge{P'}{\hyp{G}'}\regh}
	&\plto{\judge{Q}{\hyp{H}}\regh}{l'}{\judge{Q'}{\hyp{H}'}\regh}
	& \phl{\bn(l) \cap \bn(l') = \emptyset}}
}\end{gather*}
\end{spreadlines}
\end{highlight}%
\Cref{rule:mix-l,rule:mix-r} transform one of the two parallel components given that the 
transformation preserves non-interference, \ie, disjointness of names.
This condition follows from the requirement of distinct names in hyperenvironments, and gives 
the usual side-condition for \cref{rule:mix-l,rule:mix-r} that one would expect for the 
internal $\pi$-calculus (\cf \citep{S93}).
\Cref{rule:mix-sync} synchronises transformations of parallel components into a single transformation labelled with the unordered pair of the respective labels---we assume that $l$ and $l'$ are not pairs themselves, as interactions in \HCP have two parties. We write these unordered pairs of labels as $\lsync{l}{l'}$, to evoke the parallel combination of two transformations. Formally, for all $l$ and $l'$, $\lsync{l}{l'} = \lsync{l'}{l}$. Again, the condition on disjointness of bound names arises from the well-formedness of the resulting hyperenvironments.
There are no transitions for \cref{rule:mix_0}. Indeed, its corresponding term $\nil$ is the terminated program.

The rule below captures the standard propagation of unrestricted actions of the $\pi$-calculus. The extracted side-condition $\separate{P'}{x}{y}$ (which does not look at types) is induced by the name partitioning required for typing $P'$ in the target.
\begin{highlight}
\[\rulescalebox{
\infer[\rlabel{\rname{Res}}{rule:cut-res}]
	{\plto
		{\infer[\ref{rule:cut}]
			{\judge{\res{xy}{P}}{\hyp{G} \pp \Gamma,\Delta}}
			{\judge{P}{\hyp{G} \pp \Gamma,\cht{x}{A} \pp \Delta,\cht{y}{\dual{A}}}}}
		{l}
		{\infer[\ref{rule:cut}]
			{\judge{\res{xy}{P'}}{\hyp{G}' \pp \Gamma',\Delta'}}
			{\judge{P'}{\hyp{G}' \pp \Gamma',\cht{x}{A} \pp \Delta',\cht{y}{\dual{A}}}}}}
	{\plto
		{\judge{P}{\hyp{G} \pp \Gamma,\cht{x}{A} \pp \Delta,\cht{y}{\dual{A}}}}
		{l}
		{\judge{P'}{\hyp{G}' \pp \Gamma',\cht{x}{A} \pp \Delta',\cht{y}{\dual{A}}}}
	&\phl{x,y \notin \cn(l)}
	&\phl{\separate{P'}{x}{y}}}}
\]
\end{highlight}

\paragraph{Communication}
Communication is captured by simplifying applications of \cref{rule:cut}, given by the transformations below, one for each type of dual actions.
\begin{highlight}
\begin{spreadlines}{\proofltsskipamount}
\begin{gather*}
\rulescalebox{
\infer[\rlabel{\ref*{rule:one}\ref*{rule:bot}}{rule:cut-one}]
	{\plto
	  {\infer[\ref{rule:cut}]
     {\judge{\res{xy}{P}}{\hyp{G} \pp \Gamma,\Delta}}
     {\judge{P}{\hyp{G}\pp \Gamma,\cht{x}{\one} \pp \Delta, \cht{y}{\bot}}}}
	  {\tau}
	  {\judge{P'}{\hyp{G}'\pp \Gamma'}}}
	{\plto
		{\judge{P}{\hyp{G}\pp \Gamma,\cht{x}{\one} \pp \Delta, \cht{y}{\bot}}}
		{\lsync
      {\plbl{\lclose{x}}{\lone}}
      {\plbl{\lwait{y}}{\lbot}}}
		{\judge{P'}{\hyp{G}' \pp \Gamma'}}}
}\\\rulescalebox{
\infer[\rlabel{\ref*{rule:tensor}\ref*{rule:parr}}{rule:cut-tensor}]
	{\plto
		{\infer[\ref{rule:cut}]
		{\judge{\res{xy}{P}}{\hyp{G} \pp \Gamma,\Delta, \Theta}}
		{\judge{P}{\hyp{G}\pp\Gamma,\Delta, \cht{x}{A \tensor B} \pp \Theta, \cht{y}{\dual{A} \parr \dual{B}}}}}
	{\tau}
	{\infer[\ref{rule:cut}]
		{\judge{\res{xy}{\res{x'y'}{P'}}}{\hyp{G} \pp \Gamma, \Delta, \Theta}}
		{\infer[\ref{rule:cut}]
			{\judge{\res{x'y'}{P'}}{\hyp{G} \pp \Gamma,\cht{x}{B} \pp \Delta, \Theta,\cht{y}{\dual{B}}}}
			{\judge{P'}{\hyp{G} \pp \Gamma, \cht{x}{B} \pp \Delta,\cht{x'}{A} \pp \Theta, \cht{y}{\dual{B}}, \cht{y'}{\dual{A}}}}}}
	}{\plto
		{\judge{P}{\hyp{G}\pp\Gamma,\Delta, \cht{x}{A \tensor B} \pp \Theta, \cht{y}{\dual{A} \parr \dual{B}}}}
		{\lsync
      {\plbl{\lsend{x}{x'}}{\ltensor{A}{B}}}
      {\plbl{\lrecv{y}{y'}}{\lparr{\dual{A}}{\dual{B}}}}}
		{\judge{P'}{\hyp{G} \pp \Gamma, \cht{x}{B} \pp \Delta,\cht{x'}{A} \pp \Theta, \cht{y}{\dual{B}}, \cht{y'}{\dual{A}}}}}
}\end{gather*}
\end{spreadlines}
\end{highlight}%
These transformations do not interact with the context nor have any effect on the types of the conclusions besides shuffling (\cf \cref{ex:tau-shuffling,thm:hcp-lts-subject-reduction}).
Hence, they represent internal actions and we label them with $\tau$, as common for process calculi.

\begin{example}\label{ex:multiplicatives-1}
Let $P = \res {xy}{\left(x[x'].Q \pp y(y').z().R\right)}$ for some $Q$ and $R$ such that $P$ is well-typed. Then, we have the following transitions.
\begin{align*}
P & \lto\tau \res {xy}{\res {x'y'}{\left( Q \pp z().R\right)}} &
& \mbox{by \rname{$\tensor\parr$}, \rname{Syn}, and the axioms for $x[y]$ and $x(y)$}
\\
& \lto{z()} \res {xy}{\res {x'y'}{\left( Q \pp R\right)}} &
& \mbox{by \rname{Res}, \rname{Par$_2$}, and the axiom for $z()$}.
\end{align*}
\end{example}

\begin{remark}
The reader familiar with linear logic might recognise that our transition rules for communications evoke cut reductions in \CLL: the way in which types are matched and deconstructed is similar.
The key difference is that we do not need to permute cuts in proofs (commuting conversions) until they reach the rule applications that formed the types being deconstructed.
This is because we can \emph{observe} what we need from our transition labels, rather than having to inspect the structure of the proofs for the premises of our transition rules.
\end{remark}

\paragraph{Delayed actions}
\HCP supports the notion of ``delayed actions'', originally introduced for the $\pi$-calculus to formulate non-blocking I/O actions \cite{MS04}. Delayed actions allow actions under a prefix to be executed (observed), as long as they do not interfere with the prefix. The following rule delays an input action.
\begin{highlight}
\[\rulescalebox{
\infer[\rlabel{\delayedrule{rule:parr}}{rule:parr-delayed}]
	{\plto
		{\infer[\ref{rule:parr}]
	    {\judge{\recv x{x'} P}{\hyp{G} \pp \Gamma, \cht{x}{A \parr B}}}
	    {\judge{P}{\hyp{G} \pp \Gamma, \cht{x'}{A}, \cht{x}{B}}}}
		{l}
		{\infer[\ref{rule:parr}]
			{\judge{\recv x{x'}{P'}}{\hyp{G}' \pp \Gamma', \cht{x}{A \parr B}}}
			{\judge{P'}{\hyp{G}' \pp \Gamma', \cht{x'}{A}, \cht{x}{B}}}}}
	{\plto
		{\judge{P}{\hyp{G} \pp \Gamma, \cht{x'}{A}, \cht{x}{B}}}
		{l}
		{\judge{P'}{\hyp{G}' \pp \Gamma', \cht{x'}{A}, \cht{x}{B}}}
	&\phl{x,x' \notin \cn(l)}
	&\phl{\together{P'}{x}{x'}}}
}\]
\end{highlight}
Any transition that does not depend on or separates the parameters of \cref{rule:parr} (names $x$ and $x'$ and their types $A$ and $B$) is propagated.
This condition is verified by checking in the premise that $A$ and $B$ are still available in the hyperenvironment after the transition. At the process level, this corresponds to checking that the names $x$ and $x'$ are not in the label $l$ and that $P'$ supports a partition that does not separate $x$ and $x'$.

\HCP supports also a generalised version of self-synchronisation, originally introduced by \citet{MS04} together with delayed actions to model self-communication. This captures that prefixes are truly non-blocking.
The idea is to execute a prefix and a non-interfering action from its continuation at the same time.
This is the self-synchronisation rule for input actions.
\begin{highlight}
\[\rulescalebox{
\infer[\rlabel{\selfsyncrule{rule:parr}}{rule:parr-sync}]
	{\plto
		{\infer[\ref{rule:parr}]
	    {\judge{\recv{x}{x'}{P}}{\hyp{G} \pp \Gamma, \cht{x}{A \parr B}}}
	    {\judge{P}{\hyp{G} \pp \Gamma, \cht{x'}{A}, \cht{x}{B}}}}
		{\lsync{l}{\plbl{\lrecv{x}{x'}}{\lparr{A}{B}}}}
		{\judge{P'}{\hyp{G}' \pp \Gamma', \cht{x'}{A}, \cht{x}{B}}}}
	{\plto
		{\judge{P}{\hyp{G} \pp \Gamma, \cht{x'}{A}, \cht{x}{B}}}
		{l}
		{\judge{P'}{\hyp{G}' \pp \Gamma', \cht{x'}{A}, \cht{x}{B}}}
		&\phl{x,x' \notin \cn(l)}
		&\phl{\separate{{P}}{x}{\fn(l)}}
		&\phl{\together{P'}{x}{x'}}}
}\]
\end{highlight}
The rule is essentially a combination of the transition axiom for the input prefix and the rule for 
delaying its execution.

The rules for delayed execution and self-synchronisation of the remaining prefixes are obtained 
likewise, below (we omit the rules for $\one$\IfLabelExistsT{sec:omitted-rules}{, see \cref{sec:omitted-rules}}). In \cref{rule:bot-delayed}, the extracted premise $\fn(P') \neq \emptyset$ ensures, as a consequence of the proof theory, that $\Gamma'$ is not empty and that $P'$ is not a parallel composition  of $\nil$.
\begin{highlight}
\begin{spreadlines}{\proofltsskipamount}
\begin{gather*}\rulescalebox{
\infer[\rlabel{\delayedrule{rule:bot}}{rule:bot-delayed}]
	{\plto
		{\infer[\ref{rule:bot}]
	    {\judge{\wait xP}{\hyp{G} \pp \Gamma, \cht{x}{\bot}}}
	    {\judge{P}{\hyp{G}\pp \Gamma}}}
		{l}
		{\infer[\ref{rule:bot}]
	    {\judge{\wait x{P'}}{\hyp{G'} \pp \Gamma', \cht{x}{\bot}}}
	    {\judge{P'}{\hyp{G'} \pp \Gamma'}}}}
	{\plto
		{\judge{P}{\hyp{G} \pp \Gamma}}
		{l}
		{\judge{P'}{\hyp{G'} \pp \Gamma'}}
		&\phl{\fn(P') \neq \emptyset}}
}\qquad \rulescalebox{
\infer[\rlabel{\selfsyncrule{rule:bot}}{rule:bot-sync}]
	{\plto
		{\infer[\ref{rule:bot}]
	    {\judge{\wait xP}{\hyp{G} \pp \Gamma, \cht{x}{\bot}}}
	    {\judge{P}{\hyp{G}\pp \Gamma}}}
		{\lsync{l}{\plbl{\lwait{x}}{\lbot}}}
		{\judge{P'}{\hyp{G'}}}}
	{\plto
		{\judge{P}{\hyp{G} \pp \Gamma}}
		{l}
		{\judge{P'}{\hyp{G'}}}
	&\phl{\separate{\wait{x}{P}}{x}{\fn{(l)}}}}
}\\ \rulescalebox{
\infer[\rlabel{\delayedrule{rule:tensor}}{rule:tensor-delayed}]
	{\plto
		{\infer[\ref{rule:tensor}]
			{\judge{\send{x}{x'}{P}}{\hyp{G} \pp \Gamma,\Delta,\cht{x}{A \tensor B}}}
    	{\judge{P}{\hyp{G} \pp \Gamma,\cht{x'}{A} \pp \Delta,\cht{x}{B}}}}
		{l}
		{\infer[\ref{rule:tensor}]
			{\judge{\send{x}{x'}{P'}}{\hyp{G'} \pp \Gamma',\Delta',\cht{x}{A \tensor B}}}
    	{\judge{P'}{\hyp{G'} \pp \Gamma',\cht{x'}{A} \pp \Delta',\cht{x}{B}}}}}
	{\plto
		{\judge{P}{\hyp{G} \pp \Gamma,\cht{x'}{A} \pp \Delta,\cht{x}{B}}}
		{l}
		{\judge{P'}{\hyp{G'} \pp \Gamma',\cht{x'}{A} \pp \Delta',\cht{x}{B}}}
		&\phl{x,x' \notin \cn(l)}
		&\phl{\separate{P'}{x}{x'}}}
}\\\rulescalebox{
\infer[\rlabel{\selfsyncrule{rule:tensor}}{rule:tensor-sync}]
	{\plto
		{\infer[\ref{rule:tensor}]
			{\judge{\send{x}{x'}{P}}{\hyp{G} \pp \Gamma,\Delta,\cht{x}{A \tensor B}}}
    	{\judge{P}{\hyp{G} \pp \Gamma,\cht{x'}{A} \pp \Delta,\cht{x}{B}}}}
		{\lsync{\plbl{\lsend{x}{x'}}{\ltensor{A}{B}}}{l}}
   	{\judge{P'}{\hyp{G'} \pp \Gamma',\cht{x'}{A} \pp \Delta',\cht{x}{B}}}}
	{\plto
		{\judge{P}{\hyp{G} \pp \Gamma,\cht{x'}{A} \pp \Delta,\cht{x}{B}}}
		{l}
		{\judge{P'}{\hyp{G'} \pp \Gamma',\cht{x'}{A} \pp \Delta',\cht{x}{B}}}
		&\phl{x,x' \notin \cn(l)}
		&\phl{\separate{\send{x}{x'}{P}}{x}{\fn{(l)}}}
		&\phl{\separate{P'}{x}{x'}}}
}\end{gather*}%
\end{spreadlines}%
\end{highlight}%

\begin{example}
\label{ex:multiplicatives-2}
The \lts of \HCP recalls full $\beta$-reduction for the $\lambda$-calculus.
Consider again the process from \cref{ex:multiplicatives-1}: $P = \res{xy}{\left(\send{x}{x'}{Q} \pp \recv{y}{y'}{\wait{z}{R}}\right)}$ for some $Q$ and $R$ such that $P$ is well-typed. Because of delayed actions, we might observe the action on $z$ first. By \crefv{rule:cut-res,rule:mix-r,rule:parr-delayed}, and the axiom for $\lwait{z}$, we get the transition $P \lto{\lwait{z}} P'$ where $P' = \res{xy}{\left(\send{x}{x'}{Q} \pp \recv{y}{y'}{R}\right)}$.
Self-synchronisation can give rise to self-communication.
Consider the self-communicating process $S = \res{wz}{\close wP}$, where $P$ is as above. By \ref{rule:cut-one}, \ref{rule:tensor-sync}, and the transition above we get $S \lto\tau P'$.
\end{example}

\begin{example}
\label{ex:tau-shuffling}
Consider $\judge{P}{\cht{v}{\bot},\cht{w}{\one} \pp \cht{x}{\one} \pp \cht{y}{\bot},\cht{z}{\one}}$ and
$\judge{P}{\cht{w}{\one} \pp \cht{v}{\bot},\cht{x}{\one} \pp \cht{y}{\bot},\cht{z}{\one}}$ for $P = \wait{v}{\close{w}{\close{x}{\nil}}} \pp \wait{y}{\close{z}{\nil}}$. Both have a transition synchronising $x$ and $y$ (derived using \cref{rule:cut-one,rule:mix-sync}, and the axioms for \ref{rule:bot} and \ref{rule:one}) leading to $\judge{P'}{\cht{v}{\bot},\cht{w}{\one} \pp \cht{z}{\one}}$ for $P' = \wait{v}{\close{w}{\nil}} \pp \close{z}{\nil}$.
Let $Q = \res{xy}{P}$. $\judge{Q}{\cht{v}{\bot},\cht{w}{\one} \pp \cht{z}{\one}}$ has a $\tau$-transition to $\judge{P'}{\cht{v}{\bot},\cht{w}{\one} \pp \cht{z}{\one}}$ derived using \cref{rule:cut-one}; observe that types are preserved.
$\judge{Q}{\cht{w}{\one} \pp \cht{v}{\bot},\cht{z}{\one}}$ can also perform the same synchronisation and reach $\judge{P'}{\cht{v}{\bot},\cht{w}{\one} \pp \cht{z}{\one}}$; here types are preserved but $v$ has been shuffled.
\end{example}

\subsection{Additives}
The derivation rules for selection (\ref{rule:oplus_1},\ref{rule:oplus_2}) and choice (\ref{rule:with}) are given below and are obtained with the same technique as for multiplicatives.
There are left and right rules for actions, delayed actions, and communications. They are all symmetric. We omit the right cases%
\IfLabelExistsTF{sec:omitted-rules}
	{, which are \cref{sec:omitted-rules}.}
	{ here.}

\begin{highlight}
\begin{spreadlines}{\proofltsskipamount}
\begin{gather*}
\rulescalebox{\plto
	{\infer[\ref{rule:oplus_1}]
		{\judge{\inl xP}{\hyp G \pp \Gamma, \cht{x}{A \oplus B}}}
		{\judge{P}{\hyp G \pp \Gamma, \cht{x}{A}}}}
	{\plbl{\linl{x}}{\loplusl{A}{B}}}
	{\judge{P}{\hyp G \pp \Gamma, \cht{x}{A}}}
}\\ \rulescalebox{
\plto
	{\infer[\ref{rule:with}]
		{\judge{\choice{x}{P}{Q}}{\Gamma, \cht{x}{A \with B}}}
		{\judge{P}{\Gamma, \cht{x}{A}} &
		\judge{Q}{\Gamma, \cht{x}{B}}}}
	{\plbl{\lcoinl{x}}{\lwithl{A}{B}}}
	{\judge{P}{\Gamma, \cht{x}{A}}}
}\\\rulescalebox{
\infer[\rlabel{\ref*{rule:oplus_1}\ref*{rule:with}}{rule:cut-oplus_1}%
	\IfLabelExistsF{sec:omitted-rules}{\phantomlabel{infrule}{rule:cut-oplus_2}{\ref*{rule:oplus_2}\ref*{rule:with}}}%
	]
	{\plto
	  {\infer[\ref{rule:cut}]
     {\judge{\res{xy}{P}}{\hyp{G} \pp \Gamma, \Delta}}
     {\judge{P}{\hyp{G}\pp\Gamma,\cht{x}{A \tensor B} \pp \Delta, \cht{y}{\dual{A} \parr \dual{B}}}}}
	  {\tau}
	  {\infer[\ref{rule:cut}]
     {\judge{\res{xy}{P'}}{\hyp{G} \pp \Gamma, \Delta}}
     {\judge{P'}{\hyp{G} \pp \Gamma, \cht{x}{A} \pp \Delta, \cht{y}{\dual{A}}}}}
  }{\plto
		{\judge{P}{\hyp{G}\pp\Gamma, \cht{x}{A \oplus B} \pp \Delta, \cht{y}{\dual{A} \with \dual{B}}}}
		{\lsync
      {\plbl{\linl{x}}{\loplusl{A}{B}}}
      {\plbl{\lcoinl{y}}{\lwithl{\dual{A}}{\dual{B}}}}}
		{\judge{P'}{\hyp{G} \pp \Gamma, \cht{x}{A} \pp \Delta, \cht{y}{\dual{A}}}}}
}\end{gather*}
\end{spreadlines}
\end{highlight}
The rules for delayed and self-synchronising selection are straightforward.
\begin{highlight}
\begin{spreadlines}{\proofltsskipamount}
\begin{gather*}\rulescalebox{
\infer[\rlabel{\delayedrule{rule:oplus_1}}{rule:oplus_1-delayed}%
	\IfLabelExistsF{sec:omitted-rules}{\phantomlabel{infrule}{rule:oplus_2-delayed}{\delayedrule{rule:oplus_2}}}%
	]
	{\plto
		{\infer[\ref{rule:oplus_1}]
	    {\judge{\inl{x}{P}}{\hyp{G} \pp \Gamma, \cht{x}{A \oplus B}}}
	    {\judge{P}{\hyp{G} \pp \Gamma, \cht{x}{A}}}}
		{l}
		{\infer[\ref{rule:oplus_1}]
			{\judge{\inl{x}{P'}}{\hyp{G'} \pp \Gamma', \cht{x}{A \oplus B}}}
			{\judge{P'}{\hyp{G'} \pp \Gamma', \cht{x}{A}}}}}
	{\plto
		{\judge{P}{\hyp{G} \pp \Gamma, \cht{x}{A}}}
		{l}
		{\judge{P'}{\hyp{G'} \pp \Gamma', \cht{x}{A}}}
		&\phl{x \notin \cn(l)}}
}\\ \rulescalebox{
\infer[\rlabel{\selfsyncrule{rule:oplus_1}}{rule:oplus_1-sync}%
	\IfLabelExistsF{sec:omitted-rules}{\phantomlabel{infrule}{rule:oplus_2-sync}{\selfsyncrule{rule:oplus_2}}}%
	]
	{\plto
		{\infer[\ref{rule:oplus_1}]
	    {\judge{\inl{x}{P}}{\hyp{G} \pp \Gamma, \cht{x}{A \oplus B}}}
	    {\judge{P}{\hyp{G} \pp \Gamma, \cht{x}{A}}}}
		{\lsync{\plbl{\linl{x}}{\loplusl{A}{B}}}{l}}
		{\judge{P'}{\hyp{G'} \pp \Gamma, \cht{x}{A}}}}
	{\plto
		{\judge{P}{\hyp{G} \pp \Gamma, \cht{x}{A}}}
		{l}
		{\judge{P'}{\hyp{G'} \pp \Gamma, \cht{x}{A}}}
		&\phl{x \notin \cn(l)}
		&\phl{\separate{\inl{x}{P}}{x}{\fn(l)}}}
}\end{gather*}%
\end{spreadlines}%
\end{highlight}%
We choose not to define rules for delayed or self-synchronising choices, since \cref{rule:with} does not allow for internal independent components ($\hyp G$).

\begin{remark}
If we wished to allow for delayed choices, we could add the following rule. The rule allows for delaying a choice if its two branches simultaneously undergo transformations with the same label and to targets with no parallel components. (We omit the rule for self-syn\-chron\-isa\-tion.) 
\begin{highlight}
\begin{spreadlines}{\proofltsskipamount}
\begin{gather*}\rulescalebox{
\infer[\rlabel{\delayedrule[_0]{rule:with}}{rule:with_0-delayed}]
  {\plto
  	{\infer[\ref{rule:with}]
  		{\judge{\choice x{P_1}{P_2}}{\Gamma, \cht{x}{A_1 \with A_2}}}
  		{\judge{P_1}{\Gamma, \cht{x}{A_1}}
  		&\judge{P_2}{\Gamma, \cht{x}{A_2}}}}
  	{l}
  	{\infer[\ref{rule:with}]
  		{\judge{\choice x{P_1'}{P_2'}}{\Gamma, \cht{x}{A_1 \with A_2}}}
  		{\judge{P'}{\Gamma', \cht{x}{A_1}}
  		&\judge{Q'}{\Gamma', \cht{x}{A_2}}}}}
  {\plto
  	{\judge{P_i}{\Gamma', \cht{x}{A_i}}}
  	{l}
  	{\judge{P'_i}{\Gamma', \cht{x}{A_i}}}
  &\phl{x \notin \cn(l)}
& \phl{|\np(P'_i)| = 1}
	&\phl{\text{for } i \in \{1,2\}}}
}\end{gather*}%
\end{spreadlines}%
\end{highlight}%
\end{remark}

\begin{remark}\label{rem:lifting-with}
Keeping the analogy with full $\beta$-reduction (\cref{ex:multiplicatives-1}), we could add the following transition rule for lifting internal actions by a choice branch (we omit the symmetric rule).
\begin{highlight}
\begin{spreadlines}{\proofltsskipamount}
\begin{gather*}\rulescalebox{
\infer[\rlabel{\delayedrule[_1]{rule:with}}{rule:with_1-delayed}]
  {\plto
  	{\infer[\ref{rule:with}]
  		{\judge{\choice x{P}{Q}}{\Gamma, \cht{x}{A \with B}}}
  		{\judge{P}{\Gamma, \cht{x}{A}}
  		&\judge{Q}{\Gamma, \cht{x}{B}}}}
  	{\tau}
  	{\infer[\ref{rule:with}]
  		{\judge{\choice x{P'}{Q}}{\Gamma, \cht{x}{A \with B}}}
  		{\judge{P'}{\Gamma, \cht{x}{A}}
  		&\judge{Q}{\Gamma, \cht{x}{B}}}}}
  {\plto
  	{\judge{P}{\Gamma, \cht{x}{A}}}
  	{\tau}
  	{\judge{P'}{\Gamma, \cht{x}{A}}}}
}\end{gather*}%
\end{spreadlines}%
\end{highlight}%
We choose not to, purely because it is unintuitive that a branch may perform any kind of computation before it is selected. Moreover, the rule does not change the expressiveness of \HCP and its behavioural theory (our semantic equivalences abstract from internal actions, \cf \cref{sec:behavioural-theory}).
\end{remark}

\subsection{Links}\label{sec:hcp-lts-links}
There are two transitions for \ref{rule:axiom} and are given by the (symmetric) axioms below.
\begin{highlight}
\[\rulescalebox{
  \plto
  	{\infer[\ref{rule:axiom}]
  		{\judge{\forward{x}{y}}{\cht{x}{\dual{A}}, \cht{y}{A}}}
  		{\phantom{\pp}}}
  	{\plbl{\lforward{x}{y}}{\laxiom{A}}}
  	{\judge{\nil}{\emptyhyp}}
}\qquad\rulescalebox{
  \plto
   	{\infer[\ref{rule:axiom}]
   		{\judge{\forward{x}{y}}{\cht{x}{\dual{A}}, \cht{y}{A}}}
   		{\phantom{\pp}}}
   	{\plbl{\lforward{y}{x}}{\laxiom{\dual{A}}}}
   	{\judge{\nil}{\emptyhyp}}
}\]
\end{highlight}
The two transitions differ only for the order of names in the label to capture the symmetry of the link.
\Cref{rule:cut-forward} below corresponds to the cut of \cref{rule:axiom}.
\begin{highlight}
\begin{spreadlines}{\proofltsskipamount}
\begin{gather*}\rulescalebox{
\infer[\rlabel{\rname{\ref*{rule:axiom}\rname{Cut}}}{rule:cut-forward}]
	{\plto
	  {\infer[\ref{rule:cut}]
     {\judge{\res{yz}{P}}{\hyp{G} \pp \Gamma, \Delta, \cht{x}{\dual{A}}}}
     {\judge{P}{\hyp{G}\pp \Gamma,\cht{x}{\dual{A}},\cht{y}{A} \pp \Delta, \cht{z}{\dual{A}}}}}
	  {\tau}
	  {\judge{P'\{x/z\}}{\hyp{G}' \pp \Gamma', \cht{x}{\dual{A}}}}
  }{\plto
		{\judge{P}{\hyp{G}\pp \Gamma,\cht{x}{\dual{A}},\cht{y}{A} \pp \Delta, \cht{z}{\dual{A}}}}
		{\plbl{\lforward{x}{y}}{\laxiom{A}}}
		{\judge{P'}{\hyp{G}' \pp \Gamma',\cht{z}{\dual{A}}}}}
}\end{gather*}
\end{spreadlines}
\end{highlight}

\subsection{Exponentials}

In \HCP, clients can interact with servers in three ways: requesting an instance of the service provided by the server, duplicating a server, or disposing of a server.

\paragraph{Requesting an instance} Client requests for server instances are modelled by the following rules, for the prefixes and their communication.
\begin{highlight}
\begin{spreadlines}{\proofltsskipamount}
\begin{gather*}\rulescalebox{
\plto
	{\infer[\ref{rule:?}]
		{\judge{\clientuse x{x'}P}{\hyp{G} \pp \Gamma, \cht{x}{\query A}}}
		{\judge{P}{\hyp{G} \pp \Gamma, \cht{x'}{A}}}}
	{\plbl{\luse{x}{x'}}{\lquery{A}}}
	{\judge{P}{\hyp{G} \pp \Gamma, \cht{x'}{A}}}
}\quad\rulescalebox{
\plto
	{\infer[\ref{rule:!}]
		{\judge{\server{x}{x'}{P}}{\query\Gamma, \cht{x}{\bang A}}}
		{\judge{P}{\query\Gamma, \cht{x'}{A}}}}
	{\plbl{\lcouse{x}{x'}}{\lbang{A}}}
	{\judge{P}{\query\Gamma, \cht{x'}{A}}}
}\\\rulescalebox{
\infer[\rlabel{\ref*{rule:!}\ref*{rule:?}}{rule:cut-?}]
{\plto
	{\infer[\ref{rule:cut}]
	{\judge
		{\res{xy}{P}}
		{\hyp{G} \pp \Gamma, \Delta}}
	{\judge
		{P}
		{\hyp{G} \pp \Gamma, \cht{x}{\query A} \pp \Delta,\cht{y}{\bang \dual A}}}}
	{\tau}
	{\infer[\ref{rule:cut}]
	{\judge
		{\res{x'y'}{P'}}
		{\hyp{G} \pp \Gamma, \Delta}}
	{\judge
		{P'}
		{\hyp{G} \pp \Gamma, \cht{x'}{A} \pp \Delta,\cht{y'}{\dual A}}}}}
{\plto
	{\judge
		{P}
		{\hyp{G} \pp \Gamma, \cht{x}{\query A} \pp \Delta,\cht{y}{\bang \dual A}}}
	{\lsync
    {\plbl{\luse{x}{x'}}{\lquery{A}}}
    {\plbl{\lcouse{y}{y'}}{\lbang{\dual{A}}}}}
	{\judge{P'}
		{\hyp{G} \pp \Gamma, \cht{x'}{A} \pp \Delta, \cht{y'}{\dual A}}}}
}\end{gather*}
\end{spreadlines}
\end{highlight}
Instance requests can be delayed and self-synchronise as follows.
\begin{highlight}
\begin{spreadlines}{\proofltsskipamount}
\begin{gather*}\rulescalebox{
	\infer[\rlabel{\delayedrule{rule:?}}{rule:?-delayed}]
		{\plto
		{\infer[\ref{rule:?}]
			{\judge{\clientuse{x}{x'}{P}}{\hyp{G} \pp \Gamma, \cht{x}{\query A}}}
			{\judge{P}{\hyp{G} \pp \Gamma, \cht{x'}{A}}}}
		{l}
		{\infer[\ref{rule:?}]
			{\judge{\clientuse{x}{x'}{P'}}{\hyp{G'} \pp \Delta, \cht{x}{\query A}}}
			{\judge{P'}{\hyp{G'} \pp \Gamma, \cht{x'}{A}}}}}
		{\plto
			{\judge{P}{\hyp{G} \pp \Gamma, \cht{x'}{A}}}
			{l}
			{\judge{P'}{\hyp{G'} \pp \Gamma, \cht{x'}{A}}}
			&\phl{x,x' \notin \cn(l)}}
}\\\rulescalebox{
	\infer[\rlabel{\selfsyncrule{rule:?}}{rule:?-sync}]
		{\plto
		{\infer[\ref{rule:?}]
			{\judge{\clientuse{x}{x'}{P}}{\hyp{G} \pp \Gamma, \cht{x}{\query A}}}
			{\judge{P}{\hyp{G} \pp \Gamma, \cht{x'}{A}}}}
		{\lsync{\plbl{\luse{x}{x'}}{\lquery{A}}}{l}}
		{\judge{P'}{\hyp{G'} \pp \Gamma, \cht{x'}{A}}}}
		{\plto
			{\judge{P}{\hyp{G} \pp \Gamma, \cht{x'}{A}}}
			{l}
			{\judge{P'}{\hyp{G'} \pp \Gamma, \cht{x'}{A}}}
			&\phl{x,x' \notin \cn(l)}
			&\phl{\separate{\clientuse{x}{x'}{P}}{x}{\fn(l)}}}
}\end{gather*}
\end{spreadlines}
\end{highlight}

\paragraph{Server duplication}
Server duplication is captured by the following axioms and synchronisation rule. Since servers might contain dependencies, \ie, client terms (the free names in $\query\Gamma$ in \cref{rule:!}), their duplication must be propagated to their dependencies before the new copies can be used.
For a substitution $\sigma$, we write $P\sigma$, $\Gamma\sigma$, and $x\sigma$ for the application of $\sigma$ to a process, environment, and name, respectively.

\begin{highlight}
\begin{spreadlines}{\proofltsskipamount}
\begin{gather*}
\rulescalebox{
	{\plto
  	{\infer[\ref{rule:contract}]
  		{\judge{\clientdup{x}{x_1}{x_2}{P}}{\hyp{G} \pp \Gamma, \cht{x}{\query A}}}
  		{\judge{P}{\hyp{G} \pp \Gamma, \cht{x_1}{\query A},\cht{x_2}{\query A}}}}
  	{\plbl{\ldup{x}{x_1}{x_2}}{\lcontract{A}}}
  	{\infer[\ref{rule:parr}]
  		{\judge{\recv{x_1}{x_2}{P}}{\hyp{G} \pp \Gamma, \cht{x_1}{\query A \parr \query A}}}
  		{\judge{P}{\hyp{G} \pp \Gamma, \cht{x_1}{\query A},\cht{x_2}{\query A}}}}}
}\\ \rulescalebox{
\infer
	{\plto*
  	{\infer[\ref{rule:!}]
  		{\judge{\server{x}{x'}{P}}{\query\Gamma, \cht{x}{\bang A}}}
  		{\judge{P}{\bang\Gamma, \cht{x'}{A}}}}
  	{\plbl{\lcodup{x}{x_1}{x_2}}{\lcocontract{A}}}
  	{\infer=[\ref{rule:contract}]
  		{\judge
  			{\clientdup{z_1}{z_1\sigma_1}{z_1\sigma_2}{\dots\clientdup{z_n}{z_n\sigma_1}{z_n\sigma_2}{\send{x_1}{x_2}{(\server{x_1}{x'\sigma_1}{P_1} \pp \server{x_2}{x'\sigma_2}{P_2})}}}}
  			{\query\Gamma, \cht{x_1}{\bang A \tensor \bang A}}}
	  	{\infer[\ref{rule:tensor}]
	  		{\judge{\send{x_1}{x_2}{(\server{x_1}{x'\sigma_1}{P} \pp \server{x_2}{x'\sigma_2}{P})}}{\query\Gamma_1,\query\Gamma_2, \cht{x_1}{\bang A \tensor \bang A}}}
	  		{\infer[\ref{rule:mix}]
 					{\judge
 						{\server{x_1}{x'\sigma_1}{P_1} \pp \server{x_2}{x'\sigma_2}{P_2}}
 						{\query\Gamma_1,\cht{x_1}{\bang A} \pp \query\Gamma_2,\cht{x_2}{\bang A}}}
 					{\infer[\ref{rule:!}]
 						{\judge{\server{x_1}{x'\sigma_1}{P_1}}{\query\Gamma_1, \cht{x_1}{\bang A}}}
 						{\judge{P_1}{\query\Gamma_1, \cht{x'\sigma_1}{A}}}
 					&\infer[\ref{rule:!}]
 						{\judge{\server{x_2}{x'\sigma_2}{P_2}}{\query\Gamma_2, \cht{x_2}{\bang A}}}
 						{\judge{P_2}{\query\Gamma_2, \cht{x'\sigma_2}{A}}}}}}}}
  {\phl{P_i = P\sigma_i}
 	&\thl{\Gamma_i = \Gamma\sigma_i}
 	&\phl{\fn(P) = \{x',z_1,\dots,z_n\}}
 	&\phl{\text{for } i \in \{1,2\}}}
}\\ \rulescalebox{
\infer[\rlabel{\ref*{rule:!}\ref*{rule:contract}}{rule:cut-contract}]
{\plto
	{\infer[\ref{rule:cut}]
    {\judge{\res{xy}{P}}{\hyp{G} \pp \Gamma, \query \Delta}}
    {\judge{P}{\hyp{G} \pp \Gamma, \cht{x}{\query A} \pp \query \Delta,\cht{y}{\bang \dual A}}}}
	{\tau}
	{\infer[\ref{rule:cut}]
		{\judge{\res{x_1y_1}{P'}}{\hyp{G} \pp \Gamma, \query \Delta}}
		{\judge{P'}{\hyp{G} \pp \Gamma, \cht{x_1}{\query A\parr\query A} \pp \query \Delta,\cht{y_1}{\bang \dual A \tensor\bang \dual A}}}}}
{\plto
	{\judge{P}{\hyp{G} \pp \Gamma, \cht{x}{\query A} \pp \query \Delta,\cht{y}{\bang \dual A}}}
	{\lsync
   {\plbl{\ldup{x}{x_1}{x_2}}{\lcontract{A}}}
   {\plbl{\lcodup{y}{y_1}{y_2}}{\lcocontract{\dual A}}}}
	{\judge{P'}{\hyp{G} \pp \Gamma, \cht{x_1}{\query A\parr\query A} \pp \query \Delta,\cht{y_1}{\bang \dual A \tensor\bang \dual A}}}}
}\end{gather*}
\end{spreadlines}
\end{highlight}

Server duplication is done slightly differently compared to \HCP's predecessor, Classical Processes (\CP) \citep{W14}. Because in \CP \cref{rule:contract,rule:weaken} lack proof terms (they can be applied ``silently''), the duplication of all the resources that a server needs to introduce a new copy is not visible at the process level, whereas it is in \HCP. Our semantics has thus a clearer computational interpretation.
Also, the duplication of a server in \CP is handled by the synchronisation rule for cut of \cref{rule:!} with \cref{rule:contract} (\cite[Fig.~3]{W14}), whereas our semantics respects locality: the new servers appears exactly where the original server was, and our transition rule is independent of the presence of any cut in the context.

Another difference with \CP's server duplication is the exchange following a duplication request (terms $\recvs{y_1}{y_2}$ and $\sends{y_1}{y_2}$ in the targets of the transitions). This exchange is suggested by the typing: server copies must be in parallel components (they might be used concurrently) whereas \cref{rule:contract} assumes copies are in the same environment.
We could do away with these exchanges by adding \emph{external contraction} to our proof theory, a variant of \cref{rule:contract} where the two names to appear in separate environments \citep{A91}.
We chose not to, for simplicity of our proof theory.
Because the exchange can only take place after the server sends all requests to duplicate its dependencies and because server copies can only be used after the exchange has been done, it has a clear operational interpretation: it is an acknowledgement.

Duplication requests can be delayed and self-synchronise.
\begin{highlight}
\begin{spreadlines}{\proofltsskipamount}
\begin{gather*}
\rulescalebox{
\infer[\rlabel{\delayedrule{rule:contract}}{rule:contract-delayed}]
	{\plto
  	{\infer[\ref{rule:contract}]
  		{\judge{\clientdup{x}{x_1}{x_2}{P}}{\hyp{G} \pp \Gamma, \cht{x}{\query A}}}
  		{\judge{P}{\hyp{G} \pp \Gamma, \cht{x_1}{\query A},\cht{x_2}{\query A}}}}
  	{l}
   	{\infer[\ref{rule:contract}]
   		{\judge{\clientdup{x}{x_1}{x_2}{P'}}{\hyp{G'} \pp \Gamma', \cht{x}{\query A}}}
   		{\judge{P'}{\hyp{G'} \pp \Gamma', \cht{x_1}{\query A},\cht{x_2}{\query A}}}}}
	{\plto
		{\judge{P}{\hyp{G} \pp \Gamma, \cht{x_1}{\query A},\cht{x_2}{\query A}}}
  	{l}
  		{\judge{P'}{\hyp{G'} \pp \Gamma', \cht{x_1}{\query A},\cht{x_2}{\query A}}}
  	&\phl{x,x_1,x_2 \notin \cn(l)}
    &\phl{\together{P'}{x_1}{x_2}}}
}\\ \rulescalebox{
\infer[\rlabel{\selfsyncrule{rule:contract}}{rule:contract-sync}]
	{\plto
  	{\infer[\ref{rule:contract}]
  		{\judge{\clientdup{x}{x_1}{x_2}{P}}{\hyp{G} \pp \Gamma, \cht{x}{\query A}}}
  		{\judge{P}{\hyp{G} \pp \Gamma, \cht{x_1}{\query A},\cht{x_2}{\query A}}}}
  	{\lsync{\plbl{\ldup{x}{x_1}{x_2}}{\lcontract{A}}}{l}}
   	{\infer[\ref{rule:parr}]
   		{\judge{\recv{x_1}{x_2}{P'}}{\hyp{G}' \pp \Gamma', \cht{x_1}{\query A \parr \query A}}}
   		{\judge{P'}{\hyp{G}' \pp \Gamma', \cht{x_1}{\query A},\cht{x_2}{\query A}}}}}
	{\plto
		{\judge{P}{\hyp{G} \pp \Gamma, \cht{x_1}{\query A},\cht{x_2}{\query A}}}
  	{l}
  	{\judge{P'}{\hyp{G}' \pp \Gamma', \cht{x_1}{\query A},\cht{x_2}{\query A}}}
    &\!\phl{x,x_1,x_2 \!\notin\! \cn(l)}
    &\!\phl{\together{P'}{x_1\!}{\!x_2}}
    &\!\phl{\separate{\clientdup{x}{x_1}{x_2}{P}}{x\!}{\!\fn(l)}}}
}\end{gather*}
\end{spreadlines}
\end{highlight}
Duplication requests by a server being duplicated cannot be delayed after the operation acknowledgement: $\sends{y_1}{y_2}$ will always be the last operation before the new instances become available.

\begin{example}
Continuing \cref{ex:and-typing}, we can formally observe that using different clients that send different bits yields different results.
\[
\begin{array}{lcl}
\res{xy}{\left( Client^{01}_{xz} \pp Server_y \right)}
& \lto{\tau} \cdots \lto{\tau} \lto{\linl{z}} \lto{\lclose{z}} & \nil \pp \nil
\\
\res{xy}{\left( Client^{11}_{xz} \pp Server_y \right)}
& \lto{\tau} \cdots \lto{\tau} \lto{\linr{z}} \lto{\lclose{z}} & \nil \pp \nil
\end{array}
\]
We can use the service multiple times by duplicating it.
\[
\res{xy}{\left(
\clientdup{x}{x_1}{x_2}{\left(
	Client^{01}_{x_1z_1} \pp
	Client^{11}_{x_2z_2}
\right)}
\pp Server_y
\right)}
\lto{\tau} \cdots \lto{\tau} \lto{\linl{z_1}}
\lto{\linr{z_2}}
\lto{\lclose{z_1}}
\lto{\lclose{z_2}}
\nil \pp \nil \pp \nil \pp \nil
\]
\end{example}

\begin{remark}\label{rem:lifting-!}
\IfLabelExistsTF{rem:lifting-with}
	{In connection to \cref{rem:lifting-with}, }%
	{Keeping the analogy with full $\beta$-reduction (\cref{ex:multiplicatives-1}), }%
we could add the following transition rule for lifting internal actions by servers.
\begin{highlight}
\begin{spreadlines}{\proofltsskipamount}
\begin{gather*}\rulescalebox{
	\infer[\rlabel{\delayedrule{rule:!}}{rule:!-delayed}]
		{\plto
		{\infer[\ref{rule:!}]
				{\judge{\server{x}{x'}{P}}{\query\Gamma, \cht{x}{\bang A}}}
				{\judge{P}{\query\Gamma, \cht{x'}{A}}}}
		{\tau}
		{\infer[\ref{rule:!}]
			{\judge{\server{x}{x'}{P'}}{\query\Gamma', \cht{x}{\bang A}}}
			{\judge{P}{\query\Gamma', \cht{x'}{A}}}}}
		{\plto
			{\judge{P}{\query\Gamma, \cht{x'}{A}}}
			{\tau}
			{\judge{P'}{\query\Gamma', \cht{x'}{A}}}}
}\end{gather*}
\end{spreadlines}
We choose to exclude it because it might be surprising that a server performs any kind of computation before it is triggered. However, in general, the rule might be interesting to allow servers to ``optimise'' themselves before being run.
\end{highlight}
\end{remark}

\paragraph{Server disposal}
Server disposal is captured by the following rules, again for the dual prefixes and their communication. Since servers might contain dependencies, their disposal must be propagated before the operation is acknowledged to the client. 
\begin{highlight}
\begin{spreadlines}{\proofltsskipamount}
\begin{gather*}
\rulescalebox{
\plto*
	{\infer[\ref{rule:weaken}]
		{\judge{\clientdisp{x}{P}}{\hyp{G} \pp \Gamma, \cht{x}{\query A}}}
		{\judge{P}{\hyp{G} \pp \Gamma}}}
	{\plbl{\ldisp{x}}{\lweaken{A}}}
	{\infer[\ref{rule:bot}]
		{\judge{\wait{x}{P}}{\hyp{G} \pp \Gamma, \cht{x}{\bot}}}
		{\judge{P}{\hyp{G} \pp \Gamma}}}
}\quad \rulescalebox{
  \infer
	{\plto
		{\infer[\ref{rule:!}]
			{\judge{\server{x}{x'}{P}}{\query \Gamma, \cht{x}{\bang A}}}
			{\judge{P}{\query \Gamma, \cht{x'}{A}}}}
		{\plbl{\lcodisp{x}}{\lcoweaken{A}}}
		{\infer=[\ref{rule:weaken}]
			{\judge
				{\clientdisp{z_1}{\dots\clientdisp{z_n}{\close{x}{\nil}}}}
				{\query \Gamma, \cht{x}{\one}}}
			{\infer[\ref{rule:one}]
				{\judge{\close{x}{\nil}}{\cht{x}{\one}}}
				{\infer[\ref{rule:mix_0}]
					{\judge{\nil}{\emptyhyp}}
					{}}}}}
	{\phl{\fn(P) = \{x',z_1,\dots,z_n\}}}
}\\ \rulescalebox{
\infer[\rlabel{\ref*{rule:!}\ref*{rule:weaken}}{rule:cut-weaken}]
{\plto
	{\infer[\ref{rule:cut}]
		{\judge{\res{xy}{P}}{\hyp{G} \pp \Gamma, \query \Delta}}
		{\judge{P}{\hyp{G} \pp \Gamma, \cht{x}{\query A} \pp \query \Delta,\cht{y}{\bang \dual A}}}}
	{\tau}
	{\infer[\ref{rule:cut}]
		{\judge{\res{xy}{P'}}{\hyp{G} \pp \Gamma, \query \Delta}}
		{\judge{P}{\hyp{G} \pp \Gamma, \cht{x}{\bot} \pp \query \Delta,\cht{y}{\one}}}}}
{\plto
	{\judge{P}{\hyp{G} \pp \Gamma, \cht{x}{\query A} \pp \query \Delta,\cht{y}{\bang \dual A}}}
	{\lsync
		{\plbl{\ldisp{x}}{\lweaken{A}}}
		{\plbl{\lcodisp{y}}{\lcoweaken{\dual{A}}}}}
	{\judge{P}{\hyp{G} \pp \Gamma, \cht{x}{\bot} \pp \bang\Delta,\cht{y}{\one}}}}
}\end{gather*}
\end{spreadlines}
\end{highlight}%
\looseness=-1
As for instance requests, disposal requests can be delayed and self-synchronise (we omit the proof  transformations%
\IfLabelExistsTF{sec:omitted-rules}%
{, see \cref{sec:omitted-rules}}%
{\phantomlabel{infrule}{rule:weaken-delayed}{\delayedrule{rule:weaken}}%
\phantomlabel{infrule}{rule:weaken-sync}{\selfsyncrule{rule:weaken}}}%
).
Akin to duplication, disposal requests by a server being disposed cannot be delayed after the operation acknowledgement: $\waits{x}$ will always be the last operation.

\subsection{Subject reduction and Progress}

All transition rules are derived from proof transformations that preserve provability. Rules for $\tau$-transitions preserve also types: the types in (the judgement in) the conclusion remain unchanged. Thus, we immediately obtain the theorem below. Given a proof transition label $l$, we write $\strip{l}$ for the same label where the logical part (the blue part) is stripped, \eg $\strip{\plbl{\lclose{x}}{\lone}} = \lclose{x}$.

\begin{theorem}[Subject reduction]
	\label{thm:hcp-lts-subject-reduction}
  Let $P$ and $Q$ be well-typed. Then:
	\begin{itemize}[nosep,noitemsep]
    \item If\,  $\plto{\judge{P}{\hyp{G}}}{l}{\judge{Q}{\hyp{H}}}$ then, $P \lto{\strip{l}} Q$ and $\hyp{G} \shuffle \hyp{H}$ iff $l = \tau$.
 		\item If $\judge{P}{\hyp{G}}$ and $P \lto{l} Q$, then $\plto{\judge{P}{\hyp{G}}}{l'}{\judge{Q}{\hyp{H}}}$ for some $\hyp{H}$ and $l'$ such that $l = \strip{l'}$.
	\end{itemize}
\end{theorem}

\begin{proofatend}
By induction on the derivation of the transition relation on proofs and processes.
The cases follow essentially by the fact that our transition rules are extracted from proofs transformations, which as we showed manipulate types correctly.
\end{proofatend}

\Cref{thm:hcp-lts-subject-reduction} formalises that $\tau$-transitions of a process $P$ have no dependencies on the context, since they do not influence the types of $P$. This matches the intuition of \lts semantics for the $\pi$-calculus, where $\tau$-transitions capture internal ``unobservable'' moves. In \HCP, performing unobservable moves coincides with type-preserving proof transitions.

The next results use ``saturated'' transitions (Milner's double arrow).
Formally, $\slto{}$ is the smallest relation such that: $P \slto{\tau} P$ for all $P$; and if $P 
\slto{\tau} P'$, $P' \lto{l} Q'$ and $Q' \slto{\tau} Q$, then $P \slto{l} Q$.

We write $l_x$ to range over labels that denote actions on a name $x$: $\lclose x$, $\lwait{x}$, $\lsend xy$, $\lrecv xy$, \etc. Actions typed by separate environments can always be observed in parallel.
\begin{lemma}\label{lemma:parallel-actions}
If $\judge{P}{\hyp G \pp \Gamma, \cht{x}{A} \pp \Delta, \cht{y}{B}}$,
$P \slto{l_x} P'$, and $P \slto{l_y}{P''}$, then there exists $Q$ such that
$P \slto{\lsync{l_x}{l_y}} Q$ (up to $\alpha$-renaming).
\end{lemma}
\begin{proofatend}
By induction on the structure of $P$.
If $P = R_1 \pp R_2$, then either: the two transitions come from $R_i$ for $i\in \{1,2\}$, and then we use the induction hypothesis on $R_i$ and apply \cref{rule:mix-l}; if the two transitions come one from $R_1$ and the other from $R_2$, then we apply \cref{rule:mix-sync}.
If $P$ is a restriction, then the thesis follows by applying \cref{rule:cut-res} to the induction hypothesis.
If $P = \pi.R$ and $x,y\not\in\fn(\pi)$, then the thesis follows by induction hypothesis and by applying the delayed transition rule for $\pi$.
If $P = \pi.R$ and $\{x,y\}\cap\fn(\pi)\neq \emptyset$, then the thesis follows by applying the self-synchronisation rule for $\pi$.
\end{proofatend}

Well-typed processes enjoy a notion of readiness on their free channels, captured by the separation of environments. Assume that $\judge{P}{\hyp G \pp \Gamma}$. Then, there is always a name $x$ in $\cn(\Gamma)$ such that $P \slto{l_x} P'$. In other words, every environment in the hyperenvironment used to type a process contains at least one name where the process can perform an action (possibly after $\tau$-transitions).
\begin{theorem}[Readiness]
\label{thm:readiness}
Let $\judge{P}{\Gamma_1 \pp \cdots \pp \Gamma_n}$. For every $i \in [1,n]$, there exist $x \in \cn(\Gamma_i)$, $l_x$, and $P'$ such that $P \slto{l_x} P'$.
\end{theorem}

\begin{proofatend}
Let $\hyp G = \Gamma_1 \pp \cdots \pp \Gamma_n$, and thus $\judge{P}{\hyp G}$.
We proceed by nested induction on the number of prefixes $\ldup{x}{x_1}{x_2}$ not under a server prefix $\bang x(y)$ (this is needed for dealing with contraction, as usual in classical linear logic), the derivation of $\judge{P}{\hyp G}$ and the size of the types in $\hyp G$. We write $\pi_x$ for a prefix that acts on name $x$.

\begin{enumerate}
\item If $P$ has the form $\forward{x}{y}$, $\choice{x}{Q}{R}$, or $\server{x}{y}{Q}$, then from typing we know that $\hyp G = \Gamma$ for some $\Gamma$ and that $x \in \cn(\Gamma)$. The thesis follows by applying the corresponding transition rules for consuming the head of each of these terms.

\item If $P$ has the form $\pi_x.Q$, where $\pi_x \neq \bang x(y)$, then we know that $\hyp G = \hyp H \pp \Gamma, \cht{x}{A}$ for some $\hyp H$ and $\Gamma$. For readiness on $\Gamma, \cht{x}{A}$, we can apply the transition rule for $\pi_x$. For readiness on each environment in $\hyp H$, we invoke the induction hypothesis on $Q$ and apply the corresponding transition rule for delaying $\pi_x$. We know that the appropriate transition rule for delaying $\pi_x$ can always be applied, because disjointness of names between $\hyp H$ and $\Gamma, \cht{x}{A}$ ensures that the premises are satisfied.

\item If $P$ has the form $Q \pp R$, then $\hyp G = \hyp H \pp \hyp I$, $\judge{Q}{\hyp H}$, and $\judge{R}{\hyp I}$. We know that
at least one between $\hyp H$ and $\hyp I$ is a non-empty hyperenvironment.
If $\hyp H$ is non-empty, we apply the induction hypothesis to $Q$ and lift the transition using \cref{rule:mix-l}. Otherwise, if $\hyp I$ is non-empty, we follow the same reasoning for $R$ and apply \cref{rule:mix-r}.

\item If $P = \res{xy}{Q}$, then by typing $\judge{P}{\hyp H \pp \Gamma,\Delta}$
and $\judge{Q}{\hyp H \pp \Gamma, \cht{x}{A} \pp \Delta, \cht{y}{\dual A}}$.
To show readiness for each environment in $\hyp H$, we invoke the induction hypothesis on $Q$ and lift the resulting transition using \cref{rule:cut-res}---this is possible because we know that these transitions are on channels that are not $x$ or $y$, by disjointness of names in hyperenvironments.

To show readiness for $\Gamma, \Delta$, we invoke the induction hypothesis on $Q$ and get two subcases.
\begin{enumerate}
\item $Q \slto{l_z} Q'$ for some $z \in \cn(\Gamma,\Delta)$. The thesis follows by lifting the transition with \cref{rule:cut-res}; it suffices to show that the separation premise $\separate{P'}{x}{y}$ of the rule is satisfied.
Assume \wl that $z$ is in $\Gamma$ and that is is the only ready name.
If $z$ triggers the shuffling of $x$, then the action on $z$ is $\lclose{z}$ or $\lforward{z}{z'}$ since \cref{rule:one,rule:axiom} are the only ones to introduce new environments.
It follows from $Q$ being well-typed that actions on $x$ are delayed to activate that on $z$ and hence $x$ must also ready.
Likewise, if there is a ready name in $\Delta$ and its actions triggers the shuffling of $y$, then also $y$ must be ready.
Therefore, if there is no $l_z$ such that the separation premise $\separate{P'}{x}{y}$ of the rule is satisfied both $x$ and $y$ are ready and the next case applies.
\item 
$Q \notslto{l_z} Q'$ for any $z \in \cn(\Gamma,\Delta)$.
Then, readiness on $\Gamma, \cht{x}{A}$ for $Q$ dictates that $Q \slto{l_x} Q'$ and readiness on $\Delta, \cht{y}{\dual A}$ for $Q$ dictates that $Q \slto{l_y} Q''$.
By \cref{lemma:parallel-actions}, we get $Q \slto{\lsync{l_x}{l_y}} R$ for some $R$.
Since labels are extracted from proof transformations, where there is a correspondence between actions (red part) and their types (blue part), and the types of $x$ and $y$ are dual of each other, we know that $l_x$ and $l_y$ are compatible actions (up to $\alpha$-renaming) and thus we can apply exactly one transition rule for a communication.
If we apply a communication rule for multiplicatives or additives, or \cref{rule:cut-?}, then the thesis follows by induction hypothesis on the result of the transition.
If we apply \cref{rule:cut-weaken}, then we have two cases: if the receiving server term contained clients, then the transition produced new prefixes for which we can immediately apply the transition rule for $\lquery{z}$ and the thesis follows; otherwise, the thesis follows by induction hypothesis on the result of the transition.
If we apply \cref{rule:cut-contract}, then we have two cases: if the receiving server term contained clients, then the transition produced new prefixes for which we can immediately apply the transition rule for duplication requests;
otherwise, the thesis follows by induction hypothesis on the result of the transition (the number of duplication request terms not in a server decreased).
\end{enumerate}
Showing readiness for $\Delta, \cht{y}{\dual A}$ follows the same reasoning as for $\Gamma, \cht{x}{A}$, thanks to the involution of duality.
\qedhere
\end{enumerate}
\end{proofatend}

As a corollary of \cref{thm:readiness}, we get that well-typed processes that are not terminated can always progress, because the hyperenvironment used to type a process $P\not\equiv\nil$ can never be empty.
\begin{corollary}[Progress]
If $P$ is well-typed and $P \not\equiv \nil$, then $P \lto{l} P'$ for some $l$ and $P'$.
\end{corollary}

\section{Behavioural Theory}
\label{sec:behavioural-theory}

We study the behavioural theory of \HCP under the lenses of two classical notions of behavioural equivalence: bisimulation and barbed congruence.
The first has emerged as a powerful operational method for proving equivalence of programs in various kinds of languages, due to the associated coinductive proof method.
The second is the equivalent of contextual equivalence in concurrency.

\paragraph{Bisimilarity}
The standard notion of strong bisimilarity can be instantiated on the \lts of \HCP.

\begin{definition}[Strong bisimilarity]
	\label{def:strong-bisimilarity}
	A symmetric relation $\crel{R}$ on processes is a strong bisimulation if $P \crel{R} Q$ implies that
	if $P \lto{l} P'$ then  $Q \lto{l} Q'$ for some $Q'$ such that $P' \crel{R} Q'$.
	Strong bisimilarity is the largest relation $\sbis$ that is a strong bisimulation.
\end{definition}

Strong bisimilarity can discriminate processes whose behaviours differs only by $\tau$-transitions. For instance, the processes $\res{xy}{(\close{x}{\nil} \pp \wait{y}{\close{z}{\nil}})}$ and $\close{z}{\nil}$ are not bisimilar exclusively because the first has a $\tau$-transition. Instead, no \HCP process composed in parallel with either of them would be able to tell the difference: $\tau$-transitions have no effect on or interaction with the context. This deficiency of strong bisimilarity is well-known and has been widely studied since Milner's seminal works on CCS \cite{M89}. The solution is to define bisimilarity on the saturated transition relation $\slto{}$.
\begin{definition}[Bisimilarity]
	\label{def:bisimilarity}
	A symmetric relation $\crel{R}$ on processes is a bisimulation if $P \crel{R} Q$ implies that if $P \slto{l} P'$ then  $Q \slto{l} Q'$ for some $Q'$ such that $P' \crel{R} Q'$.
	Bisimilarity is the largest relation $\bis$ that is a bisimulation.
\end{definition}

It follows from the inclusion of \cref{rule:alpha} in the specification of the \lts of \HCP that ${\aleq} \subsetneq {\sbis}$. Also, from the definition of saturation it follows that ${\sbis} \subsetneq {\bis}$, as usual for saturation-based behavioural equivalences \cite{BMP15}.

\begin{fact}
\label{ex:hcp-bis-laws}
Assume processes are well-typed.
As in standard process algebras, parallel composition and $\nil$ obey the
laws of an abelian monoid under (strong) bisimilarity. Formally, for any $P$, $Q$, and $R$:
\begin{equation*}
  P \sbis P \pp \nil
  \qquad
  P \pp Q \sbis Q \pp P
  \qquad
  P \pp (Q \pp R) \sbis (P \pp Q) \pp R
\end{equation*}
Restriction distributes over actions, parallel composition, and restriction, provided that they do not depend on the restricted channel. For $x,y \notin (\fn(R) \cup \cn(\pi) \cup \{x',y'\})$:
\begin{equation*}
  \res{xy}{(P \pp R)} \sbis \res{xy}{(P)} \pp R
  \quad
  \res{xy}{\prefixed{\pi}{P}} \sbis \prefixed{\pi}{\res{xy}{P}}
  \qquad
  \res{xy}{\res{x'y'}{P}} \sbis \res{x'y'}{\res{xy}{P}}
\end{equation*}
Closing a channel is non-blocking: $\close{x}{P} \sbis \close{x}{\nil} \pp P$.
Links are symmetric: $\forward{x}{y} \sbis \forward{y}{x}$.
\end{fact}

Bisimilarity and strong bisimilarity are congruences, in the sense that they are preserved by all syntactic operators of \HCP.
\begin{theorem}[Congruence]
	\label{thm:hcp-bis-congruence}
	For ${\asymp} \in \{\sbis,\bis\}$, if $P \asymp Q$, then
	\begin{enumerate}[nosep, noitemsep]
 		\item ${P \pp R} \asymp {Q \pp R}$ for any $R$ (and the symmetric);
		\item $\prefixed{\pi}{P} \asymp \prefixed{\pi}{Q}$ for any prefix $\pi$;
		\item $\choice{x}{R}{P} \asymp \choice{x}{R}{Q}$ for any $x$ and $R$ (and the symmetric);
		\item $\res{xy}{P} \asymp \res{xy}{Q}$ for any $x$, $y$.
	\end{enumerate}
\end{theorem}

\begin{proofatend}
	We prove each statement for bisimilarity, the ones for strong bisimilarity can be proven likewise with minor changes. We use the well-known alternative characterisation of bisimulation that follows: a symmetric relation $\crel{R}$ is a bisimulation if $P \crel{R} Q$ implies that if $P \lto{l} P'$ then $Q \slto{l} Q'$ and $P' \crel{R}  Q'$.
  \begin{enumerate}
  \item 
    Let $\crel{R}$ be $\{(P \pp R, Q \pp R) \mid P \bis Q, R\}$. We prove that $\crel{R}$ is a bisimulation hence included in $\bis$.
    For $(P \pp R) \crel{R} (Q \pp R)$,
    \begin{enumerate}
    \item if $(P \pp R) \lto{l} (P \pp R')$ then $R \lto{l} R'$, $(Q \pp R) \lto{l} (Q \pp R')$, and $(P \pp R') \crel{R} (Q \pp R')$;
    \item if $(P \pp R) \lto{l} (P' \pp R)$ then $P \lto{l} P'$, there is $Q' \bis P'$ s.t.~$Q \slto{l} Q'$, $(Q \pp R) \slto{l} (Q' \pp R)$, and $(P' \pp R) \crel{R} (Q' \pp R)$;
    \item if $(P \pp R) \lto{l} (P' \pp R')$, then $l = \lsync{l'}{l''}$, $P \lto{l'} P'$, $R \lto{l''} R'$, $Q \slto{l'} Q'$ for $Q' \bis P'$, $(Q \pp R) \slto{\lsync{l'}{l''}} (Q' \pp R')$, and $(P' \pp R') \crel{R} (Q' \pp R')$.
    \end{enumerate}
  \item 
    Let $\pi$ be any prefix but a server.
    We prove that $\crel{R} = {\bis} \cup \{(\prefixed{\pi}{P}, \prefixed{\pi}{Q}) \mid P \bis Q\}$ is a bisimulation hence equal to $\bis$.
    Assume $\prefixed{\pi}{P} \lto{l} {P'}$, we proceed by case analysis on the transition derivation.
    \begin{enumerate}
    	\item
    		If the head of the derivation is an application of the axiom for $\pi$, then there are two cases.
    		\begin{enumerate}
    			\item If $\pi$ does not have form $\ldisp{x}$ or $\ldup{x}{x_1}{x_2}$, then $P' = P$ and $l$ is the action of $\pi$. By the same axiom $\prefixed{\pi}{Q} \lto{l} {Q}$ and, since $P \bis Q$ by hypothesis, $P \crel{R} Q$.
    			\item Otherwise, $P'$ has form $\prefixed{\pi'}{P}$ where $\pi'$ is the opportune acknowledgement action. By the same axiom $\prefixed{\pi}{Q} \lto{l} \prefixed{\pi'}{Q}$ and, since $P \bis Q$ by hypothesis, $\prefixed{\pi'}{P} \crel{R} \prefixed{\pi'}{Q}$.
    		\end{enumerate}
    	\item
    		If the head of the derivation is an application of the delayed action rule for $\pi$, then $P' = \prefixed{\pi}{P''}$, $P \lto{l} P''$,
    		$Q \slto{l} Q''$ for some $Q'' \bis P''$ (by hypothesis on $P$ and $Q$), $\prefixed{\pi}{Q} \slto{l} \prefixed{\pi}{Q''}$ (by delaying $\pi$ after $l$ and all $\tau$s), and hence $\prefixed{\pi}{P''} \crel{R} \prefixed{\pi}{Q''}$.
			\item
    		If the head of the derivation is a self synchronisation rule, then $l$ has form $\lsync{l'}{l''}$ where \wl $l'$ is the label for $\pi$. By hypothesis $P \bis Q$ and since $P \lto{l''} P'$, $Q \slto{l''} Q'$ for $Q' \bis P'$. By delaying $\pi$ past all the $\tau$s until $l''$, $\prefixed{\pi}{Q} \slto{\lsync{l'}{l''}} Q'$ and $P' \crel{R} Q'$ since $P' \bis Q'$.
    \end{enumerate}
    Let $\pi$ be a server prefix. 
    
    We prove that $\crel{R} = {\bis} \cup {\bis} \cup \{(\server{x}{x'}{P},\server{x}{x'}{Q})  \mid P \bis Q\}$ is a bisimulation hence equal to $\bis$.
    Let $\server{x}{x'}{P} \crel{R} \prefixed{x}{x'}{Q}$. There are three cases to consider: server use, disposal, and duplication (server prefixes do not allow for delaying or self-synchronisation).
    \begin{enumerate}
      \item
      if $\server{x}{x'}{P} \lto{\lcouse{x}{x'}} {P'}$, then $P' = P$, $\server{x}{x'}{Q} \lto{\lcouse{x}{x'}} {Q}$, $P \bis Q$ by hypothesis hence $P \crel{R} Q$;
      \item
      if $\server{x}{x'}{P} \lto{\lcodisp{x}} P'$ then $P' = \clientdisp{z_1}{\dots\clientdisp{z_n}{\close{x}{\nil}}}$ for $\fn(P) = \{z_1,\dots,z_n,x'\}$.
      By hypothesis on $P$ and $Q$, $\fn(P) = \fn(Q)$. It follows that $\server{x}{x'}{Q} \lto{\lcodisp{x}} Q'$ for $Q' = \clientdisp{z_1}{\dots\clientdisp{z_n}{\close{x}{\nil}}}$ and, since $\bis$ is reflexive, $P' \crel{R} Q'$;
      \item
      if $\server{x}{x'}{P} \lto{\lcodup{x}{x_1}{x_2}} P'$ then \[P' = \clientdup{z_1}{\sigma_1z_1}{\sigma_2z_1}{\dots\clientdup{z_n}{\sigma_1z_n}{\sigma_2z_n}{\send{x_1}{x_2}{(\server{x_1}{\sigma_1 x'}{P_1} \pp \server{x_2}{\sigma_2 x'}{P_2})}}}\text{,}\]
      where $\fn(P) = \{z_1,\dots,z_n,x'\}$  and $P_i = P\sigma_i$ for $\sigma_i$ replacing all free names of $P$ with fresh ones.
      By hypothesis on $P$ and $Q$, $\fn(P) = \fn(Q)$. It follows that there is a transition $\server{x}{x'}{Q} \lto{\lcodup{x}{x_1}{x_2}} Q'$ where $Q'$ is the process $\clientdup{z_1}{\sigma_1z_1}{\sigma_2z_1}{\dots\clientdup{z_n}{\sigma_1z_n}{\sigma_2z_n}{\send{x_1}{x_2}{(\server{x_1}{\sigma_1 x'}{Q_1} \pp \server{x_2}{\sigma_2 x'}{Q_2})}}}$. It follows from the fact that bisimulation is preserved by parallel composition and by send and duplication prefixes that $P' \bis Q'$ hence that $P' \crel{R} Q'$.
    \end{enumerate}
  \item 
    It suffices to prove that $\crel{R} = {\bis} \cup \{(\choice{x}{P}{R}, \choice{x}{Q}{R}) \mid P \bis Q\}$ is a bisimulation and this can be done by case analysis on the transition derivation with minor adaptations to the arguments used in the case of prefixes.
  \item 
    Let $\crel{R}$ be the union of
    \begin{gather*}
    {\crel{R}_0} = {\bis} \quad
    {\crel{R}_1} = \{(\res{xy}{P},\res{xy}{Q}) \mid P \crel{R}_0 Q\} \qquad
    {\crel{R}_2} = \{(\res{xy}{P},\res{xy}{Q}) \mid P \crel{R}_1 Q\}\text{.}
    \end{gather*}
    We prove that $\crel{R}$ is a strong bisimulation.
    There are three cases, one for each $\crel{R}_i$.
    \begin{itemize}
    \item
    The first, ${P} \crel{R}_0 {Q}$, follows by construction.
    \item
    The second, $\res{xy}{P} \crel{R}_1 \res{xy}{Q}$, has the subcases below (given by inspection of the SOS specification of \HCP).
    \begin{itemize}
    \item if $\res{xy}{P} \lto{l} \res{xy}{P'}$ and $l \neq \tau$, then $x,y \notin \cn(l)$, $P \lto{l} P'$, $Q \slto{l} Q'$ for $Q' \bis P'$, $\res{xy}{Q} \slto{l} \res{xy}{Q'}$, and $\res{xy}{P'} \crel{R} \res{xy}{Q'}$;
    \item if $\res{xy}{P} \lto{\tau} P'$ and $P \lto{\lsync{\lclose{x}}{\lwait{y}}} P'$, then $Q \slto{\lsync{\lclose{x}}{\lwait{y}}} Q'$ for $Q' \bis P'$ and $P' \crel{R} Q'$;
    \item if $\res{xy}{P} \lto{\tau} P'$ and $P \lto{\lforward{x}{x'}} P'$, then $Q \slto{\lforward{x}{x'}} Q'$ for $Q' \bis P'$, $\res{xy}{Q} \slto{\tau} Q'$, and $P' \crel{R} Q'$;
    \item if $\res{xy}{P} \lto{\tau} \res{xy}{P'}$, then either $P \lto{\lsync{\linl{x}}{\lcoinl{y}}} P'$ or $P \lto{\lsync{\linr{x}}{\lcoinr{y}}} P'$. Assume \wl the first, $Q \slto{\lsync{\linl{x}}{\lcoinl{y}}} Q'$ for $Q' \bis P'$, $\res{xy}{Q} \slto{\tau} \res{xy}{Q'}$, and $\res{xy}{P'} \crel{R} \res{xy}{Q'}$;
    \item if $\res{xy}{P} \lto{\tau} \res{x'y'}{P'}$ then $P \lto{\lsync{\luse{x}{x'}}{\lcouse{y}{y'}}} P'$, $Q \slto{\lsync{\luse{x}{x}}{\lcouse{y'}{y'}}} Q'$ for $Q' \bis P'$, $\res{xy}{Q} \slto{\tau} \res{x'y'}{Q'}$, and $\res{x'y'}{P'} \crel{R} \res{x'y'}{Q'}$;
    \item if $\res{xy}{P} \lto{\tau} \res{xy}{P'}$ and $P \lto{\lsync{\ldisp{x}}{\lcodisp{y}}} P'$, then $Q \slto{\lsync{\ldisp{x}}{\lcodisp{y}}} Q'$  for $Q' \bis P'$, $\res{xy}{Q} \slto{\tau} \res{xy}{Q'}$, and $\res{xy}{P'} \crel{R} \res{xy}{Q'}$;
    \item if $\res{xy}{P} \lto{\tau} \res{x_1y_1}{P'}$ and $P \slto{\lsync{\ldup{x}{x_1}{x_2}}{\lcodup{y}{y_1}{y_2}}} P'$, then
    $Q \slto{\lsync{\ldup{x}{x_1}{x_2}}{\lcodup{y}{y_1}{y_2}}} Q'$, for $Q' \bis P'$, $\res{xy}{Q} \slto{\tau} \res{x_1y_1}{Q'}]$, and $\res{x_1y_1}{P'} \crel{R} \res{x_1y_1}{Q'}$;
    \item if $\res{xy}{P} \lto{\tau} \res{xy}{\res{x'y'}{P'}}$ then $P \lto{\lsync{\lsend{x}{x'}}{\lrecv{y}{y'}}} P'$, $Q \slto{\lsync{\lsend{x}{x'}}{\lrecv{y}{y'}}} Q'$ for $Q' \bis P'$, $\res{xy}{Q} \slto{\tau} \res{xy}{\res{x'y'}{Q'}}$, and $\res{xy}{\res{x'y'}{P'}} \crel{R} \res{xy}{\res{x'y'}{Q'}}$.
    \end{itemize}
    \item
    The third, $\res{xy}{P} \crel{R}_2 \res{xy}{Q}$, follows from the same case analysis we used in the second case ($\crel{R}_1$) and from the fact that $\res{xy}{\res{x'y'}{P}} \sbis \res{xy}{\res{x'y'}{P}}$.
 		\qedhere
    \end{itemize}
  \end{enumerate}
\end{proofatend}

Bisimilarity implies type equivalence on well-typed processes.
\begin{proposition}
If $P$ and $Q$ are well-typed and $P \bis Q$, then
	$\judge{P}{\hyp{G}}$ iff $\judge{Q}{\hyp{G}}$.
\end{proposition}

\begin{proofatend}
  The statement can be proved directly by induction on the height ot the typing derivations. However, it follows from \cref{thm:hcp-bis-bcong,def:bcong}.
\end{proofatend}

\paragraph{Barbed congruence}
Contextual equivalence defines as equivalent all programs that ``behave in the same manner'' in any given ``context'' \cite{M68}.
In typed languages, a context is a typed program $C$ with a typed ``hole'' $\hole$ where we can plug a program $P$ and obtain the program $C[P]$. In \HCP this instantiates to the following.
\begin{definition}[Typed context]
	A context $C$ is a \emph{($\hyp{G}/\hyp{H}$)-context} if $\judge{C}{\hyp{H}}$ is valid  when the hole $\hole$ of $C$ is considered as a process and the following rule is added to the theory of \HCP:
	\[
		\infer
			{\judge{\hole}{\hyp{G}}}
			{}
	\]
\end{definition}

In concurrency, behaviours are characterised in terms of the observable interactions between a process and its execution environment.
We start by borrowing the notion of observability used by the $\pi$-calculus.
\begin{definition}[Observability predicates]
	Given a name $x$, the \emph{observability predicate} $P\barb{x}$ holds iff $P \slto{l} Q$ for some $Q$ and $l$ such that $x \in \fn(l)$.
\end{definition}

\begin{example}
	Consider $Q_1 = \recv{x}{x'}{\recv{y}{y'}{P}}$ and $Q_2 = \recv{y}{y'}{\recv{x}{x'}{P}}$ from the \nameref{sec:introduction}.
	In \HCP they exhibit exactly the same observations:
	$Q_1\barb{x}$, $Q_1\barb{y}$, $Q_2\barb{x}$, and $Q_2\barb{y}$.
\end{example}

\citet{A17} argues that a suitable notion of observational equivalence must discriminate clients that use non-linear resources in the environment in different quantities.
This requires some care in the definition of barbed congruence for \HCP. We illustrate the reason below.
\begin{example}
	\label{ex:hcp-bcong-exponentials-1}
	Consider type equivalent processes that consume a different number of instances of the same server: $P_0 = \clientdisp{x}{Q_0}$, $P_1 = \clientuse{x}{x'}{Q_1}$, $P_2 = \clientdup{x}{x_1}{x_2}{\clientuse{x_1}{x'}{\clientuse{x_2}{x''}{Q_2}}}$, and $P_3 = \clientdup{x}{x_1}{x_2}{\clientdisp{x_1}{\clientuse{x_2}{x''}{Q_3}}}$.
	The observability predicate does not distinguish these processes, since $\barb{x}$ holds for all of them. Thus, distinguishing them requires using a process context. Due to typing, this context can only connect $x$ to a server, and the only way that a server has to relay information to an external observer is via client requests. Thus, we end up in the same problem we started from. Consider for instance the context
	$\res{xy}{(\hole \pp \server{y}{y'}{\clientdisp{u}{\clientdisp{v}{\clientuse{z}{w}{\forward{y'}{w}}}}})}$. For any $i \in \{0,1,2\}$, $C[P_i]\barb{u}$ and  $C[P_i]\barb{v}$.
	(For the case of $P_0$, recall that disposing of a server triggers the disposal of its dependencies.)
\end{example}
This kind of deficiency of the standard observability predicate is not new for session-typed calculi: \citet{YHB07} faced it for selections and choices.
Likewise, we introduce predicates for observing how a program may use or dispose of a non-linear resource from the context.
\begin{definition}[Non-linear observability predicates]
	For $x$ a name and $t$ a binary tree with leaves in $\{\useleaf,\displeaf\}$ (for use and dispose, respectively), the \emph{non-linear observability predicate} $\expbarb{t}{x}$ (read ``$x$ is used as in $t$'') holds on $P$ whenever any of the following holds:
	\begin{itemize}[nosep,noitemsep]
		\item $t = \dispnode$, $P \lto{l} Q$, $\ldisp{x}$ occurs in $l$, and $x \notin \fn(P)$;
		\item $t = \usenode$, $P \lto{l} Q$, $\luse{x}{x'}$ occurs in $l$ for some $x'$, and $x \notin \fn(P)$;
		\item $t = \dupnode{t_1}{t_2}$, $P \lto{l} Q$, $\ldup{x}{x_1}{x_2}$ occurs in $l$ for some $x_1$ and $x_2$, $Q\expbarb{t_1}{x_1}$ and $Q\expbarb{t_2}{x_2}$;
		\item $P \lto{l} Q$, $\ldisp{x}$, $\luse{x}{x'}$, and $\ldup{x}{x_1}{x_2}$ do not occur in $l$, and $Q\expbarb{t}{x}$.
	\end{itemize}
\end{definition}

\begin{example}
	\label{ex:hcp-bcong-exponentials-2}
	Consider processes $P_0$, $P_1$, and $P_2$ from  \cref{ex:hcp-bcong-exponentials-1}. Then, 
	$P_0\expbarb{t}{x}$ iff $t=\dispnode$,
	$P_1\expbarb{t}{x}$ iff $t=\usenode$,
	$P_2\expbarb{t}{x}$ iff $t=\dupnode{\usenode}{\usenode}$,
	$P_3\expbarb{t}{x}$ iff $t=\dupnode{\dispnode}{\usenode}$.	
\end{example}

When \citet{YHB07} faced a similar problem for selections, they extended observability to observe the payload of selections (left or right): the new predicates $\barb[\minl]{x}$ and $\barb[\minr]{x}$ hold on $P$ whenever $P$ sends on $x$ a left or a right selection, respectively. In \HCP, we do not need this extension because we can subsume these predicates using our non-linear observability predicates and appropriate contexts, as we exemplify below.

\begin{example} 
	\label{ex:hcp-bcong-additives} 
	Consider a pair of processes that differ exclusively for the selection they make, $P_l = \inl x{Q}$ and $P_r = \inr x{Q}$. There is no direct observation that distinguishes them: for any $w$ and $t$, $P_l\barb{w}$ iff $P_r\barb{w}$ and $P_l\expbarb{t}{w}$ iff $P_r\expbarb{t}{w}$. However, we can make an indirect observation using any context that offers a choice between two observationally distinct branches: for
	$C = \res{xy}{(\hole \pp \choice{y}{\clientdisp{w}{\forward{y}{z}}}{\clientuse{w}{w'}{\wait{w'}{\forward{y}{z}}}})}$, $C[P_0]\expbarb{t}{w}$ iff $t = \dispnode$ and $C[P_1]\expbarb{t}{w}$ iff $t = \usenode$.
\end{example}

\begin{definition}[Barbed congruence]
	\label{def:bcong}
	Barbed congruence is the largest symmetric relation $\bcong$ on well-typed \HCP processes that is \begin{itemize}[nosep,noitemsep]
		\item typed ($P \bcong Q$ implies $\judge{P}{\hyp{G}}$ iff $\judge{Q}{\hyp{G}}$);
		\item context-closed (\ie if $P \bcong Q$ then $C[P] \bcong C[Q]$ for any typed context);
		\item barb preserving (\ie if $P \bcong Q$ and $P\barb{x}$ then $Q\barb{x}$; if $P \bcong Q$ and $P\expbarb{t}{x}$ then $Q\expbarb{t}{x}$).
	\end{itemize}
\end{definition}

On well-typed processes, bisimilarity implies typed barbed congruence.
\begin{theorem}
	\label{thm:hcp-bis-bcong}
	${\bis} \subseteq {\bcong}$.
\end{theorem}

\begin{proofatend}
	The thesis follows from \cref{thm:hcp-bis-congruence}.
\end{proofatend}

We anticipate that also the converse of \cref{thm:hcp-bis-bcong} holds but we do not prove it directly. Instead, we show in \cref{sec:full-abstraction} that this holds by comparison with the denotational semantics.

\section{Denotational Semantics}
\label{sec:hcp-denotational}

Inspired by the relational semantics for proofs in linear logic by \citet{B96}, \citet{A17} developed a denotational semantics for \CP that interprets well-typed processes as sets of possible observable interactions on each of their free channels.
In this section, we show that a similar denotational semantics can be also developed for \HCP.

The main novelty, besides extending denotations to hyperenvironments, is that we rediscover denotations from a very different angle, based on \HCP's notion of observable actions.
Specifically, inspired by Brzozowski derivatives of regular expressions \cite{B64}, we introduce ``derivatives of denotations'' \wrt\ \HCP actions. As the derivative of a set of strings \wrt a character is the set of strings that can can follow that character, the derivative of a denotation \wrt an action is the denotation that can follow that action. This approach allows us to abstract from the syntax of \HCP and study actions and interactions solely in terms of denotations. For instance, we characterise non-interference, usually a topic of operational semantics, as Fubini's equality for double antiderivatives \cite{F07}.

\citeauthor{A17}'s denotations do not distinguish outputs from inputs.
For instance, (typed) processes $\judge{\inl{x}{\close{x}{\nil}}}{\one \oplus \one}$ and $\judge{\choice{x}{\close{x}{\nil}}{\close{x}{\nil}}}{\one \with \one}$ are given the same denotation.
To characterise observational equivalence, Atkey resorts to the intersection of type equivalence with denotational equivalence: two processes are observationally equivalent if they have the same types and denotations.
Differently, our denotations distinguish inputs from outputs, and our notion of denotational equivalence does not require type equivalence.

\paragraph{Interpretation of types}\looseness=-1
We define the domain of denotations for a name by recursion on the structure of its type. We tag denotations with natural numbers and use these identifiers in certain denotations to describe dependencies with other channels:
\begin{itemize}
\item in the denotations of $\tensor$, we use identifiers to specify which observations must be in which of two parallel components merged by a tensor;
\item in the denotations of $\with$ and $\bang$, we use identifiers to specify which observations are blocked until a choice is performed or a server is used, respectively---which are the only constructs in \HCP without delayed actions.
\end{itemize}
We use $\denout$ and $\denin$ to distinguish between outputs and inputs. We write $\wpf$ for the finite powerset.

We interpret multiplicative units as singletons because they are the types that characterise the empty output and input, respectively.
\begin{highlight}
\begin{align*}
	\dendom{\thl{\one}}        & = \{\denout\} \times \iddom \times \{\ast\}
&
	\dendom{\thl{\bot}}        & = \{\denin \} \times \iddom \times \{\ast\}
\intertext{We interpret multiplicatives as cartesian products pairing denotations of their components.}
	\dendom{\thl{A \tensor B}} & = \{\denout\} \times \iddom \times (\wpf\iddom \times \dendom{\thl{A}} \times \wpf\iddom \times \dendom{\thl{B}})
&
	\dendom{\thl{A \parr B}}   & = \{\denin \} \times \iddom \times (\dendom{\thl{A}} \times \dendom{\thl{B}})
\intertext{We interpret additives as coproducts and represent them as dependent pairs indexed over $\{\minl,\minr\}$.}
	\dendom{\thl{A \oplus B}}  & = \{\denout\} \times \iddom \times (\dendom{\thl{A}} + \dendom{\thl{B}})
&
	\dendom{\thl{A \with B}}   & = \{\denin \} \times \iddom \times \wpf\iddom \times (\dendom{\thl{A}} + \dendom{\thl{B}})
\intertext{We interpret exponentials as sets of binary trees of denotations (written $\mathcal{T}$) since they characterise both unbounded (but finite) interactions of a given type and resource allocation (duplication via branchings and disposal via leaves).}
	\dendom{\thl{\query A}}     & = \{\denout\} \times \iddom \times (\mathcal{T}(\dendom{\thl{A}}+\{\displeaf\}))
&
	\dendom{\thl{\bang A}}    & = \{\denin \} \times \iddom \times \wpf\iddom \times (\mathcal{T}(\dendom{\thl{A}}+\{\displeaf\}))
\end{align*}
\end{highlight}\looseness=-1
If we ignore identifiers in denotations, we can turn $a \in \dendom{A}$ into an element of $\dendom{\dual{A}}$ by simultaneously replacing $\denin$ with $\denout$ and \viceversa. We extend the notion of duality from types to their interpretation and write $a \dendual b$ whenever $a \in\dendom{A}$, $b \in \dendom{\dual{A}}$, and reversing the direction of triangles in $a$ yields $b$ up to differences in identifiers---sometimes, we abuse the terminology and call $a$ and $b$ dual.
\begin{example}
Consider $(\denout,n,\ast) \in \dendom{\one}$,
$(\denin,n',\ast) \in \dendom{\bot}$ and write $a$ and $b$ for them, respectively.
We have that $a \dendual b$, for any $n,n' \in \iddom$. Also, $(\denout,m,\minl,a) \dendual (\denin,m',M,\minl,b)$ for any $m,m' \in \iddom$ and $M \in \wpf\iddom$.
\end{example}
We sometimes use the wildcard $\_$ when pattern matching denotations, \eg, we write $(\denin,\_,\ast) = a$ instead of $(\denin,n,\ast) = a$ if binding $n$ is superfluous.

We define the denotation domain of an environment as the set of all assignments that take each name to observations of its associated type. We write $\langle x_1 \mapsto a_1,\dots, x_n \mapsto a_n \rangle$ for the assignment taking each name $x_i$ of type $A_i$ to its denotation $a_i \in \dendom{A_i}$. 
As usual, we work up to exchange and hence we interpret ``$,$'' as the cartesian product and assignments as tuples.
\begin{highlight}
\begin{align*}
\dendom{\thl{\emptyseq}}          & = \{\langle\rangle\} &
\dendom{\thl{\Gamma, \cht{x}{A}}} & = \dendom{\thl{\Gamma}} \times \{ \langle\phl{x} \mapsto a\rangle \mid a \in \dendom{\thl{A}}\}
\intertext{Likewise, we interpret hyperenvironments as collections of assignments from the domains of their environments and map ``$\pp$'' to the cartesian product. We use $\gamma$, $\delta$ to range over elements of $\dendom{\thl{\Gamma}}$.}
\dendom{\thl{\emptyhyp}}          & = \{\langle\rangle\} &
\dendom{\thl{\hyp{G} \pp \Gamma}} & = \dendom{\thl{\hyp{G}}} \times \{ \langle\gamma\rangle \mid  \gamma \in \dendom{\thl{\Gamma}}\}
\end{align*}
\end{highlight}
We write $\id(a)$ for the outermost identifier in $a$ and $\ids(a)$ for the set of all identifiers in $a$ (excluding dependencies used in the interpretations of $\tensor$, $\with$, and $\bang$), and
extend these notations to the interpretations of environments and hyperenvironments ($\id(\gamma)$ is the set of all outermost identifiers in $\gamma$ and $\ids(\gamma)$ is the set of all identifiers in $\gamma$).

Since we are using numeric identifiers to encode dependencies, it is natural to assume that 
identifiers are unique within a denotation and that dependencies in the interpretations of $\tensor$, $\with$, and $\bang$ are limited to the same environment and have no cycles. We refer to denotations that respect this simple identifier discipline as \emph{well-formed}.
\begin{itemize}
\item We call $a \in \dendom{A}$ well-formed if its identifiers are unique and decrease along its structure (\eg, if $a = (\denout,n,\minl,a')$, then $n > \ids(a')$).
\item We call $\gamma = \langle x_1 \mapsto a_1,\dots, x_k \mapsto a_k \rangle \in \dendom{x_1:A_1, \dots, x_k:A_k}$  well-formed if its identifiers are unique and for every $i \in \{1,\dots,k\}$:
\begin{enumerate}
	\item $a_i$ is well-formed;
	\item if $A_i = B_1 \tensor B_2$, then
	$N_1, N_2 \subseteq \ids(\gamma)$ and $n > N_1, N_2$, where $a_i = (\denout,n,N_1,\_,N_2,\_)$;
	\item if $A_i = B_1 \with B_2$ or $A_i = \bang B$, then
	$N \subseteq \ids(\gamma)$ and $n > N$, where $a_i = (\denout,n,N,\_,\_)$ or $a_i = (\denout,n,N,\_)$.
\end{enumerate}
\item We call $g = \langle\gamma_1,\dots,\gamma_k\rangle$ well-formed if $\gamma_1$, \dots, $\gamma_k$ are well-formed and its identifiers are unique.
\end{itemize}	
From now on, unless otherwise specified, we assume that denotations are well-formed.

\paragraph{Derivatives and Antiderivatives}\looseness=-1
Recall from \cite{B64} that the derivative \wrt a character $l$ of a set of strings $D$ is the set $\{ a \mid la \in D \}$ of strings $a$ that can follow $l$ to yield a string $la$ in $D$. In particular, if we consider the language accepted by a state $p$ of a deterministic automaton and take its derivative \wrt an input symbol $l$, then we obtain the accepted language for the state reached by the automaton if we input $l$ while in $p$. In light of this correspondence, Brzozowski derivatives define an automaton on accepted languages, not any automaton, but the one that is final in the coalgebraic sense thus marrying denotational and observational equivalence of automata \cite{BBHPRS14}.

Inspired by Brzozowski derivatives, we define the derivative of a set of
denotations $D$ \wrt to an action $l$ as the set of the denotations that can follow $l$. For instance, derivation \wrt $\lclose{x}$ takes $\langle\langle
\phl{x} \mapsto (\denout,n,\ast)\rangle\rangle \in \dendom{\cht{x}{\one}}$ to
$\langle\rangle \in \dendom{\emptyhyp}$ since there are no actions left;
derivation \wrt $\lsend{x}{y}$ takes $\langle\langle \phl{x} \mapsto
(\denout,n,\emptyset,a,\emptyset,b)\rangle\rangle \in \dendom{\cht{x}{A \tensor B}}$ to $\langle\langle \phl{x} \mapsto b\rangle,\langle \phl{y} \mapsto a\rangle\rangle \in \dendom{\cht{x}{B} \pp \cht{y}{A}}$ since after $y$ is output on $x$ the two have separate behaviours. 
Below, we write $\deldep{a}{n}$ for the denotation where $n$ is removed from all interpretations of $\tensor$ in $a$, \eg, $\deldep{(\denin,m,\{n\},a)}{n} = (\denin,m,\{n\},\deldep a n)$ and $\deldep{(\denout,m,\{n\},a,\emptyset,a')}{n} = (\denout,m,\emptyset,\deldep{a}{n},\emptyset,\deldep{a'}{n})$); likewise for environments. We write $\tdot$ for juxtaposition, \eg, $\langle \langle x \mapsto a \rangle\rangle \tdot \langle\langle y \mapsto a', z \mapsto a'' \rangle \rangle$ is $\langle \langle x \mapsto a \rangle, \langle y \mapsto a', z \mapsto a'' \rangle \rangle$ from $\dendom{\cht{x}{A} \pp \cht{y}{A'},\cht{z}{A''}}$. 

\begin{definition}[Derivatives]
\label{def:derivative}
For $D$ a set of denotations, the derivative $\der{l}{D}$ of $D$ \wrt an action $l$ is the set defined by:
\begin{spreadlines}{0pt}
\begin{align*}
	\der{\lclose{x}}{D} & = \left\{
		g
  \,\middle|\,
		g \tdot \langle \langle \phl{x} \mapsto (\denout,\_,\ast) \rangle \rangle
	\in D\right\}
\\
	\der{\lwait{x}}{D} & = \left\{
		g \tdot \langle \deldep{\gamma}{n} \rangle
  \,\middle|\,
		g \tdot \langle \gamma \tdot \langle \phl{x} \mapsto (\denin,n,\ast) \rangle \rangle
	\in D\right\}
\\
	\der{\lsend{x}{y}}{D} & = \left\{
		g \tdot \langle \deldep{\gamma}{n,N'} \tdot \langle \phl{x} \mapsto b\rangle, \deldep{\delta}{n,N} \tdot \langle \phl{y} \mapsto a \rangle \rangle
  \,\middle|\,
  	\array{l}
		g \tdot \langle \gamma \tdot \delta \tdot \langle \phl{x} \mapsto (\denout,n,N,a,N',b) \rangle \rangle
	\in D\text,\\
	N \subseteq \ids(\gamma), N'\subseteq \ids(\delta)
	\endarray
	\right\}
\\
	\der{\lrecv{x}{y}}{D} & = \left\{
		g \tdot \langle \deldep{\gamma}{n} \tdot \langle \phl{x} \mapsto b, \phl{y} \mapsto a \rangle\rangle
  \,\middle|\,
		g \tdot \langle \gamma \tdot \langle \phl{x} \mapsto (\denin,n,a,b) \rangle \rangle
	\in D\right\}
\\
	\der{\linl{x}}{D} & = \left\{
		g \tdot \langle \deldep{\gamma}{n} \tdot \langle \phl{x} \mapsto a \rangle \rangle
  \,\middle|\,
		g \tdot \langle \gamma \tdot \langle \phl{x} \mapsto (\denout,n,\minl,a) \rangle \rangle
	\in D\right\}
\\
	\der{\lcoinl{x}}{D} & = \left\{
		g \tdot \langle \deldep{\gamma}{n} \tdot \langle \phl{x} \mapsto a \rangle\rangle
  \,\middle|\,
		g \tdot \langle \gamma \tdot \langle \phl{x} \mapsto (\denin,n,\_,\minl,a) \rangle \rangle
	\in D\right\}
\\
	\der{\linr{x}}{D} & = \left\{
		g \tdot \langle \deldep{\gamma}{n} \tdot \langle \phl{x} \mapsto b \rangle \rangle
  \,\middle|\,
		g \tdot \langle \gamma \tdot \langle \phl{x} \mapsto (\denout,n,\minr,b) \rangle \rangle
	\in D\right\}
\\
	\der{\lcoinr{x}}{D} & = \left\{
		g \tdot \langle \deldep{\gamma}{n} \tdot \langle \phl{x} \mapsto b \rangle\rangle
  \,\middle|\,
		g \tdot \langle \gamma \tdot \langle \phl{x} \mapsto (\denin,n,\_,\minr,b) \rangle \rangle
	\in D\right\}
\\
	\der{\luse{x}{y}}{D} & = \left\{
		g \tdot \langle \deldep{\gamma}{n} \tdot \langle \phl{x} \mapsto a \rangle \rangle
  \,\middle|\,
		g \tdot \langle \gamma \tdot \langle \phl{x} \mapsto (\denout,n,\usenode[a]) \rangle \rangle
	\in D\text{, } a \neq \displeaf\right\}
\\
	\der{\lcouse{x}{y}}{D} & = \left\{
		g \tdot \langle \deldep{\gamma}{n} \tdot \langle \phl{x} \mapsto a \rangle \rangle
  \,\middle|\,
		g \tdot \langle \gamma \tdot \langle \phl{x} \mapsto (\denin,n,\_,\usenode[a]) \rangle \rangle
	\in D\text{, } a \neq \displeaf\right\}
\\
	\der{\ldisp{x}}{D} & = \left\{
		g \tdot \langle \gamma  \tdot \langle \phl{x} \mapsto (\denin,n,\ast) \rangle \rangle
  \,\middle|\,
		g \tdot \langle \gamma \tdot \langle \phl{x} \mapsto (\denout,n,\dispnode) \rangle \rangle
	\in D\right\}
\\
	\der{\lcodisp{x}}{D} & = \left\{
		g \tdot \langle \gamma \tdot \langle \phl{x} \mapsto (\denout,n,\ast) \rangle \rangle
  \,\middle|\,
		g \tdot \langle \gamma \tdot \langle \phl{x} \mapsto (\denin,n,\_,\dispnode) \rangle \rangle
	\in D\right\}
\\
	\der{\ldup{x}{x_1}{x_2}}{D} & = \left\{
		g \tdot \langle \gamma \tdot \langle x_1 \mapsto (\denin, n, a_1,a_2) \rangle \rangle
  \,\middle|\,
		g \tdot \langle \gamma \tdot \langle \phl{x} \mapsto (\denout,n,\dupnode{a_1}{a_2}) \rangle \rangle
	\in D\right\}
\\
	\der{\lcodup{x}{x_1}{x_2}}{D} & = \left\{
		g \tdot \langle \gamma \tdot \langle x_1 \mapsto (\denout, n, N_1,a_1,N_2,a_2) \rangle \rangle
  \,\middle|\,
  \mspace{-7mu}\array{l}
		g \tdot \langle \gamma \tdot \langle \phl{x} \mapsto (\denin,n,N,\dupnode{a_1}{a_2}) \rangle \rangle
	\in D,\\
  N_i = \left\{\id(b_i) \,\middle|\, (\denout,m,\dupnode{b_1}{b_2}) \in \gamma, m \in N\right\}
  \endarray\mspace{-7mu}
  \right\}
\\
	\der{\lforward{x}{y}}{D} & = \left\{
		g
  \,\middle|\,
		g \tdot \langle \langle \phl{x} \mapsto a, \phl{y} \mapsto a' \rangle\rangle
	\in D, a \dendual a'\right\}
\\
	\der{\lsync{l}{l'}}{D} & = \left\{
		g' \tdot h'
	\,\middle|\,
		g \tdot h \in D\text{, }
		g' \in \der{l}{\{g\}}\text{, and }
    h' \in \der{l'}{\{h\}}
	\right\}
\\
	\der{\tau}{D} & = D
\end{align*}
\end{spreadlines}
\end{definition}
\noindent Likewise, we define antiderivation reversing the reasoning above.
\begin{definition}[Antiderivatives]
\label{def:antiderivative}
For $D$ a set of denotations, the antiderivative $\antider{l}{D}$ of $D$ \wrt an action $l$ is the set defined by:
\begin{spreadlines}{0pt}
\begin{align*}
	\antider{\lclose{x}}{D} & = \left\{
    g \tdot \langle \langle \phl{x} \mapsto (\denout,n,\ast) \rangle \rangle
  \,\middle|\,
		g
	\in D\right\}
\\
	\antider{\lwait{x}}{D} & = \left\{
		g \tdot \langle \gamma \tdot \langle \phl{x} \mapsto (\denin,n,\ast) \rangle \rangle
  \,\middle|\,
		g \tdot \langle \gamma \rangle
	\in D\right\}
\\
	\antider{\lsend{x}{y}}{D} & = \left\{
		g \tdot \langle \gamma \tdot \delta \tdot \langle \phl{x} \mapsto (\denout,n,\ids(\gamma),a,\ids(\delta),b) \rangle \rangle
  \,\middle|\,
		g \tdot \langle \gamma \tdot \langle \phl{x} \mapsto b\rangle, \delta \tdot \langle \phl{y} \mapsto a \rangle \rangle
	\in D\right\}
\\
	\antider{\lrecv{x}{y}}{D} & = \left\{
		g \tdot \langle \gamma \tdot \langle \phl{x} \mapsto (\denin,n,a,b) \rangle \rangle
  \,\middle|\,
		g \tdot \langle \gamma \tdot \langle \phl{x} \mapsto b, \phl{y} \mapsto a \rangle\rangle
	\in D\right\}
\\
	\antider{\linl{x}}{D} & = \left\{
		g \tdot \langle \gamma \tdot \langle \phl{x} \mapsto (\denout,n,\minl,a) \rangle \rangle
  \,\middle|\,
		g \tdot \langle \gamma \tdot \langle \phl{x} \mapsto a \rangle \rangle
	\in D\right\}
\\
	\antider{\lcoinl{x}}{D} & = \left\{
		g \tdot \langle \gamma \tdot \langle \phl{x} \mapsto (\denin,n,\id(\gamma),\minl,a) \rangle \rangle
  \,\middle|\,
		g \tdot \langle \gamma \tdot \langle \phl{x} \mapsto a \rangle\rangle
	\in D\right\}
\\
	\antider{\linr{x}}{D} & = \left\{
		g \tdot \langle \gamma \tdot \langle \phl{x} \mapsto (\denout,n,\minr,b) \rangle \rangle
  \,\middle|\,
		g \tdot \langle \gamma \tdot \langle \phl{x} \mapsto b \rangle \rangle
	\in D\right\}
\\
	\antider{\lcoinr{x}}{D} & = \left\{
		g \tdot \langle \gamma \tdot \langle \phl{x} \mapsto (\denin,n,\id(\gamma),\minr,b) \rangle \rangle
  \,\middle|\,
		\langle \gamma \tdot \langle \phl{x} \mapsto b \rangle\rangle
	\in D\right\}
\\
	\antider{\luse{x}{y}}{D} & = \left\{
		g \tdot \langle \gamma \tdot \langle \phl{x} \mapsto (\denout,n,\usenode[a]) \rangle
  \,\middle|\,
		g \tdot \langle \gamma \tdot \langle \phl{x} \mapsto a \rangle \rangle  \rangle
	\in D\right\}
\\
	\antider{\lcouse{x}{y}}{D} & = \left\{
		g \tdot \langle \gamma \tdot \langle \phl{x} \mapsto (\denin,n,\id(\gamma),\usenode[a]) \rangle\rangle
  \,\middle|\,
	  g \tdot \langle \gamma \tdot \langle \phl{x} \mapsto a \rangle \rangle  \rangle \in D	 
  \right\}
\\
	\antider{\ldisp{x}}{D} & = \left\{
		g \tdot \langle \gamma \tdot \langle \phl{x} \mapsto (\denout,n,\dispnode) \rangle \rangle
  \,\middle|\,
		g \tdot \langle \gamma  \tdot \langle \phl{x} \mapsto (\denin,n,\ast) \rangle \rangle
	\in D\right\}
\\
	\antider{\lcodisp{x}}{D} & = \left\{
		g \tdot \langle \gamma \tdot \langle \phl{x} \mapsto (\denin,n,\id(\gamma),\dispnode) \rangle \rangle
  \,\middle|\,
		g \tdot \langle \gamma  \tdot \langle \phl{x} \mapsto (\denout,n,\ast) \rangle \rangle
	\in D\right\}
\\
	\antider{\ldup{x}{x_1}{x_2}}{D} & = \left\{
		g \tdot \langle \gamma \tdot \langle \phl{x} \mapsto (\denout,n,\dupnode{a_1}{a_2}) \rangle \rangle
  \,\middle|\,
		g \tdot \langle \gamma \tdot \langle x_1 \mapsto (\denin, n, a_1,a_2) \rangle \rangle
	\in D\right\}
\\
	\antider{\lcodup{x}{x_1}{x_2}}{D} & = \left\{
		g \tdot \langle \gamma \tdot \langle \phl{x} \mapsto (\denin,n,\id(\gamma),\dupnode{a_1}{a_2}) \rangle \rangle
  \,\middle|\,
		g \tdot \langle \gamma \tdot \langle x_1 \mapsto (\denout, n, N_1,a_1,N_2,a_2) \rangle \rangle
	\in D 
  \right\}
\\
	\antider{\lforward{x}{y}}{D} & = \left\{
		g \tdot \langle \langle \phl{x} \mapsto a, \phl{y} \mapsto \dual{a} \rangle\rangle
  \,\middle|\,
	  g
	\in D\right\}
\\
	\antider{\lsync{l}{l'}}{D} & = \left\{
		g \tdot h \,\middle|\,
		g' \tdot h' \in D\text{, }
 		g \in \antider{l}{\{g'\}}\text{, and }
    h \in \antider{l'}{\{h'\}}
 	\right\}
\\
	\antider{\tau}{D} & = D
\end{align*}
\end{spreadlines}
\end{definition}

\paragraph{Interpretation of typed processes}
We use sets of denotations for hyperenvironments ($D$) as denotations for well-typed processes.
We define the denotational semantics of a well-typed process $P$ as the subset $\densem{P}$ of $\bigcup_{\judge{P}{\hyp{G}}}\dendom{\hyp{G}}$ given below by recursion on the structure of ${P}$.

Denotations for links are all pairs of dual name denotations.
\begin{gather*}
	\densem{\forward{x}{y}} = \{\langle\langle \phl{x} \mapsto a,\phl{y} \mapsto a'\rangle\rangle \mid a \dendual a'\}
\intertext{Denotations for restriction are obtained by removing the restricted names and requiring that their observations are dual of each other.}
  \densem{\res{xy}{P}} = \left\{ g \tdot \langle \gamma \tdot \delta \rangle \mid g \tdot \langle \gamma \tdot \langle \phl{x} \mapsto a \rangle, \delta \tdot \langle \phl{y} \mapsto a' \rangle\rangle \in \densem{P}\text{ and } a \dendual a'\right\}
\intertext{Denotations for parallel composition reflect the denotation domain construction; they are given by cartesian product (restricted to pairs with disjoint identifiers since we assume well-formedness) and its unit, respectively.}
  \densem{P \pp Q} = \densem{P} \times \densem{Q}
  \qquad
  \densem{\judge{\nil}{\emptyhyp}} = \{\langle\rangle\}
\intertext{Denotations for choice are the antiderivatives of denotations for each branch.}
  \densem{\choice{x}{P}{Q}} = \antider{\lcoinl{x}}{\densem{P}} + \antider{\lcoinr{x}}{\densem{Q}}
\intertext{All remaining terms are action prefixes. For those, denotations are given by the associated antiderivatives of the denotation of the continuation---with some care for the cases of $\clientdisps{x}$, $\clientdups{x}{x_1}{x_2}$, $\serveruses{x}{x'}$, which have additional operations for managing server dependencies. For conciseness, let $\pi$ range over prefixes except $\clientdisps{x}$, $\clientdups{x}{x_1}{x_2}$, $\serveruses{x}{x'}$ in the following definitions.}
  \densem{\prefixed{\pi}{P}} = \antider{\pi}{\densem{P}}
  \qquad
  \densem{\clientdisp{x}{P}} = \antider{\ldisp{x}}{{\densem{\wait{x}P}}}
  \qquad
  \densem{\clientdup{x}{x_1}{x_2}{P}} = \antider{\ldup{x}{x_1}{x_2}}{{\densem{\recv{x_1}{x_2}P}}}
\intertext{Servers are the only prefix with more than one axiom yielding three cases.}
	\densem{\server{x}{x'}{P}} 
	\array[t]{rl}
	  {=} & 
	  \antider{\lcouse{x}{x'}}{\densem{P}} + \antider{\lcodisp{x}}{\densem{\clientdisp{z_1}{\cdots\clientdisp{z_n}{\close{x}{\nil}}}}} {} + \\
	  {+} & 
		\antider{\lcodup{x}{x_1}{x_2}}{\densem{\clientdup{z_1}{z_1\sigma_1}{z_1\sigma_2}{\dots\clientdup{z_n}{z_n\sigma_1}{z_n\sigma_2}{\send{x_1}{x_2}{(\server{x_1}{x'\sigma_1}{P_1} \pp \server{x_2}{x'\sigma_2}{P_2})}}}}}
	\endarray
\end{gather*}
for $\{x',z_1,\ldots,z_n\} = \fn(P)$ and any $P_1$ and $P_2$ such that 
$P_1 = P\sigma_1$, $P_2 = P\sigma_2$, and $\fn(P_1) \cap \fn(P_2) = \emptyset$.

Antiderivation also serves as a sanity check for the rule giving the denotations of links:
\[
  \densem{\forward{x}{y}} = \antider{\lforward{x}{y}}{\densem{\nil}}
  \text{.}
\]

\begin{definition}[Denotational equivalence]
Denotational equivalence $\deneq$ is the relation on well-typed processes s.t.~$P \deneq Q$ whenever $\densem{P} = \densem{Q}$.
\end{definition}

Denotational equivalence implies type equivalence.
\begin{proposition}
 \label{thm:denotational-equivalence-type-sound}
If $P \deneq Q$, then $\judge{P}{\hyp{G}}$ iff $\judge{Q}{\hyp{G}}$.
\end{proposition}

\paragraph{Discussion}
In general, our antiderivatives respect Fubini's theorem for double antiderivation \citep{F07}. This formalises the intuition that the order of independent actions is not discriminated.
\begin{theorem}[Order-invariance]
\label{thm:antiderivation-fubini}
  For $l_x$, $l_y$, and $P$ s.t.~$\separate{P}{x}{y}$
  $\antider{l_x}{\antider{l_y}{\densem{P}}} = \antider{l_y}{\antider{l_x}{\densem{P}}} = \antider{\lsync{l_x}{l_y}}{\densem{P}}$.\par
\end{theorem}

\begin{proofatend}
  By inspection of each case in \cref{def:antiderivative}, if antiderivation \wrt to $l_z$ affects any channel $z'$ besides $z$, then $z'$ belongs to the environment of $z$.
\end{proofatend}

\begin{example}
  Consider the processes described in \cref{ex:and-server}. The type of bits is $\one \oplus \one$. The terms introduced for bit transmission have the following denotations, parametric in those of $P$ and $Q$ (we omit types, since they do not matter for our argument here).
  \begin{gather*}
    \densem{x[0].P}
    = \antider{\lsend{x}{y}}{
      \antider{\linl{y}}{
        \antider{\lclose{y}}{
          \densem{P}}}}
    =  \{
    g \tdot \langle \gamma \tdot \langle x \mapsto (\denout, \_, (\denout, \_, \minl, (\denout, \_, \ast) ), a )\rangle\rangle \mid
    g \tdot \langle \gamma \tdot \langle x \mapsto a \rangle\rangle \in \densem{P}\}\\
    \densem{x[1].P}
    = \antider{\lsend{x}{y}}{
      \antider{\linr{y}}{
        \antider{\lclose{y}}{
          \densem{P}}}}
    = \{
    g \tdot \langle \gamma \tdot \langle x \mapsto (\denout, \_, (\denout, \_, \minr, (\denout, \_, \ast), a ))\rangle\rangle \mid
    g \tdot \langle \gamma \tdot \langle x \mapsto a \rangle\rangle \in \densem{P}\}\\
    \densem{\choicebit{x}{P}{Q}} =
    \antider{\linl x}{\densem{P}} + \antider{\linr x}{\densem{Q}}
  \end{gather*}
  Below is the semantics of a client sending $0$ and $1$:
  \[\densem{Client^{01}_{xz}} = \left\{
    \mspace{2mu}\left\langle\left\langle\mspace{-5mu}\array{l}
    z \mapsto (\denout,n, b, (\denout, \ast) )\text{, }
    x \mapsto (\denout,\_,  \langle 
      (\denout,\_,  (\denout,\_,  \minl, (\denout, \ast)), 
      	\\
        (\denout,\_,  (\denout,\_,  \minr, (\denout, \ast)), 
          (\denin, \_, \{n\}, b, (\denin, \ast)) 
      )) \rangle )
   \endarray\mspace{-5mu}\right\rangle\right\rangle\mspace{2mu}
   \,\middle|\, b \in \{\minl, \minr\}
   \right\}\]
   Observe how the order of transmission of each bit is preserved in the structure of the denotation for $x$, that the reply $b$ appears in the denotations of both $x$ and $z$, and that the semantics includes all possible replies the client may receive. As a consequence, the semantics captures the fact that the client echoes the server reply, whatever that may be.
\end{example}

\begin{remark}
Consider $P_1 = \clientdup{x}{x_1}{x_2}{\clientdisp{x_1}{\clientuse{x_2}{x'}{Q}}}$
and 
$P_2 = \clientdup{x}{x_1}{x_2}{\clientuse{x_1}{x'}{\clientdisp{x_2}{Q}}}$.
Our behavioural equivalences (bisimilarity, barbed congruence, and denotational equivalence) discriminate these two processes, because we can observe how each copy ($x_1$ and $x_2$) is used.
We could have a coarser behavioural theory where copies of the same server are not distinguished, by adopting the following transition rules that non-deterministically assign the names $x_1$ and $x_2$ to the copies.
\begin{highlight}
\begin{gather*}
	\infer
		{\phl{\clientdup{x}{x_1}{x_2}{P}} \lto{\phl{\ldup{x}{y_1}{y_2}}} \phl{\recv{y_1}{y_2}{P}}}
		{\phl{\{x_1,x_2\} = \{y_1,y_2\}}}
	\quad
	\infer
		{\phl{\clientdup{x}{x_1}{x_2}{P}} \lto{\lsync{\phl{\ldup{x}{y_1}{y_2}}}{\phl{l}}} \phl{\recv{y_1}{y_2}{P'}}}
		{\phl{\clientdup{x}{x_1}{x_2}{P}} \lto{\phl l} \phl{\recv{y_1}{y_2}{P'}}
		&\phl{\{x_1,x_2\} = \{y_1,y_2\}}
		&\phl{\separate{\clientdup{x}{x_1}{x_2}{P}}{x}{\fn(l)}}}
\end{gather*}
\end{highlight}
Likewise, it suffices to adopt trees with unordered branching in non-linear observability predicates and interpretations of exponentials in the denotational semantics---just regard $\dupnode{a_1}{a_2}$ as an unordered pair.
\end{remark}

\begin{remark}
\citet{A17} proposed a denotational semantics for Classical Processes (CP) \citep{W14} that is coarser than ours when it comes to observing client requests.
We give a representative example. Consider the processes $P_1 = \clientuse{x}{x'}{Q}$
and $P_2 = \clientdup{x}{x_1}{x_2}{\clientdisp{x_1}{\clientuse{x_2}{x'}{Q}}}$. The difference is that $P_1$ uses a single instance of the server at $x$, whereas $P_2$ first duplicates the server at $x$ and then uses only one of the two copies ($x_2$), discarding the other copy.
In \citep{A17}, $P_1$ and $P_2$ are observationally and denotationally equivalent. This makes sense because weakening and contraction do not have corresponding terms in CP (so $P_1$ and $P_2$ would essentially be syntactically equivalent). Differently, our explicit terms and their semantics clearly show that $P_2$ makes the context (the server) ``work more'' in \HCP, which motivates the distinction between $P_1$ and $P_2$.
If we wished for the same coarse equivalence as in \citep{A17}, we could exploit delayed actions for cancelling out unnecessary duplications whenever the continuation would dispose of one of the copies, as in the following rules. These would align bisimilarity and barbed congruence with Atkey's interpretation of exponentials (at the cost of a more complex lts).
\begin{highlight}
\[
    \infer[\rname{C1}]
        {\phl{\clientdup{x}{x_1}{x_2}{P}} \lto{\phl{\tau}} \phl{P'}}
        {\phl{P} \lto{\phl{\ldisp{y}}} \phl{P'} &\phl{y \in \{x_1,x_2\}}}
    \qquad
    \infer[\rname{C2}]
        {\phl{\clientdup{x}{x_1}{x_2}{P}} \lto{\phl{\ldup{x}{y_1}{y_2}}} \phl{\recv{y_1}{y_2}{P}}}
        {\phl{P} \notlto{\phl{\ldisp{y_i}}}
        &\phl{\{y_1,y_2\} = \{x_1,x_2\}}}
\]
\end{highlight}
\end{remark}

\section{Full Abstraction}
\label{sec:full-abstraction}

Barbed congruence implies denotational equivalence: given the element $g$ of 
$\densem{P}$, there is a family of contexts $\mathbb{C}$ with the property that $g \in \densem{Q}$ whenever $C[P]$ and $C[Q]$ satisfy the same observability predicates for every $C \in \mathbb{C}$. 
\begin{theorem}
	\label{thm:hcp-bcong-deneq}
	${\bcong} \subseteq {\deneq}$.
\end{theorem}

\begin{proofatend}
	As seen in \cref{ex:hcp-bcong-additives}, we can use client requests to observe branching. To this end we define the following ``probes'' \ie contexts that are type equivalent (both have type $\cht{s}{\bot\parr\bot},\cht{t}{\one \otimes \one}$) but observationally distinct in their use of $t$:
	\begin{align*}
		L_{s} = {}& 
			\choice{s}
				{\close{s}{\send{t}{t'}{\wait{t}{\wait{t'}{\nil}}}}}
				{\close{s}{\send{t}{t'}{\wait{t}{\wait{t'}{\nil}}}}}\\
		R_{s} = {}&
				\send{t}{t'}{\wait{t}{\choice{s}
					{\close{s}{\wait{t'}{\nil}}}
					{\close{s}{\wait{t'}{\nil}}}
				}}
	\end{align*}
	Given a denotation $g$ we build a family of typed contexts $\mathbb{C}$ with the property that if for any context $C \in \mathbb{C}$, the processes $C[P]$ and $C[Q]$ satisfy the same observability predicates then $g \in \densem{P} \cap  \densem{Q}$. We define $\mathbb{C}$ as $\chi(g)$ where the function $\chi$ is given below by recursion on $g$. Intuitively, we build a context as the parallel composition of independent ``test'' processes, one for each name in $g$ (\ie each name in the type of the hole). Test processes are built by $\xi$.
	\begin{align*}
		\chi(\langle\rangle)={}&\left\{
			\hole
		\right\} \\
		\chi(g\tdot\langle\gamma\tdot\langle x\mapsto a \rangle\rangle)={}&\left\{
			\res{xy}{(C \pp T)}
		\,\middle|\,
			C \in \chi(g\tdot\langle\gamma),
			T \in \xi(y \mapsto a'),
			a \dendual a'
		\right\} \\
		\xi(y \mapsto (\denout,\_,\ast))={}&\left\{
			\forward{y}{t}
		\right\}\\
		\xi(y \mapsto (\denin,\_,\ast))={}&\left\{
			\forward{y}{t}
		\right\}\\
		\xi(y \mapsto (\denout,\_,\_,a,\_,b))={}&\left\{
			\send{y}{y'}{(T \pp T')}
		\,\middle|\,
			T  \in \xi(t  \mapsto a),
			T' \in \xi(t' \mapsto b)
		\right\}\\
		\xi(y \mapsto (\denin,\_,a,b))={}&\left\{
			\recv{y}{y'}{(\send{t}{t'}{(\wait{t}{T} \pp \wait{t'}{T'})})}
		\,\middle|\,
			T  \in \xi(y  \mapsto a),
			T' \in \xi(y' \mapsto b)
		\right\}\\
		\xi(y \mapsto (\denout,\_,\minl,a))={}&\left\{
			\inl{y}{T}
		\,\middle|\,
			T  \in \xi(y \mapsto a),
		\right\}\\
		\xi(y \mapsto (\denin,\_,\minl,a))={}&\left\{
			\choice{y}
				{L_s[T]}
				{R_s[\forward yt]}
		\,\middle|\,
			T  \in \xi(y \mapsto a)
		\right\}\\
		\xi(y \mapsto (\denout,\_,\minr,b))={}&\left\{
			\inr{y}{T}
		\,\middle|\,
			T  \in \xi(y \mapsto b),
		\right\}\\
		\xi(y \mapsto (\denin,\_,\_,\minr,b))={}&\left\{
			\choice{y}
				{L_s[\forward yt]}
				{R_s[T]}
		\,\middle|\,
			T  \in \xi(y \mapsto b)
		\right\}\\
		\xi(y \mapsto (\denout,\_,\dispnode))={}&\left\{
			\clientdisp{y}{\close{t}{\nil}}
		\right\}\\
		\xi(y \mapsto (\denout,\_,\usenode[a]))={}&\{
			\clientuse{y}{y'}{T} \mid
			T  \in \xi(y'\mapsto a)\}\\
		\xi(y \mapsto (\denout,\_,\dupnode{a_1}{a_2}))={}&\left\{
			\clientdup{y}{y_1}{y_2}{
				(\send{t}{t'}{(\wait{t}{T_1} \pp \wait{t'}{T_2})})
			}
		\,\middle|\,
			\array{l}
			T_1 \in \xi(y_1\mapsto a_1),\\
			T_2 \in \xi(y_2\mapsto a_2)
			\endarray
		\right\}\\
		\xi(y \mapsto (\denin,\_,\_,\dispnode))={}&\left\{
			\forward{y}{t}
		\right\}\\
		\xi(y \mapsto (\denin,\_,\_,\usenode[a]))={}&\{
			\server{y}{y'}{T} \mid
			T  \in \xi(y'\mapsto a)\} \cup \{\forward{y}{t}\}\\
		\xi(y \mapsto (\denin,\_,\_,\dupnode{a_1}{a_2}))={}&\left\{
			\server{y}{y'}{T}
		\,\middle|\,
			T \in \xi(y_1\mapsto a_1) \cup \xi(y_2\mapsto a_2)
		\right\} \cup \{\forward{y}{t}\}
	\end{align*}
	We observe that every case matches the type constructor of the channel up to auxiliary non-blocking operations on fresh names, namely $\sends{t}{t'}$, $\waits{t}$, $\waits{t'}$, $\closes{t}$---needed to respect typing. Because of their non-blocking nature these do not shadow any observable for the tested name $y$. Branching is observed using probes $L_s$ and $L_r$.
	In the case of servers we generate a context for each possible use $a$ and a forwarding context which allows us to observe the server use on the free name $t$. It follows from the type theory of \HCP that single uses of servers may differ only for selects, choices, or client operations.
	\begin{itemize}
		\item
		In the first case, the process built by $\xi$ makes a select on $y$ and by construction of $g$ there there an instance with the symmetric select meaning that the other branch is covered by associated test.
		\item
		In the second case, the process built by $\xi$ offers a choice where, both branches are instrumented with probes, one branch implements the desired observation and the other forwards the channel.
		\item
		In the third case the process built by $\xi$ interacts with a with a server offered on the same channel. This server must implement every behaviour in the denotation hence covered by $\xi$.
	\end{itemize}
	Therefore, tests jointly cover all cases with observationally distinct behaviours for branches (choices and servers).
\end{proofatend}

Recall from \cref{sec:hcp-denotational} that taking derivatives of denotations with respect to an observable yields the denotations that can follow that actions.
This interpretation suggests to define a labelled transition system on process denotations where transitions correspond to derivation.

\begin{definition}
	\label{def:lts-denotations}
	The \lts of denotations has state-space the image of $\densem{-}$ and transition relation the relation $D \lto l \der{l}{D}$.
\end{definition}

There is an operational correspondence between the \lts of processes and that of denotations.
\begin{lemma} 
\label{thm:accord}
Let $P$ be well-typed. Then:
\begin{itemize}
\item	If $P \lto{l} P'$, then $\densem{P} \lto{l} \densem{P'}$.
\item If $\densem{P} \lto{l} \densem{Q}$, then $P \slto{l} P'$ for some $P' \deneq Q$.
\end{itemize}
\end{lemma}

\begin{proofatend}
	Both directions are proved by induction on the transition derivation using \cref{thm:antiderivation-fubini} to take care of self-synchronising actions.
\end{proofatend}

As a consequence, a process and its denotation are bisimilar and hence denotationally equivalent processes are bisimilar.

\begin{theorem}
	\label{thm:hcp-deneq-bis}
	${\deneq} \subseteq {\bis}$.
\end{theorem}

\begin{proofatend}
	We observe that the relation $=$ is a bisimulation on the \lts of denotations.
	Let $\crel{R}$ be the graph of $\densem{-}$ ($P \crel{R} D$ iff $\densem{P} = D$).
	We claim that $\crel{R}$ is a bisimulation.
	It follows from our claim that $\crel{R} \circ = \circ \crel{R}^{-1}$ is a bisimulation on the \lts of \HCP processes hence this relation is contained in the bisimilarity relation $\bis$---as any other bisimulation relation.
	The thesis follows readily since, from $\densem{P} = \densem{Q}$ we conclude that $P \crel{R} \circ = \circ \crel{R}^{-1} Q$ and from this and the above that $P \bis Q$.
	Finally, we observe that our claim follows from the operational correspondence stated in \cref{thm:accord}.
\end{proofatend}

By \cref{thm:hcp-bis-bcong,thm:hcp-bcong-deneq,thm:hcp-deneq-bis}, all three observational 
equivalences agree on well-typed programs.
\begin{theorem}[Full abstraction]
	\label{thm:full-abstraction}
	For well-typed processes, ${\bis} = {\bcong} = {\deneq}$.
\end{theorem}

\begin{proofatend}
 The thesis follows by combining \cref{thm:hcp-bis-bcong,thm:hcp-bcong-deneq,thm:hcp-deneq-bis}.
\end{proofatend}

\section{Related Work}
\label{sec:related}

\HCP stands on the shoulders of the prior work by
\citet{W14} on Classical Processes (\CP), which built on the works by
\citet{CP10}, \citet{A94}, and \citet{BS94} on ``Proofs as Processes''. We have already discussed the distinctive points of \HCP in the Introduction.

\citet{PCPT14} introduced \emph{typed context bisimilarity} as an
observational equivalence for processes typed with intuitionistic linear logic \citep{CP10}. The idea is to name one of the free channels of a process as ``the'' representative channel and the others as ``requirements''. Then, typed context bisimilarity equates two processes if they perform the same actions on the representative channel once all their requirements are provided (using contexts).
By contrast, our bisimilarity is standard: it does not require types or any distinction of the roles of channels, and it does not require reasoning on contexts. Also, we validated it with full abstraction.

\citet{A17} explored the first notion of observation for \CP. To cope
with the lack of an \lts, his contexts for contextual equivalence are not processes, but rather special
configuration terms designed to make communication reductions visible. Our approach is simpler, 
because we can just look at
transitions with observable actions using our \lts. Studying the behavioural theory of \HCP
does not require configurations, and we showed how a fully-abstract denotational
semantics can be formulated based on delayed actions.

In the original presentation of delayed actions by \citet{MS04}, normal
(non-delayed) action prefixes are syntactically distinguished from delayed
action prefixes (using the prefixing symbol ``$:$'' instead of ``$.$''). The
same distinction could be introduced to \HCP, adding extra machinery on top of
our proof theory to preserve full abstraction. Namely, we would need to support
enforcing sequentiality between independent channels at the same process. This
could be done by extending \HCP with safe circular dependencies, \eg, following the studies by \citep{CMSY17} and \citep{DG18}. We chose our formulation for economy of the calculus.

In connection to our delayed actions, \citet{BS94} formulated syntactic permutations of 
independent actions to match permutations of rule applications in linear logic. Later, 
\citet{DCPT12} proposed to leverage commuting conversions in intuitionistic linear logic by 
assigning asynchronous process terms to proofs, but this makes terms even more compound---their term 
for output (corresponding to rule $\tensor$), in our syntax, is $x[y] \pp P \pp Q$, denoting that 
continuations may perform some actions before $x[y]$ is executed.

\section{Conclusions and Future Work}
\label{sec:conclusions}

The keystone of process algebras is the parallel operator $P \pp Q$, but up to this work the connection between parallel and linear logic was indirect---no proof-theoretical inference rule was a straight match for parallel composition, but rather included parallel as part of more complex terms. The consequence for the agenda of ``Proofs as Processes'' \citep{A94}: either accept that processes have sound structures and equivalences that cannot be captured by the proof theory, as done by \citet{CP10}, or give up on parallel as an independent operator, as done by \citet{W14}.
\HCP cuts the head off the snake by using hyperenvironments to represent parallelism in 
sequents, internalised by the connective $\tensor$.
We can now have our cake and eat it too: the semantics of well-typed processes is completely 
captured by a proof theory rooted in linear logic, and $P \pp Q$ is an operator in its own right 
respecting the expected equational laws.
As a first step in taking advantage of this unification, we linked the ideas developed by 
\citet{A17} for a denotational semantics for Classical Processes (\CP) \citep{W14} to the standard 
theory of bisimilarity.

The design of \HCP focused on the basic features of \CP.
In the future, we intend to study principled extensions that allow for capturing more behaviours.

\citet{LM16} extended \CP with recursive types and their accompanying reductions. Adding the 
corresponding transitions to our \lts seems straightforward, and our approach with derivatives to 
the definition of denotations might offer insight in how the denotational semantics of \HCP can be 
extended to capture recursion.
Similar considerations apply for non-determinism, introduced to intuitionistic linear logic by \citet{CP17}.

Another interesting feature would be process mobility, \ie the ability to communicate code. \citet{TCP13} obtained this for intuitionistic linear logic using a monadic integration with a functional programming model. \citet{M18} extended \CP with higher-order communications directly, by using environments as types for process variables. Thus, the obvious starting point to extend \HCP to process mobility would be to type process variables with hyperenvironments.
Historically, constructing tractable and characteristic behavioural equivalences for higher-order process calculi has been nontrivial in general \citep{SKS11}. The \lts of \HCP (extended to higher-order communications) might be helpful in this regard, because we can adapt bisimilarity notions from studies on session types \citep{KPY17}.

One of the most important extensions of session types is Multiparty Session Types, by \citet{HYC16}.
\citet{CMSY17} showed that multiparty session types can be captured in \CP by generalising the notion of duality to \emph{coherence}, which checks for the compatibility of multiple types. This yields a generalised cut rule that can compose many processes instead of two. In \HCP, this would correspond to extending \cref{rule:cut} to the cutting of multiple environments.

\begin{acks}
  We are grateful to P.~Wadler, S.~Lindley, and the anonymous reviewers for their comments.
  This work was partially supported by
  the \grantsponsor{IRFD}{Independent Research Fund Denmark}{https://dff.dk/en}, grant no.\ 
  \grantnum{IRFD}{DFF-7014-00041}, and the \grantsponsor{EPSRC}{Engineering and Physical Sciences Research Council}{https://epsrc.ukri.org/}, grant no.\ \grantnum{Engineering and Physical Sciences Research Council}{EP/L01503X/1}.
\end{acks}

\appendix

\section{Omitted Rules}\label{sec:omitted-rules}

\begin{highlight}
\begin{spreadlines}{\proofltsskipamount}
\begin{gather*}
\rulescalebox{
\def\regh{\vphantom{\pp\hyp{H}'}}
\infer[\rlabel{\rname{Par$_2$}}{rule:mix-r}]
	{\plto
	  {\infer[\ref{rule:mix}]
	  	{\judge{P \pp Q}{\hyp{G} \pp \hyp{H}}\regh}
		  {\judge{P}{\hyp{G}}&\judge{Q}{\hyp{H}}\regh}}
	  {l}
	  {\infer[\ref{rule:mix}]
	  	{\judge{P \pp Q'}{\hyp{G} \pp \hyp{H}'}\regh}		  
{\judge{P}{\hyp{G}}&\judge{Q'}{\hyp{H}'}\regh}}}
	{\plto{\judge{Q}{\hyp{H}}\regh}{l}{\judge{Q'}{\hyp{H'}}\regh}
 	&\phl{\bn(l) \cap \fn(P) = \emptyset}}
}\qquad 
\rulescalebox{
\infer[\rlabel{\rname{$\aleq$}}{rule:alpha}]
	{\plto
	  {\judge{P}{\hyp{G}}}
	  {l}
	  {\judge{Q'}{\hyp{G'}}}}
	{\phl{P \aleq Q}
	&\plto
	  {\judge{Q}{\hyp{G}}}
	  {l}
	  {\judge{Q'}{\hyp{G'}}}}
}\\ \rulescalebox{
\infer[\rlabel{\delayedrule{rule:one}}{rule:one-delayed}]
	{\plto
		{\infer[\ref{rule:one}]
	    {\judge{\close xP}{\hyp{G} \pp \cht{x}{\one}}}
	    {\judge{P}{\hyp{G}}}}
		{l}
		{\infer[\ref{rule:one}]
	    {\judge{\close x{P'}}{\hyp{G'} \pp \cht{x}{\one}}}
	    {\judge{P'}{\hyp{G'}}}}}
	{\plto
		{\judge{P}{\hyp{G}}}
		{l}
		{\judge{P'}{\hyp{G'}}}
		&\phl{x \notin \cn(l)}}
}\quad\rulescalebox{
\infer[\rlabel{\selfsyncrule{rule:one}}{rule:one-sync}]
	{\plto
		{\infer[\ref{rule:one}]
	    {\judge{\close xP}{\hyp{G} \pp \cht{x}{\one}}}
	    {\judge{P}{\hyp{G}}}}
		{\lsync{\plbl{\lclose{x}}{\lone}}{l}}
		{\judge{P'}{\hyp{G'}}}}
	{\plto
		{\judge{P}{\hyp{G}}}
		{l}
		{\judge{P'}{\hyp{G'}}}
		&\phl{x \notin \cn(l)}}
}\\ 
\rulescalebox{
\plto
	{\infer[\ref{rule:oplus_2}]
		{\judge{\inr xP}{\Gamma, \cht{x}{A \oplus B}}}
		{\judge{P}{\Gamma, \cht{x}{B}}}}
	{\plbl{\linr{x}}{\loplusr{A}{B}}}
	{\judge{P}{\Gamma, \cht{x}{B}}}
}\\ \rulescalebox{
\plto
	{\infer[\ref{rule:with}]
		{\judge{\choice xPQ}{\Gamma, \cht{x}{A \with B}}}
		{\judge{P}{\Gamma, \cht{x}{A}} &
		\judge{Q}{\Gamma, \cht{x}{B}}}}
	{\plbl{\lcoinr{x}}{\lwithr{A}{B}}}
	{\judge{Q}{\Gamma, \cht{x}{B}}}
}\\ \rulescalebox{
\infer[\rlabel{\ref*{rule:oplus_2}\ref*{rule:with}}{rule:cut-oplus_2}]
	{\plto
	  {\infer[\ref{rule:cut}] 
     {\judge{\res{xy}{P}}{\hyp{G} \pp \Gamma, \Delta}}
     {\judge{P}{\hyp{G}\pp\Gamma,\cht{x}{A \tensor B} \pp \Delta, \cht{y}{\dual{A} \parr \dual{B}}}}}
	  {\tau}
	  {\infer[\ref{rule:cut}] 
     {\judge{\res{xy}{P'}}{\hyp{G} \pp \Gamma, \Delta}}
     {\judge{P'}{\hyp{G} \pp \Gamma, \cht{x}{B} \pp \Delta, \cht{y}{\dual{B}}}}}
  }{\plto
		{\judge{P}{\hyp{G}\pp\Gamma, \cht{x}{A \oplus B} \pp \Delta, \cht{y}{\dual{A} \with \dual{B}}}}
		{\lsync
      {\plbl{\linr{x}}{\loplusr{A}{B}}}
      {\plbl{\lcoinr{y}}{\lwithr{\dual{A}}{\dual{B}}}}}
		{\judge{P'}{\hyp{G} \pp \Gamma, \cht{x}{B} \pp \Delta, \cht{y}{\dual{B}}}}}
}\\ \rulescalebox{
\infer[\rlabel{\delayedrule{rule:oplus_2}}{rule:oplus_2-delayed}]
	{\plto
		{\infer[\ref{rule:oplus_2}]
	    {\judge{\inr{x}{P}}{\hyp{G} \pp \Gamma, \cht{x}{A \oplus B}}}
	    {\judge{P}{\hyp{G} \pp \Gamma, \cht{x}{B}}}}
		{l}
		{\infer[\ref{rule:oplus_2}]
			{\judge{\inr{x}{P'}}{\hyp{G'} \pp \Gamma, \cht{x}{A \oplus B}}}
			{\judge{P'}{\hyp{G'} \pp \Gamma, \cht{x}{B}}}}}
	{\plto
		{\judge{P}{\hyp{G} \pp \Gamma, \cht{x}{B}}}
		{l}
		{\judge{P'}{\hyp{G'} \pp \Gamma, \cht{x}{B}}}
		&\phl{x \notin \cn(l)}}
}\\ \rulescalebox{
\infer[\rlabel{\selfsyncrule{rule:oplus_2}}{rule:oplus_2-sync}]
	{\plto
		{\infer[\ref{rule:oplus_2}]
	    {\judge{\inr{x}{P}}{\hyp{G} \pp \Gamma, \cht{x}{A \oplus B}}}
	    {\judge{P}{\hyp{G} \pp \Gamma, \cht{x}{B}}}}
		{\lsync{\plbl{\linr{x}}{\loplusr{A}{B}}}{l}}
		{\judge{P'}{\hyp{G'} \pp \Gamma', \cht{x}{B}}}}
	{\plto
		{\judge{P}{\hyp{G} \pp \Gamma, \cht{x}{B}}}
		{l}
		{\judge{P'}{\hyp{G'} \pp \Gamma', \cht{x}{B}}}
		&\phl{x \notin \cn(l)}
		&\phl{\separate{\inr{x}{P}}{x}{\fn(l)}}}
}\\ \rulescalebox{
\infer[\rlabel{\delayedrule{rule:weaken}}{rule:weaken-delayed}]
	{\plto
  	{\infer[\ref{rule:weaken}]
  		{\judge{\clientdisp{x}{P}}{\hyp{G} \pp \Gamma, \cht{x}{\query A}}}
  		{\judge{P}{\hyp{G} \pp \Gamma}}}
  	{l}
  	{\infer[\ref{rule:weaken}]
  		{\judge{\clientdisp{x}{P'}}{\hyp{G'} \pp \Gamma', \cht{x}{\query A}}}
  		{\judge{P'}{\hyp{G'} \pp \Gamma'}}}}
	{\plto
		{\judge{P}{\hyp{G} \pp \Gamma}}
		{l}
		{\judge{P'}{\hyp{G'} \pp \Gamma'}}
		&\phl{x \notin \cn(l)}
		&\phl{\fn(P') \neq \emptyset}}
}\\ \rulescalebox{
\infer[\rlabel{\selfsyncrule{rule:weaken}}{rule:weaken-sync}]
	{\plto
  	{\infer[\ref{rule:weaken}]
  		{\judge{\clientdisp{x}{P}}{\hyp{G} \pp \Gamma, \cht{x}{\query A}}}
  		{\judge{P}{\hyp{G} \pp \Gamma}}}
  	{\lsync{\plbl{\ldisp{x}}{\lweaken{A}}}{l}}
  	{\infer[\ref{rule:bot}]
  		{\judge{\wait{x}{P'}}{\hyp{G'}\pp\Gamma',\cht{x}{\bot}}}
  		{\judge{P'}{\hyp{G'}\pp\Gamma'}}}}
	{\plto
		{\judge{P}{\hyp{G} \pp \Gamma}}
		{l}
		{\judge{P'}{\hyp{G'} \pp \Gamma'}}
		&\phl{x \notin \cn(l)}
		&\phl{\separate{\clientdisp{x}{P}}{x}{\fn(l)}}}
}\end{gather*}
\end{spreadlines}
\end{highlight}

\omittedproofs[\section{Omitted Proofs}\label{sec:proofs}]

\end{document}